\documentclass[reprint,onecolumn,12pt,secnumarabic,amssymb, nobibnotes, aps, prd, longbibliography, notitlepage, superscriptaddress, nofootinbib]{revtex4-1}
\usepackage{newfloat,algcompatible}
\usepackage{setspace}
\usepackage{etoolbox}
\usepackage{ragged2e}
\usepackage[usenames,dvipsnames,svgnames,table]{xcolor}
\usepackage{tcolorbox}
\AtBeginEnvironment{algorithm}{\noindent\hrulefill\par\nobreak\vskip-5pt}
\usepackage{indentfirst}

\usepackage{amsmath}
\usepackage[colorlinks=true,allcolors=Blue]{hyperref}

\DeclareFloatingEnvironment[
    fileext=loa,
    listname=List of Algorithms,
    name=ALGORITHM,
    placement=tbhp,
]{algorithm}

\usepackage{bm}
\usepackage[caption=false]{subfig}
\usepackage{graphicx,epstopdf}
\usepackage{cleveref}
\usepackage{natbib}
\usepackage{float}
\usepackage{placeins}
\usepackage{multirow}
\usepackage[toc,page]{appendix}

\crefformat{section}{\S#2#1#3}

\begin{document}
\title{High throughput, automated prediction of focusing patterns for inertial microfluidics}

\author{Aditya Kommajosula}
\thanks{\scriptsize These two authors contributed equally}
\affiliation{Mechanical Engineering, Iowa State University, Ames, IA 50011, USA}
\author{Jeong-ah Kim}
\thanks{\scriptsize These two authors contributed equally}
\affiliation{Graduate School of Nanoscience and Technology, Korea Advanced Institute of Science and Technology, Daejeon 34141, Republic of Korea}
\author{Wonhee Lee}
\email[\scriptsize Co-corresponding author: ]{whlee153@kaist.ac.kr}
\affiliation{Graduate School of Nanoscience and Technology, Korea Advanced Institute of Science and Technology, Daejeon 34141, Republic of Korea}
\affiliation{Department of Physics, Korea Advanced Institute of Science and Technology, Daejeon 34141, Republic of Korea}
\author{Baskar Ganapathysubramanian}
\email[\scriptsize Co-corresponding author: ]{baskarg@iastate.edu}
\affiliation{Mechanical Engineering, Iowa State University, Ames, IA 50011, USA}
\renewcommand{\abstractname}{\vspace{-\baselineskip}}

\begin{abstract}

Identifying focusing patterns in arbitrarily cross-sectioned channels is an interesting, significant, and, complex problem in applications involving microfluidic sorting, separation, and ordering. Current computational approaches involve construction of cross-sectional ``force-maps'' followed by a visual identification to confirm the presence of experimentally-observed stable points [D. Di Carlo et. al., Physical review letters, \textbf{102}, 094503 (2009)]. Such visual inspections are naturally prone to misinterpreting stable locations and focusing shifts in the case of non-trivial focusing patterns. We develop and deploy an approach for automating the calculation of focusing patterns for a general channel geometry, and thereby reduce the dependence on empirical/visual procedures to confirm the presence of stable locations. We utilize concepts from interpolation theory (to represent continuous force-fields using discrete points), and stability theory to identify ``basins of attraction" and quantitatively identify stable equilibrium points. Our computational experiments reveal that predicting equilibrium points accurately requires upto $\times$10-20 times more refined force-maps that conventionally used, which highlights the spatial resolution required for an accurate representation of cross-sectional forces. These focusing patterns are validated using experimental results for a rectangular channel, and triangular channel with an apex angle of $90^\circ$. We then apply the approach to predict and explain focusing patterns and shifts for a $90^\circ$-isosceles triangular channel across a range of Reynolds numbers for ${\color{magenta} \frac{a}{H}} = 0.4$ (particle-to-channel size ratio). We observe that the predicted focusing patterns match experiments well. The force-maps also reveal certain ``clouds'' of localized stable points, which aid in explaining the onset of bifurcation observed in experiments. The current algorithm is agnostic to channel cross-sections and straight/curved channels, which could pave way to generating a library of focusing patterns as a function of channel geometry, and {\color{magenta} $Re$}, to assist in design of novel devices for tailored particle-streams.
\end{abstract}

\maketitle


\section{Introduction}\label{sec:intro}

Lateral focusing of spherical particles to off-centre locations in inertial flow is known as the \textit{tubular pinch} effect. The experimental discovery of this phenomenon   \cite{segre1961radial} has motivated rigorous study into understanding the dynamics of dilute suspensions in flow. This behavior is particularly attractive in the field of inertial microfluidics  \cite{amini2014inertial} for manipulation of finite-sized particles. Inertial microfluidics is a laminar regime of microfluidic-physics characterized by finite {\color{magenta} $Re$}, and hence, non-zero fluid inertia. The finite {\color{magenta} $Re$} brings about an interesting interplay between competing fluid forces on a free particle, and depending on particle-size (or alternatively, {\color{magenta} $\frac{a}{H}$} for a fixed particle-size), {\color{magenta} $Re$}  \cite{yang2005migration} and release-location, the particle reaches a steady lateral location. Dominant inertial forces at play are the, \textit{shear-gradient lift}: this is attributed to the curvature of the velocity profile in the channel, which directs particles away from the channel centre, and, \textit{wall-induced lift}: this force arises due to the interaction between the particle and the adjacent wall and acts to drive particles away from the wall. This inertial migration is unique for a given channel geometry, flow speed, and {\color{magenta} $\frac{a}{H}$}, and has been widely employed for passive particle-manipulation in cell-focusing, sorting, and ordering
applications  \cite{DiCarlo2007},  \cite{DiCarlo2009},  \cite{Gossett2010}. Passive manipulation is a simple, robust, and relatively high-throughput (compared to active manipulation) class of techniques which depends solely on the hydrodynamic forces of interaction inherent to a given configuration or channel geometry.\footnote{\tiny Aside for inertial migration, other common passive techniques  \cite{zhang2016fundamentals} rely on use of co-flow or additional sheath-fluid to guide  \cite{yamada2004pinched},  \cite{gossett2012inertial}/confine  \cite{huh2005microfluidics},  \cite{lee2006hydrodynamic}/encapsulate  \cite{howell2008two} particle-samples. Active techniques on the other hand, involve the use of external manipulation forces such as, dielectrophoresis (electric fields on dielectric particles), magnetophoresis (magnetic fields on magnetic particles), acoustophoresis (sound radiation), and optical tweezers (dielectric particles in a laser beam)  \cite{ccetin2011dielectrophoresis},  \cite{forbes2012microfluidic},  \cite{wang2011recent},  \cite{grier2003revolution}.} A subset of passive techniques involves the use of boundary-induced secondary flows in addition to flow in a primary direction. The net effect of this secondary flow or vorticity serves to alter final focusing locations. Novel developments in this regard include the use of channels with a series of constrictions  \cite{chung2013three}, grooves or “herringbones”  \cite{stroock2002chaotic}, micro-structures  \cite{chung2013microstructure},  \cite{inglis2006critical},  \cite{inglis2009efficient},  \cite{huang2004continuous}, and channel curvatures  \cite{gossett2009particle},  \cite{martel2012inertial},  \cite{russom2009differential}.

Another subset of sheathless manipulation techniques relies on the channel geometry itself as a controlling-parameter, and this forms the motivation for the current study. Current experimental studies on particle-focusing so far have successfully reported the competing flow-physics pertaining to observed trends and supplemented them with numerical simulations. The approach in these studies  \cite{DiCarlo2009},  \cite{liu2015inertial},  \cite{Kim2016},  \cite{amini2014inertial} has been to validate experimentally-observed stable equilibrium points by a mere ``look-up" for their presence in corresponding numerical force-maps. This approach lacks formalism in that it does not provide sufficient information about presence of other equilibrium points, or lack thereof. This satisfies a \textit{necessary} condition but is not \textit{sufficient} to describe the system in its entirety. Furthermore, the ability to quantify the stability-attributes of an equilibrium point for an arbitrary geometry and ${\color{magenta} Re}$ is valuable and lends insight into preference of particular focusing positions. The current work thus aims to address some of the following questions:

\begin{itemize}
    \item establish a mathematical formulation for automated identification of equilibrium points and focusing locations for a general scenario
    \item define quantitative measures that characterize individual equilibrium points
    \item exhaustively predict the set of stable locations for a given configuration
    \item rank order stable points in terms of their stability
\end{itemize}

Over the past decade, there has been a fairly exhaustive study into circular, square, and rectangular channels  \cite{DiCarlo2009},  \cite{amini2014inertial},  \cite{gossett2012inertial}. Interestingly, recent results have also demonstrated the capability to fabricate unconventional channel cross-sections to control the number and locations of focusing points  \cite{Kim2016}. Additionally, it has also been shown that channels of varying geometries can be attached end-to-end to produce a stage-wise effect for focusing particles  \cite{zhou2013modulation},  \cite{Kim2016}. In light of the state-of-art for fabricating such geometries, we are faced with the possibility of enhancing this existing design-space of particle-focusing trends known to the community and enabling exploration of newer geometries, hitherto uncharted, using automated computing-tools. This motivates need for a high-throughput computational study across varied geometries, {\color{magenta} $Re$}'s, and ${\color{magenta} \frac{a}{H}}$'s, in order to generate a library of focusing locations/patterns that delineate trends, and scaling principles involved for an arbitrarily chosen geometry and flow conditions. Moreover, this library would be directly utilizable to create so-called “transition-maps” which connect particle release-locations to their final stable locations or “basins of attraction”. This can be subsequently utilized to design an array of channels of varying geometries based on individual transition-maps to create novel \textit{channel-programs}. A similar idea was reported by previous researchers in the context of deforming fluid streams in the presence of obstacles  \cite{Amini2013},  \cite{stoecklein2014micropillar},  \cite{nunes2014fabricating},  \cite{paulsen2015optofluidic},  \cite{paulsen2016non}. We lay the ground for such work by detailing a strategy for exhaustively identifying and quantifying the stability of equilibrium points in channels with arbitrary cross-sections.


\section{Equilibrium locations and stability estimates}\label{sec:methods}
We briefly outline the strategy for identifying all equilibrium points and the associated stability assessment below. Each step below is further detailed in subsequent sections. 

\begin{itemize}
    \item {\bf Step 1 \textendash} Generate cross-sectional force-maps: this step consists of calculating the lateral forces acting on a particle at different locations within the channel cross-section. This is the standard step used in evaluating force-maps   \cite{DiCarlo2009}, \cite{Kim2016}, \cite{gossett2012inertial}, \cite{liu2015inertial}. This results in a discrete force-map.
    \item {\bf Step 2 \textendash} Stability calculations which is further divided into the following steps,
    \begin{itemize}
    \item {\bf Step 2a \textendash} Find all equilibrium locations: We use the discrete force-map and utilize interpolation theory to construct a continuous interpolated force-map. We then find locations where this continuous vector field goes to zero (i.e., find the zeros of the vector field). This identifies the set of all equilibrium points.\\
    \item {\bf Step 2b \textendash}  Evaluate stability of each equilibrium locations: we utilize the continuous force-map to write the lateral motion of the particle as a first-order system. We then invoke stability theory of first-order systems to quantitatively determine (in)stability of each of the equilibrium points.
    \end{itemize}
\end{itemize}

\subsection{Generating force-maps}

\begin{figure}
\centering
\includegraphics[width=.7\linewidth]{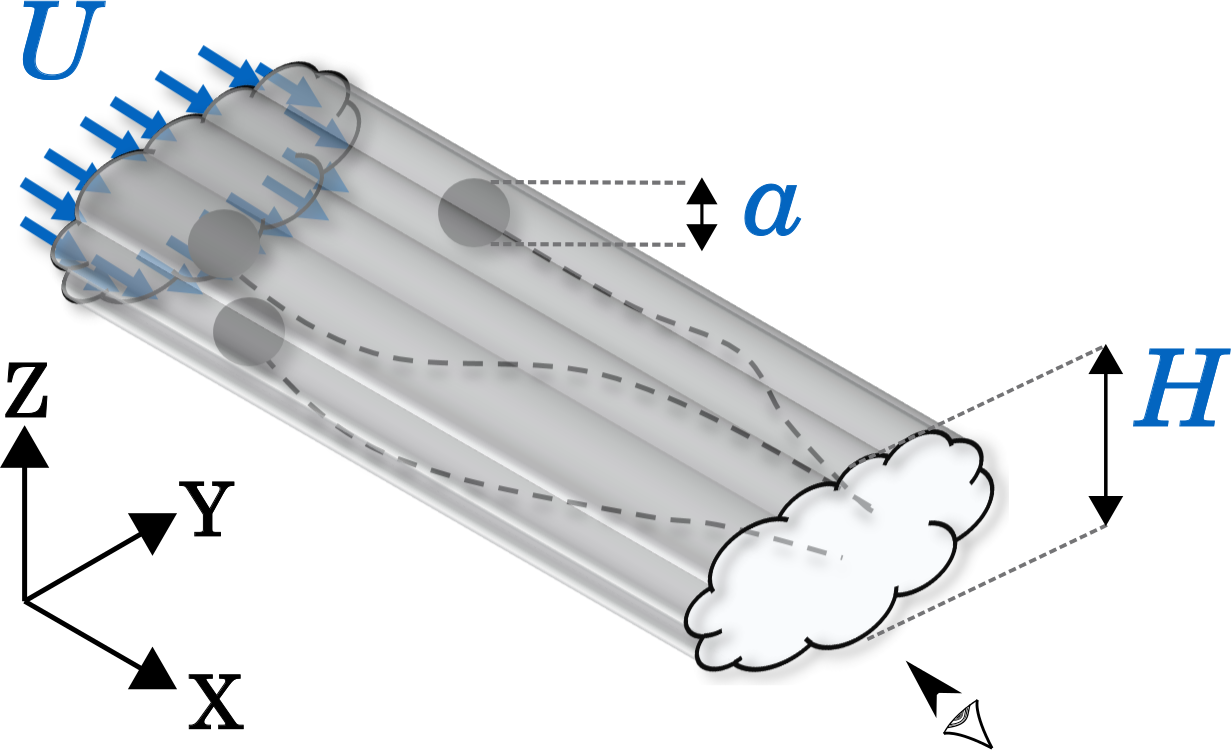}
\caption{\textit{The computational model}: A spherical particle of diameter, $a$, traverses through a channel of hydraulic diameter, $H$. The average velocity of the fluid is $U$ (view into the Y-Z cross-sectional plane).}
\label{fig:fig1}
\end{figure}

We consider a single particle in an arbitrary cross-section channel with particle-diameter ($a$), channel hydraulic-diameter ($H$), and average fluid velocity, $U$, as defined in FIG. \ref{fig:fig1}. We define a confinement ratio parameter as the ratio of the particle diameter-to-channel hydraulic diameter ($={\color{magenta}\frac{a}{H}}$). The Reynolds number, {\color{magenta} $Re$}, is based on the average fluid velocity and the channel hydraulic diameter ({\color{magenta} $Re\,$}$=\frac{{\rho}UH}{\mu}$). The governing equations (non-dimensionalized) are given as:

\begin{gather}
{\nabla} \cdot \boldsymbol{u} = 0 \label{eq:eq1} \\ 
\boldsymbol{u} \cdot \nabla{\boldsymbol{u}} = -{\nabla}p + \frac{1}{
Re}{\nabla}^2 \boldsymbol{u} \nonumber \\
\boldsymbol{m}{\frac{d \boldsymbol{V_t}}{d t}} = \boldsymbol{F}
\end{gather}

\justify
where, $\boldsymbol{u} = [u,v,w]^T$ is the fluid velocity, $p$ is the fluid pressure, and $\boldsymbol{m}$ is the mass (inertial tensor). The fluid affects the particle by imposing forces and torques on the particle. We denote the net forces and torque vectors as $\boldsymbol{F} = [F_x,F_y,F_z,\tau_x,\tau_y,\tau_z]^T$. The particle position and velocity is affected by $\boldsymbol{F}$. We denote the linear and angular velocity of the particles as $\boldsymbol{V_t} = [u_{pt},v_{pt},w_{pt},\omega_{pxt}, \allowbreak \omega_{pyt},\omega_{pzt}]^T$. The particle in turn affects the fluid via the imposition of no-slip conditions at the particle surface. 

The Navier-Stokes equations \eqref{eq:eq1} are solved using a finite element based in-house framework in a translating frame of reference attached to the particle such that the channel walls move at a velocity, $-u_{pt}$. No-slip conditions (accounting for particle angular velocity) are imposed on the particle surface. The inlet and outlet boundary conditions are chosen to have fully-developed velocity-profiles where the particle is placed sufficiently far off from the inlet and outlet. The fully-developed velocity-profiles are obtained by solving for flow in a channel without the particle, and interpolated onto the mesh containing the particle. 
\par
\begin{figure}
\centering
\subfloat[]{\includegraphics[width=1\linewidth]{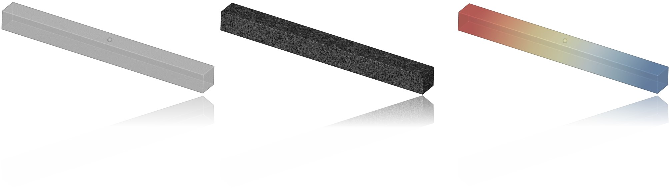}\label{fig:figa1}}
\hfill
\subfloat[]{\includegraphics[width=1\linewidth]{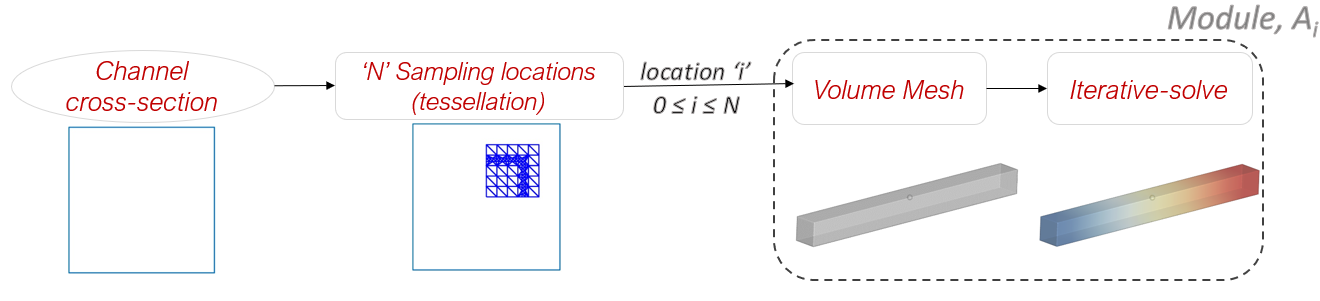}\label{fig:figa2}}
\hfill
\subfloat[]{\includegraphics[width=1\linewidth]{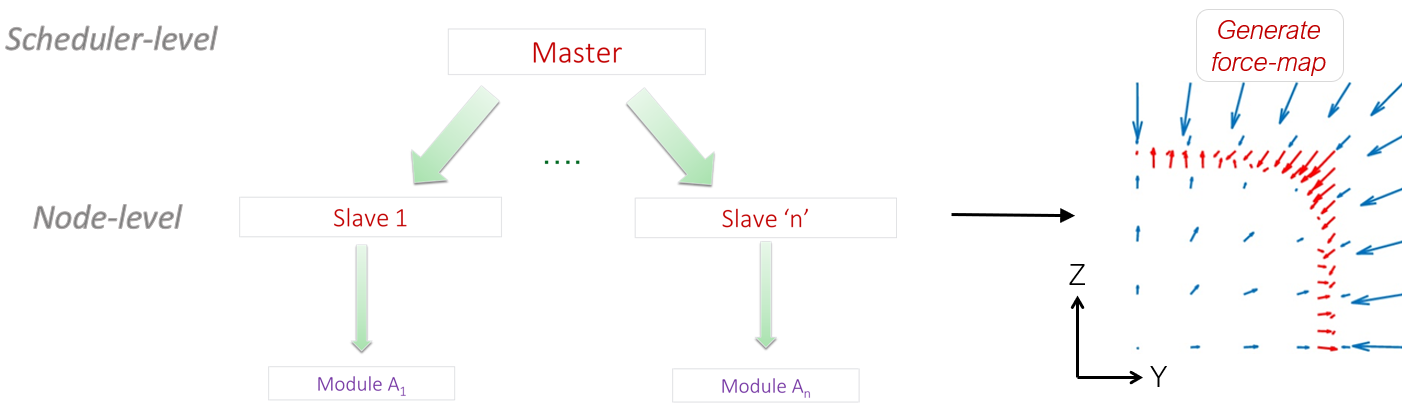}\label{fig:figa3}}
\caption{\textit{Workflow for:} \protect\subref{fig:figa1} single CFD-solve (for a given location and a set of test velocities; the Navier-Stokes equations are solved to a tolerance of $1e-8$) $\vert$ \protect\subref{fig:figa2} quasi-steady solve for a single location (the equilibrium equations are solved to a tolerance of $1e-12$) $\vert$ \protect\subref{fig:figa3} HPC-implementation}
\label{fig:figA1}
\end{figure}

To calculate the lift-forces on the particle at any location ($\tilde{y}, \tilde{z}$) in the cross-sectional plane, we use the formalized approach of constrained simulation, also called the quasi-steady (Q-S) method  \cite{DiCarlo2009}. In this procedure, the lateral velocities ($v_{pt}, w_{pt}$) are set to zero. Then, the variables ($u_{pt}, \omega_{pxt}, \omega_{pyt}, \omega_{pzt}$) are solved iteratively to ensure that the streamwise drag, $F_x$, and 3 components of torque, $\tau_x, \tau_y, \tau_z$, all go to zero (i.e., quasi-equilibrium conditions). Once these equations are self-consistently solved, the lateral drag forces at that location are computed as $(F_{z,i}, F_{y,i})$. This process is repeated at several locations in the cross-sectional plane, and a final force-map is constructed. This workflow \footnote{The force-maps were generated on our campus High-Performance Computing (HPC) facility, \textit{Condo}, which contains two 2.6 GHz 8-Core Intel E5-2640 processors per compute-node, with 8 CPU cores per processor.} is illustrated in FIG. \ref{fig:figA1}. The current application was modularized to execute each location-solve (FIG. \ref{fig:figa2}) over a single node. All locations are solved for independently and asynchronously using a master-slave paradigm (FIG. \ref{fig:figa3}) in a high-throughput fashion, thereby optimizing overall run-time for a single force-map.

\subsection{Stability calculation}
\subsubsection{Location of equilibrium points using interpolation}
\begin{figure}
\centering
\includegraphics[width=.5\linewidth]{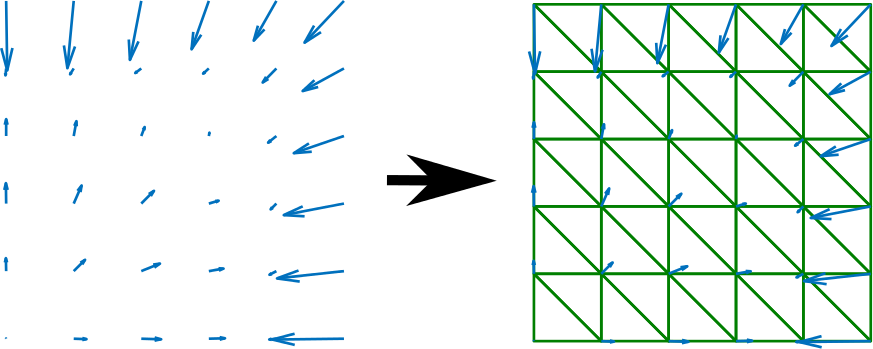}
\caption{\textit{Sample tessellation:} tessellation using particle-locations (the vectors represent cross-sectional lift-forces)}
\label{fig:tessellation}
\end{figure}

We consider a tessellation of the sampled cross-sectional particle-location (FIG. \ref{fig:tessellation}). At each of the locations, $(y_j, z_j)$ of the tessellation, we have available the net-forces, $(F_{y_j},F_{z_j})$, from the previous step. We then use interpolation theory to represent the force-map (within each element of the tessellation):
\begin{gather}
\widetilde{F}_y(\xi,\eta) = {\sum_{i=1}^{3}F_{y_i}N_i(\xi,\eta)} \label{eq:eq10} \\
\widetilde{F}_z(\xi,\eta) = {\sum_{i=1}^{3}F_{z_i}N_i(\xi,\eta)} \nonumber
\end{gather}

\justify
where $N_i$ are the basis functions localized at $(y_i, z_i)$, and, $(\xi, \eta)$ represent isoparametric coordinates. In this work, we use a tessellation consisting of triangles (FIG. \ref{fig:tessellation}) and use linear basis functions that are analytically defined within each triangle to produce a continuous interpolant across the domain. This representation is searched to identify points, $(y_i^*, z_i^*)$, where, $\widetilde{F}_y(y_i^*, z_i^*) = 0$, and $\widetilde{F}_z(y_i^*, z_i^*) = 0$. This is accomplished via a standard search through each element to identify if a solution exists to the linear equation:

\begin{gather}
\begin{bmatrix}
F_{z_1} - F_{z_2} & F_{z_3} - F_{z_2} \\
F_{y_1} - F_{y_2} & F_{y_3} - F_{y_2}
\end{bmatrix}
\begin{bmatrix}
\xi^* \\
\eta^*
\end{bmatrix}
=
\begin{bmatrix}
- F_{z_2} \\
- F_{y_2}
\end{bmatrix}
\end{gather}
\justify
where $(\xi^*,\eta^*)$ are required to satisfy, $\xi^* \ge 0, \hspace{0.2 em} \eta^* \ge 0, \hspace{0.2 em} \text{and} \hspace{0.2 em} \xi^* + \eta^* \le 1$, and transformed back into global coordinates using the equations detailed in   \cref{sec:app1}.

\subsubsection{Linearization and stability}
The force-maps, $(F_y, F_z)$, represent a dynamical system and the set of zeros, $(y^*, z^*)$, represent the equilibrium points. Viewed from this context, we can invoke formal and rigorous notions of stability from dynamical systems theory. The \textit{Hartman-Grobmann theorem}  \cite{grobman1959homeomorphism},  \cite{hartman1960lemma},  \cite{grobman1962topological} states that a non-linear dynamical system, given by 
\begin{gather}
\frac{\mathrm{d}\boldsymbol{X}}{\mathrm{d}t} = \boldsymbol{f(X)} \label{eq:eq2},
\end{gather}
\justify
is linearizable around an equilibrium point for deriving qualitative stability inferences around that point. This linearization (also called the state-space form) has the general form
\begin{gather}
\frac{\mathrm{d}\boldsymbol{X}}{\mathrm{d}t} = {A}\boldsymbol{X}  \label{eq:eq3}
\end{gather}
\justify
where,
\begin{gather}
    {A} = \bigg(\frac{\mathrm{\partial}\boldsymbol{f}}{\mathrm{\partial}\boldsymbol{X}}\bigg)_{\boldsymbol{X_0}} \label{eq:defA}
\end{gather}
\justify
here $A$ is the Jacobian of the linearized system (refer   \cref{sec:app2} for details). Stability is quantified in terms of the (real parts of) eigenvalues, $\lambda_i$, of $A$. The equilibrium location is stable if \textit{all} real-parts are negative. For our fluid-particle system, the equations of motion for the particle in the lateral direction can be approximated as,

\begin{gather}
m\frac{\mathrm{d^2}z}{\mathrm{d}t^2} = F_z(z,y) - 3{\pi}{\mu}a\frac{\mathrm{d}z}{\mathrm{d}t} \label{eq:eq6} \\
m\frac{\mathrm{d^2}y}{\mathrm{d}t^2} = F_y(z,y) - 3{\pi}{\mu}a\frac{\mathrm{d}y}{\mathrm{d}t} \nonumber,
\end{gather}

\justify
where $m$ is the mass of the particle. We include the Stokes' drag terms for the following reasons:

\begin{itemize}
    \item the resistive drag serves to eliminate unrealistic numerical oscillations (solution to equations \eqref{eq:eq6} without the drag terms); use of these drag forces has been done previously in the context of calculating channel lengths for achieving migration  \cite{di2009inertial} owing to slow migration in lateral directions
    \item the dynamical system - now in a decoupled form - includes all velocity-components of the particle as in an actual scenario, i.e., streamwise-translation (from the Q-S model), spin-velocities (from the Q-S model), and lateral translation (Stokes' drag)
\end{itemize}

\justify
The above coupled system of two second-order equations is converted to four first-order equations with the following variables, $\Big(z, \frac{dz}{dt}, y, \frac{dy}{dt}\Big)$, and the Jacobian, $A$, is evaluated from equations \eqref{eq:defA} and \eqref{eq:eq6} as:

\begin{gather}
{A} =  
{\begin{bmatrix}
0 & 1 & 0 & 0 \\
\frac{1}{m}\frac{\mathrm{\partial}F_z(z,y)}{\mathrm{\partial}z} & -\frac{3{\pi}{\mu}a}{m} & \frac{1}{m}\frac{\mathrm{\partial}F_z(z,y)}{\mathrm{\partial}y} & 0 \\
0 & 0 & 0 & 1 \\
\frac{1}{m}\frac{\mathrm{\partial}F_y(z,y)}{\mathrm{\partial}z} & 0 & \frac{1}{m}\frac{\mathrm{\partial}F_y(z,y)}{\mathrm{\partial}y} & -\frac{3{\pi}{\mu}a}{m}
\end{bmatrix}}_{\boldsymbol{X_0}} \label{eq:eq8}
\end{gather}

\justify
This representation (equations \eqref{eq:eq3} and \eqref{eq:eq8}) is a generalization to higher dimensions of the 1D-case of circular particle focusing in a straight 2D channel  \cite{yang2005migration}, where stability is interpreted in terms of the slope of the lift-versus-transverse coordinate curve at equilibrium locations. The $\frac{\mathrm{\partial}F_i(z,y)}{\mathrm{\partial}x_j}$ terms are represented by $4^{\mathrm{th}}$-order centred finite-differences as,

\begin{gather}
{\bigg(\frac{\mathrm{\partial}F_i(z,y)}{\mathrm{\partial}x_j}\bigg)}_{\boldsymbol{X_0}} = \bigg(\frac{F_{i,-2}-F_{i,2}}{12\Delta{x_j}}\bigg) + 2\bigg(\frac{F_{i,1}-F_{i,-1}}{3\Delta{x_j}}\bigg) \label{eq:eq9}
\end{gather}

\justify
where, $F_{i,1}$ denotes force in the $i^{th}$ direction for a perturbation of $\Delta{x_j}$ in the $j^{th}$ direction, $F_{i,2}$ denotes force in the $i^{th}$ direction for a perturbation of $2\Delta{x_j}$ in the $j^{th}$ direction, and so on. Once the Jacobian is constructed for an equilibrium point, the stability is calculated using the real-parts of its eigenvalues, $\lambda_i$ ($1 \leq i \leq 4$):

\begin{itemize}
    \item if $\Re(\lambda_i) < 0 \hspace{0.2 em} (1 \leq i \leq 4)$ - the given equilibrium point is \textbf{stable}, otherwise
    \item the given equilibrium point is \textbf{unstable}
\end{itemize}

\justify
Along with the focusing patterns, we also compute basins of attraction ($\omega-$limit sets)  \cite{strogatz2018nonlinear} for each stable point. Basins of attraction are characteristic of each stable point and denote the state of a system of particles after a sufficiently long time from release. They are essentially ``guiding zones" in that particles focus to that stable point whose basin contains their release-location. This feature has been exploited previously to great effect in order to create unique, ordered streams of particles  \cite{Kim2016}. The overall algorithm can be briefly summarized as follows:

\begin{algorithm}
\begin{spacing}{1}
\begin{algorithmic}
\REQUIRE{force-maps, tessellation}
\ENSURE{stable locations}\\
\textbf{initialize}: $e = 0$ (no. of equilibrium points), $s = 0$ (no. of stable points)
\WHILE{$i < N_{elem}$}
    \STATE{From equation \eqref{eq:eq10}, calculate $(\xi_0,\eta_0) \ni [F_z(\xi_0,\eta_0),F_y(\xi_0,\eta_0)]^T = \boldsymbol{0}$;}
    \IF{$(\xi_0 \ge 0) \hspace{0.2 em} \&\& \hspace{0.2 em} (\eta_0 \ge 0) \hspace{0.2 em} \&\& \hspace{0.2 em} (\xi_0 + \eta_0 \le 1)$}
        \STATE{equilibrium point - $(z(\xi_0,\eta_0),y(\xi_0,\eta_0))$ - exists in element $'i'$;}
        \IF{$A_{(z(\xi_0,\eta_0),y(\xi_0,\eta_0))}$ is stable}
            \STATE{$s++;$}
        \ENDIF
        \STATE{e++;}
    \ENDIF
    \STATE{i++;}
\ENDWHILE
\end{algorithmic}
\end{spacing}
\hrulefill\vspace{-1 em} \caption{\centering{Compute stable focusing patterns}} \label{alg:alg1}
\vspace{-1 em}\hrulefill
\end{algorithm}
\vspace{-20pt}
\section{Validation}\label{sec:val}
\begin{figure}[!htpb]
\centering
\subfloat[]{\includegraphics[width=.45\linewidth]{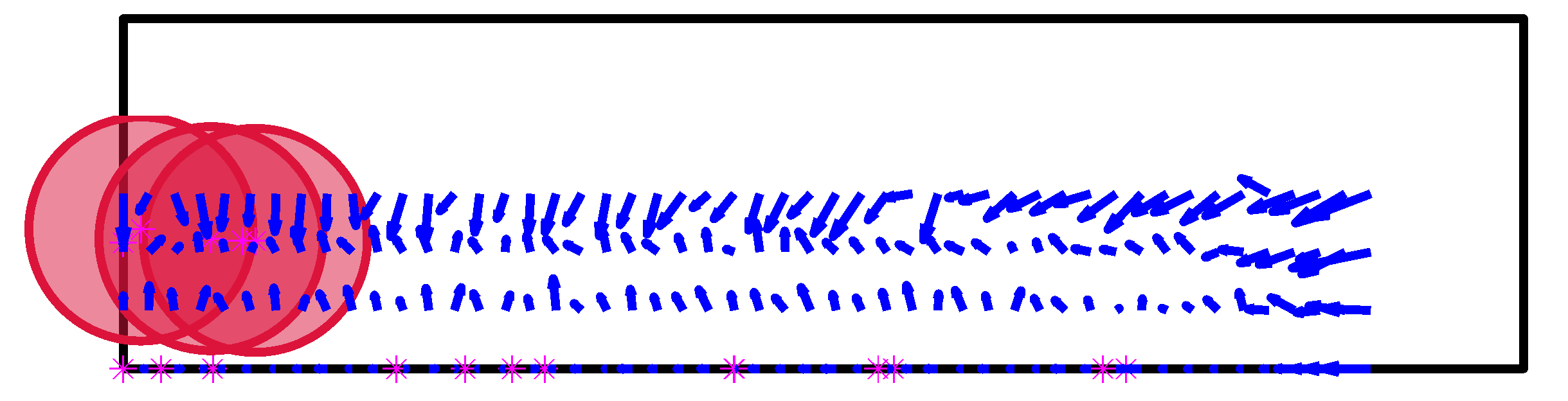}\label{fig:fig4_1}}
\hspace{4 em}
\subfloat[]{\includegraphics[width=.45\linewidth]{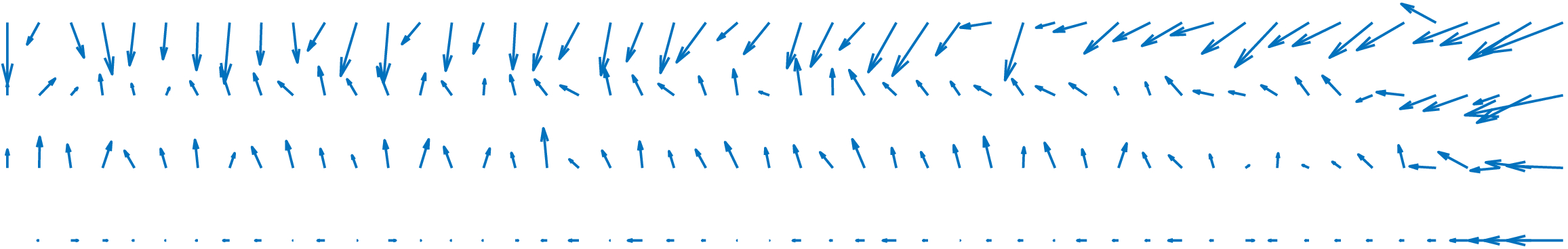}\label{fig:fig4_2}}
\hfill
\subfloat[]{\includegraphics[width=.45\linewidth]{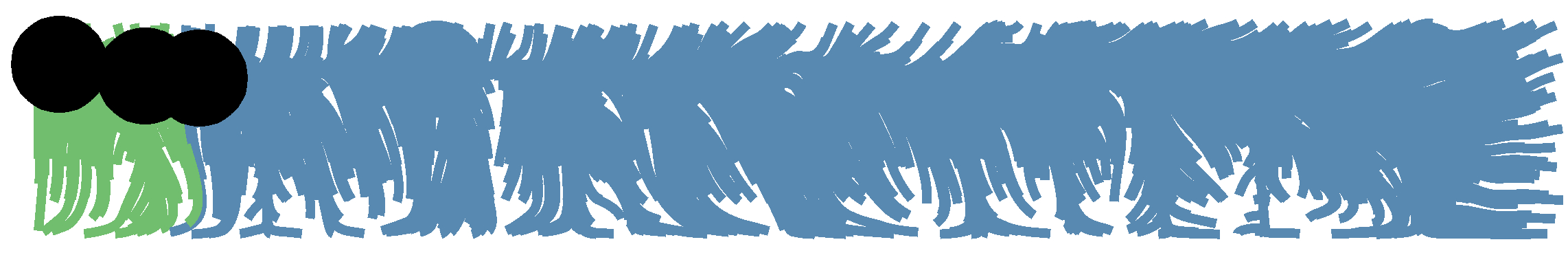}\label{fig:fig4_4}}
\hfill
\caption{\textit{Validation}: 1:4 rectangular-channel (quarter section shown) with ${\color{magenta} k} = 0.2$, and ${\color{magenta} Re} = 10$ \protect\subref{fig:fig4_1} predicted focusing-pattern (the blue region represents the sampled particle-locations/triangulation, magenta-asterisks represent all equilibrium locations, and red-circles represent the particle (to scale)) $\vert$ \protect\subref{fig:fig4_2} resultant lift-force at sampled locations $\vert$ \protect\subref{fig:fig4_4} color-coded basins of attraction for each attractor point in sampled region (black dots)}
\label{fig:fig4}
\end{figure}

\begin{figure}[!htpb]
\centering
\subfloat[]{\includegraphics[width=.4\linewidth]{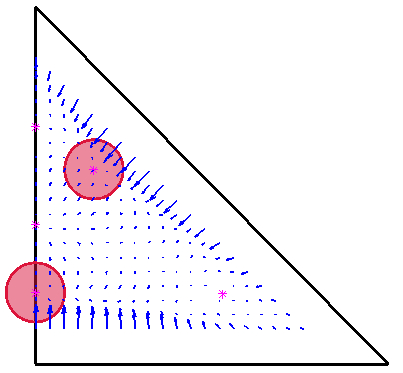}\label{fig:fig5_1}}
\subfloat[]{\includegraphics[width=.37\linewidth]{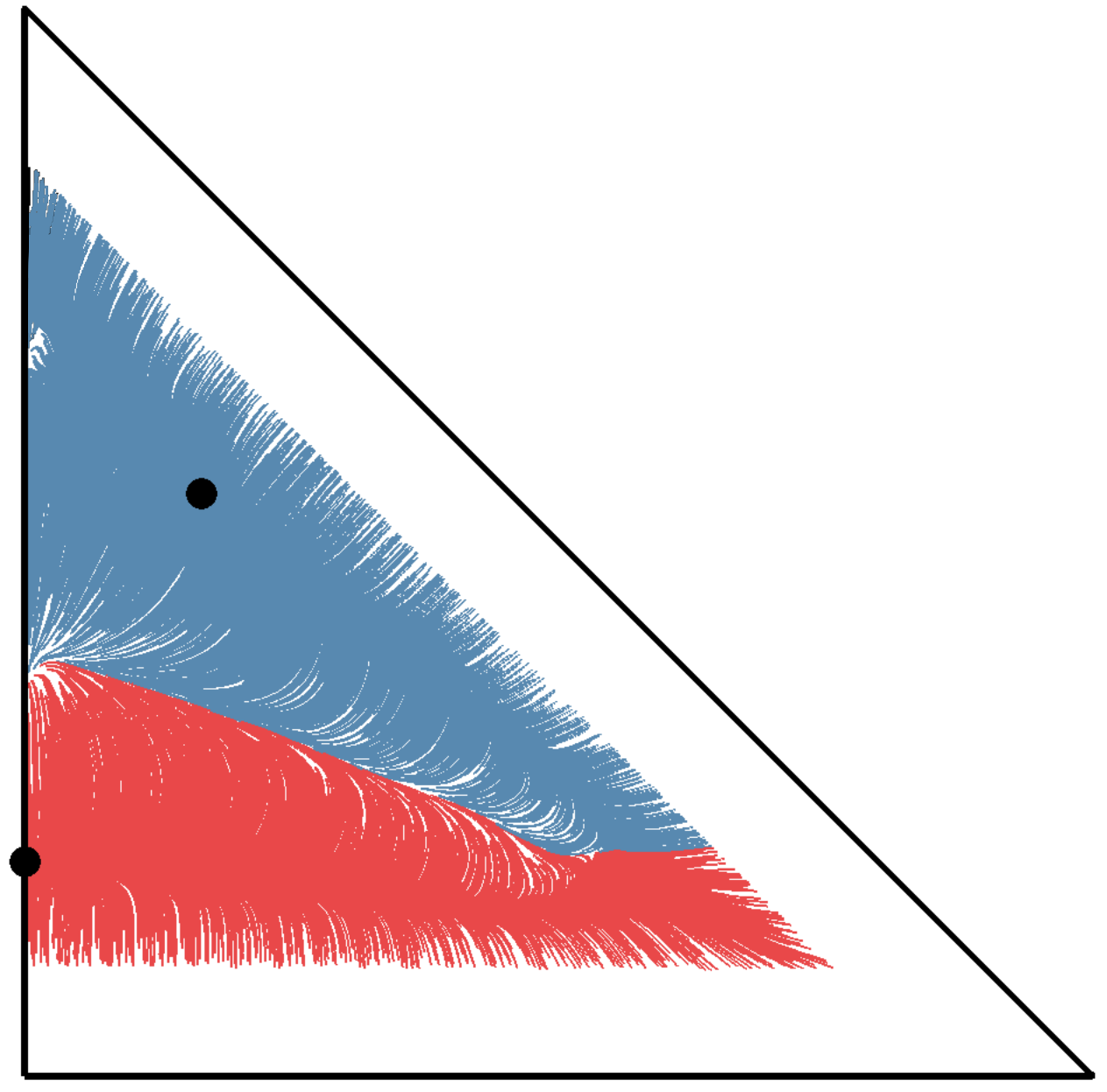}\label{fig:fig5_4}}
\caption{\textit{Validation}: isosceles right triangular-channel (half-section shown) with ${\color{magenta} k} = 0.2$, and ${\color{magenta} Re} = 100$ \protect\subref{fig:fig5_1} predicted focusing-pattern (the blue vectors represent lift-forces, magenta-asterisks represent all equilibrium locations, red-circles represent the particle, and black lines represent the half-channel boundary (sliced top-down, to scale)) $\vert$ \protect\subref{fig:fig5_4} color-coded basins of attraction for each attractor point in sampled region (black dots)}
\label{fig:fig5}
\end{figure}

The approach is validated using cases of inertial focusing that have been extensively studied in literature. The first case deals with a rectangular-channel (aspect-ratio 1:4) for ${\color{magenta} k} = 0.2$, and ${\color{magenta} Re} = 10$  \cite{hur2010sheathless}, and the second case deals with focusing in an isosceles right-triangular channel reported in a recent study  \cite{kim2017size} for ${\color{magenta} k} = 0.2$, and ${\color{magenta} Re} = 100$. We examine the possible set of equilibrium points in each of the cases, the stable locations, basins of attraction, and normalized-eigenvalues. Owing to symmetry, we sampled one-fourth of the square-channel cross-section (FIG. \ref{fig:fig4}) and one-half for the isosceles right-triangular channel (FIG. \ref{fig:fig5}).

\begin{table}
\begin{tabular}{||c|c|c|c|c|c|c||}
 \hline
 \textbf{No.} & $\bm{\frac{y}{a}}$ & $\bm{\frac{z}{a}}$ & $\bm{\lambda_{1,norm}}$ & $\bm{\lambda_{2,norm}}$ & $\bm{\lambda_{3,norm}}$ & $\bm{\lambda_{4,norm}}$ \\ 
 \hline
 1 & 0.000 & 0.000 & -0.002 & 0.002 & -1.798 & -1.802 \\ 
 \hline
 \textbf{2} & \textbf{0.589} & \textbf{0.573} & \textbf{-0.004} & \textbf{-0.004} & \textbf{-1.796} & \textbf{-1.796} \\ 
 \hline
 3 & 1.217 & 0.000 & 0.002 & 0.003 & -1.802 & -1.803 \\ 
 \hline
 4 & 2.729 & 0.000 & 0.002 & -0.003 & -1.797 & -1.802 \\ 
 \hline
 5 & 0.168 & 0.000 & 0.004 & 0.004 & -1.803 & -1.804 \\ 
 \hline
 6 & 0.000 & 0.563 & 0.004 & -0.008 & -1.792 & -1.804 \\ 
 \hline
 \textbf{7} & \textbf{0.079} & \textbf{0.624} & \textbf{-0.004} & \textbf{-0.004} & \textbf{-1.796} & \textbf{-1.796} \\ 
 \hline
 8 & 0.531 & 0.573 & 0.002 & -0.006 & -1.794 & -1.802 \\ 
 \hline
 \textbf{9} & \textbf{0.391} & \textbf{0.580} & \textbf{-0.001} & \textbf{-0.008} & \textbf{-1.792} & \textbf{-1.800} \\ 
 \hline
 10 & 2.726 & 0.000 & 0.003 & 0.003 & -1.803 & -1.803 \\ 
 \hline
\end{tabular} 
\caption{\textit{Validation}: normalized eigenvalues (real-parts) for all equilibrium points for rectangular-channel (origin taken at the centre of the channel; stable locations are shown in bold-font; we do not list all unstable points here for the sake of brevity)}
\label{tab:tab1} 
\end{table}

\begin{table}
\begin{tabular}{||c|c|c|c|c|c|c||} 
 \hline
 \textbf{No.} & $\bm{\frac{y}{a}}$ & $\bm{\frac{z}{a}}$ & $\bm{\lambda_{1,norm}}$ & $\bm{\lambda_{2,norm}}$ & $\bm{\lambda_{3,norm}}$ & $\bm{\lambda_{4,norm}}$ \\ 
 \hline
 1 & 0.000 & 1.996 & 0.025 & -0.056 & -1.744 & -1.825 \\ 
 \hline
 \textbf{2} & \textbf{0.000} & \textbf{-0.800} & \textbf{-0.023} & \textbf{-0.257} & \textbf{-1.543} & \textbf{-1.777} \\ 
 \hline
 \textbf{3} & \textbf{1.000} & \textbf{1.278} & \textbf{-0.011} & \textbf{-0.296} & \textbf{-1.504} & \textbf{-1.789} \\ 
 \hline
 4 & 0.000 & 0.349 & 0.047 & 0.088 & -1.847 & -1.889 \\ 
 \hline
 5 & 3.197 & -0.818 & -0.029 & 0.032 & -1.771 & -1.832 \\ 
 \hline
\end{tabular}
\caption{\textit{Validation}: normalized eigenvalues (real-parts) for all equilibrium points for isosceles right triangular-channel (origin taken at the centroid of the channel; stable locations are shown in bold-font)}
\label{tab:tab2}
\end{table}

\justify
It is seen that the predicted focusing patterns (FIGS. (\ref{fig:fig4_1}), (\ref{fig:fig5_1})) match well with those reported in literature  \cite{hur2010sheathless},  \cite{kim2017size}. For the rectangular-channel, two face-centred stable positions are seen along the longer faces which is in accordance with previous experimental reports, in addition to numerous unstable points. In addition, for the triangular channel we see that our prediction of an inverted triangular pattern matches qualitatively the observed patterns in experiments for low-{\color{magenta}k} ($= 0.25$), and, high-${\color{magenta}Re}$ ($= 60$)  \cite{kim2017size}. The effect of refining the force-map-sampling gives rise to crucial observations and is deferred until   \cref{sssec:convergence}. While considering the individual stabilities of various focusing locations in terms of eigenvalues, we note that the maximum of the real parts of eigenvalues needs to be taken into account. This is due to the fact that the more-negative components correspond to perturbations which decay more rapidly, and hence, the long-term behavior of the perturbations is governed by the slower decay components. In this context, we see that for the rectangular-channel stable-locations (TAB. \ref{tab:tab1}), the maximum absolute real-parts are all similar in value, whereas for the minimum components, eigenvalues for stable points \textbf{2} and  \textbf{7}, are the highest. Additionally, the basins of attraction (FIG. \ref{fig:fig4_4}) indicate that all particles released at the inlet should focus to the face-centred locations in the dilute-limit (barring particle-particle interactions). For the triangular channel (FIG. \ref{fig:fig5_4}), we see that the basins are about the same size, which gives rise to an overall inverted-triangular focusing pattern. The eigenvalues (TAB. \ref{tab:tab2}) indicate that the centre bottom-focusing position, \textbf{2}, should be more stable to perturbations in contrast to the top off-centre focusing position. 

\section{Results and Discussion}\label{sec:res}

\subsection{Test cases: ${\color{magenta} \frac{a}{H}} = 0.4, 90^\circ$-channel, {\color{magenta} $Re$} $= 20-250$}\label{sec:0.4_90_20_250}
The previous section demonstrated that the approach is able to satisfactorily predict the focusing pattern for a channel-configuration at a certain {\color{magenta} $Re$}. We now test predictions over an entire range of {\color{magenta} $Re$} as such trends are often of practical interest in identifying critical values where there is a marked-difference in observable quantities, e.g., alteration of a four-centred focusing pattern to a two-centred pattern in rectangular micro-channels with increase in channel aspect-ratio  \cite{hur2010sheathless}. In this context, it was recently reported  \cite{kim2017size} that larger particles with size-ratios, ${\color{magenta} \frac{a}{H}} = 0.4$, in a $90^\circ$-channel, display an interesting pattern of focusing beyond {\color{magenta} $Re$} $\approx 80$, where they focus to 3 positions (2 top off-centre, and 1 bottom-centre), similar to that seen earlier (FIG. \ref{fig:fig5_1}), and below which they focus to 2 positions (top and bottom) on the symmetry plane. We test to see if we can predict such a trend consistently, across {\color{magenta} $Re$} $= 20, 60, 70, 80, 90, 100, 120, 130, 140, 150, 200, \hspace{0.1em} \& \hspace{0.1em} 250$. Specifically, we attempt to explain the following observations: 

\begin{enumerate}
    \item bifurcation of top-centr stable focusing point (two-point focusing overall) to two off-centre stable locations (three-point focusing overall) near {\color{magenta} $Re$} $\approx 80$
    \item downward shifting of top focusing positions parallel to side-walls after bifurcation ({\color{magenta} $Re$} $\ge 80$)
\end{enumerate}

\subsubsection{Stable-point bifurcation}\label{sec:spb}
We start with an initial tessellation (i.e., mesh-density) consisting of 210 points. It is customary to check spatial convergence in numerical studies, but we will specifically address this aspect later to highlight its importance in the present context.

\begin{figure}[!htpb]
\centering
\subfloat[]{\includegraphics[width=.29\linewidth]{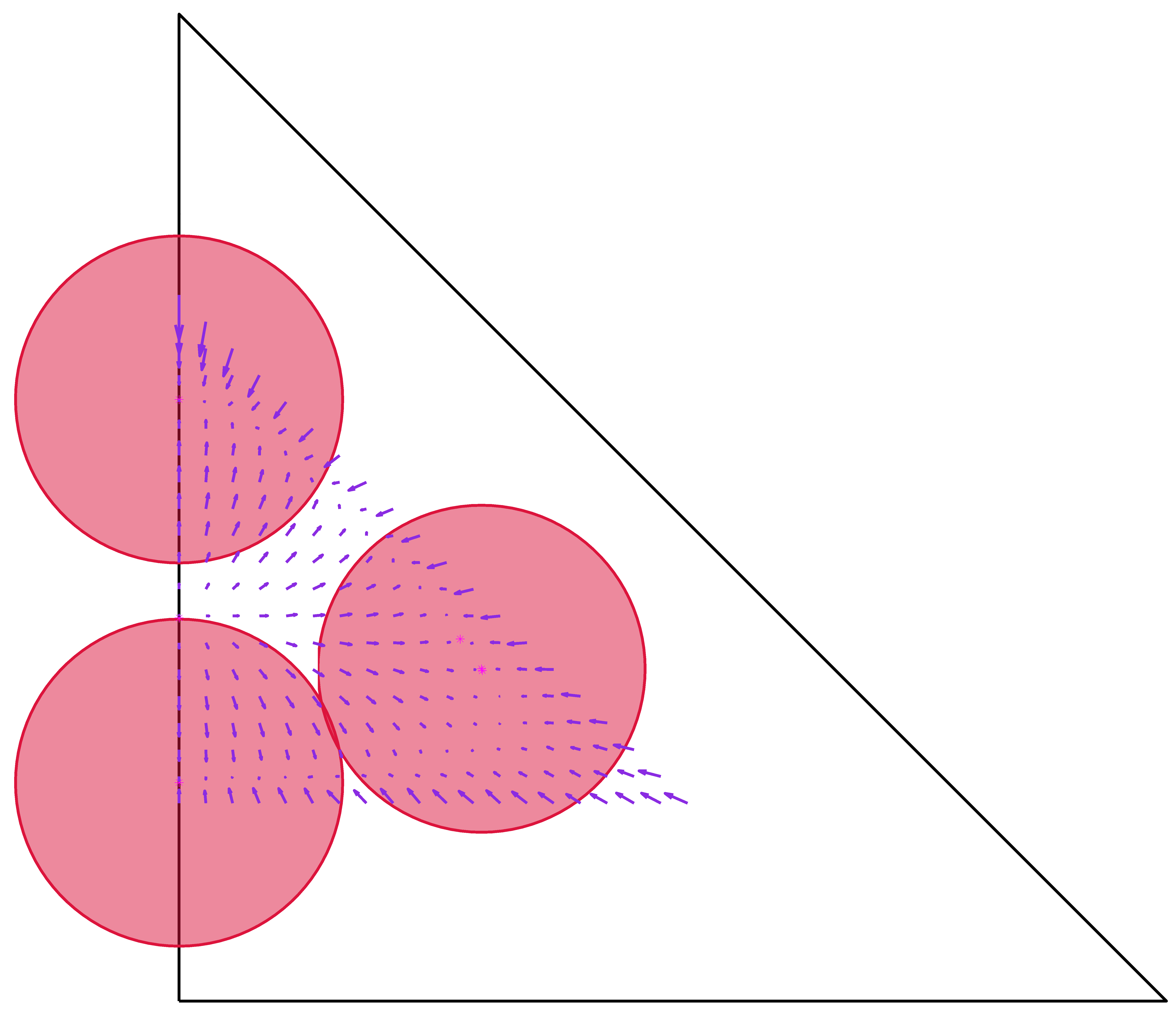}\label{fig:fig6_1}}
\subfloat[]{\includegraphics[width=.29\linewidth]{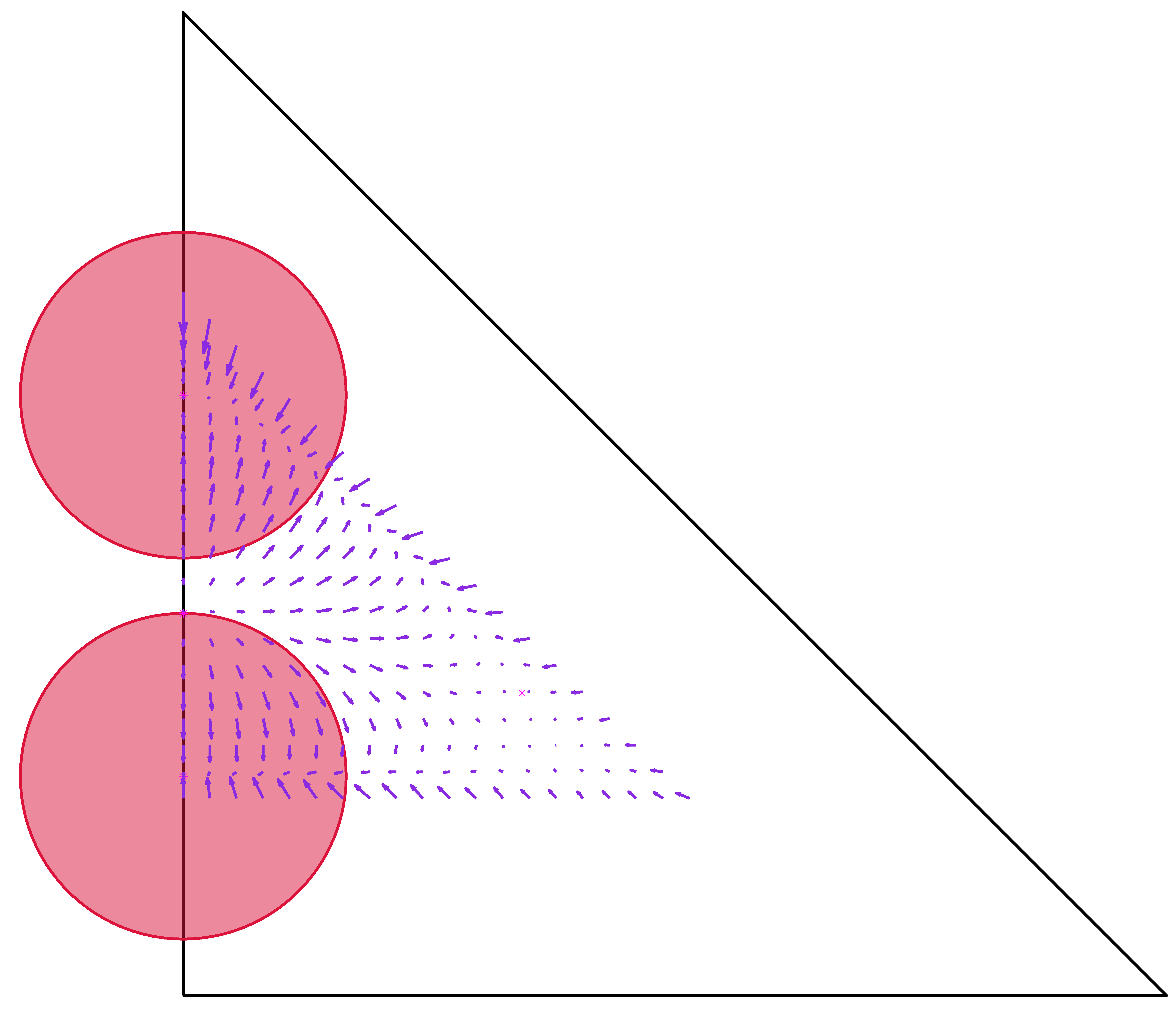}\label{fig:fig6_2}}
\subfloat[]{\includegraphics[width=.29\linewidth]{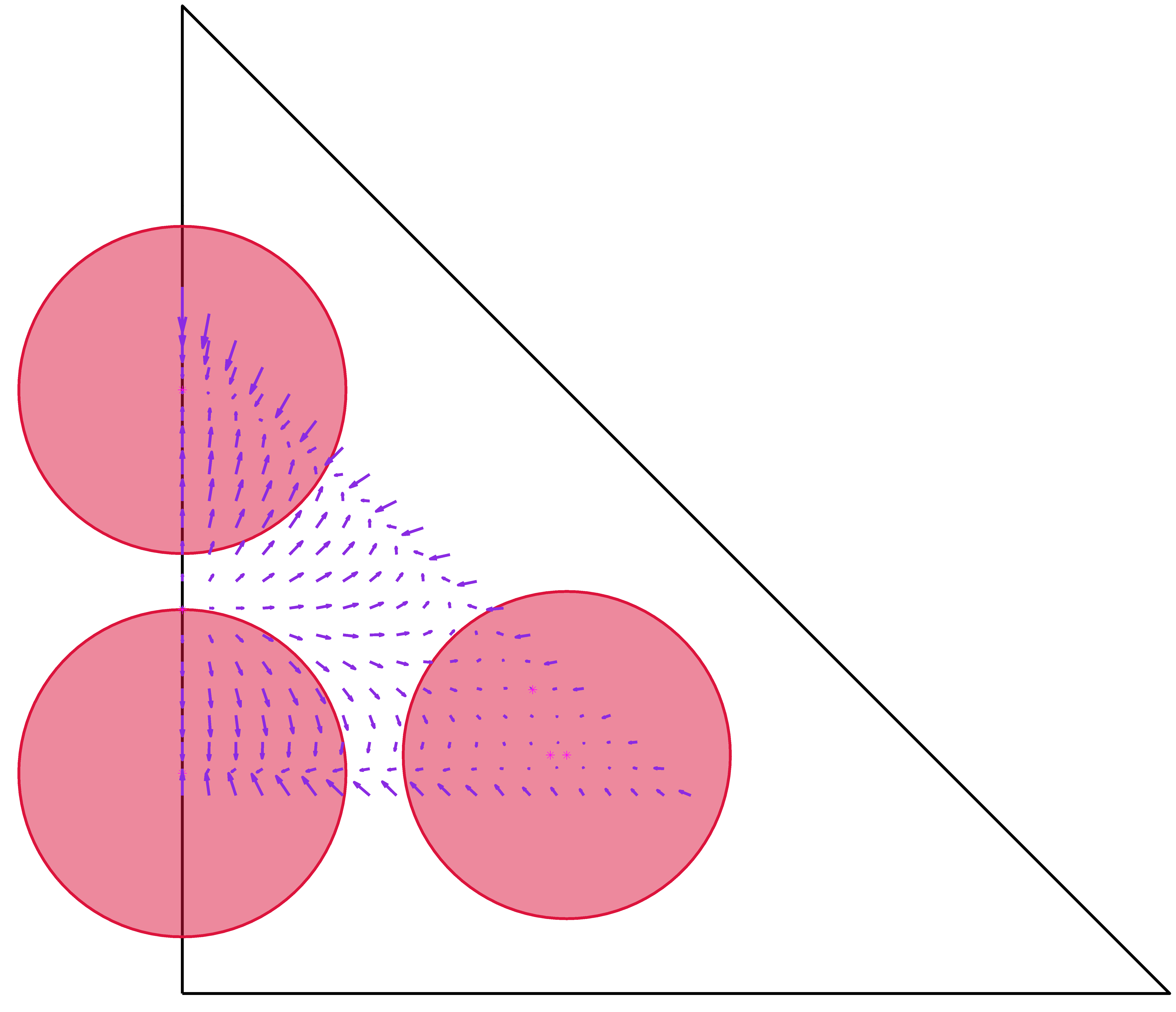}\label{fig:fig6_3}}
\hfill
\subfloat[]{\includegraphics[width=.29\linewidth]{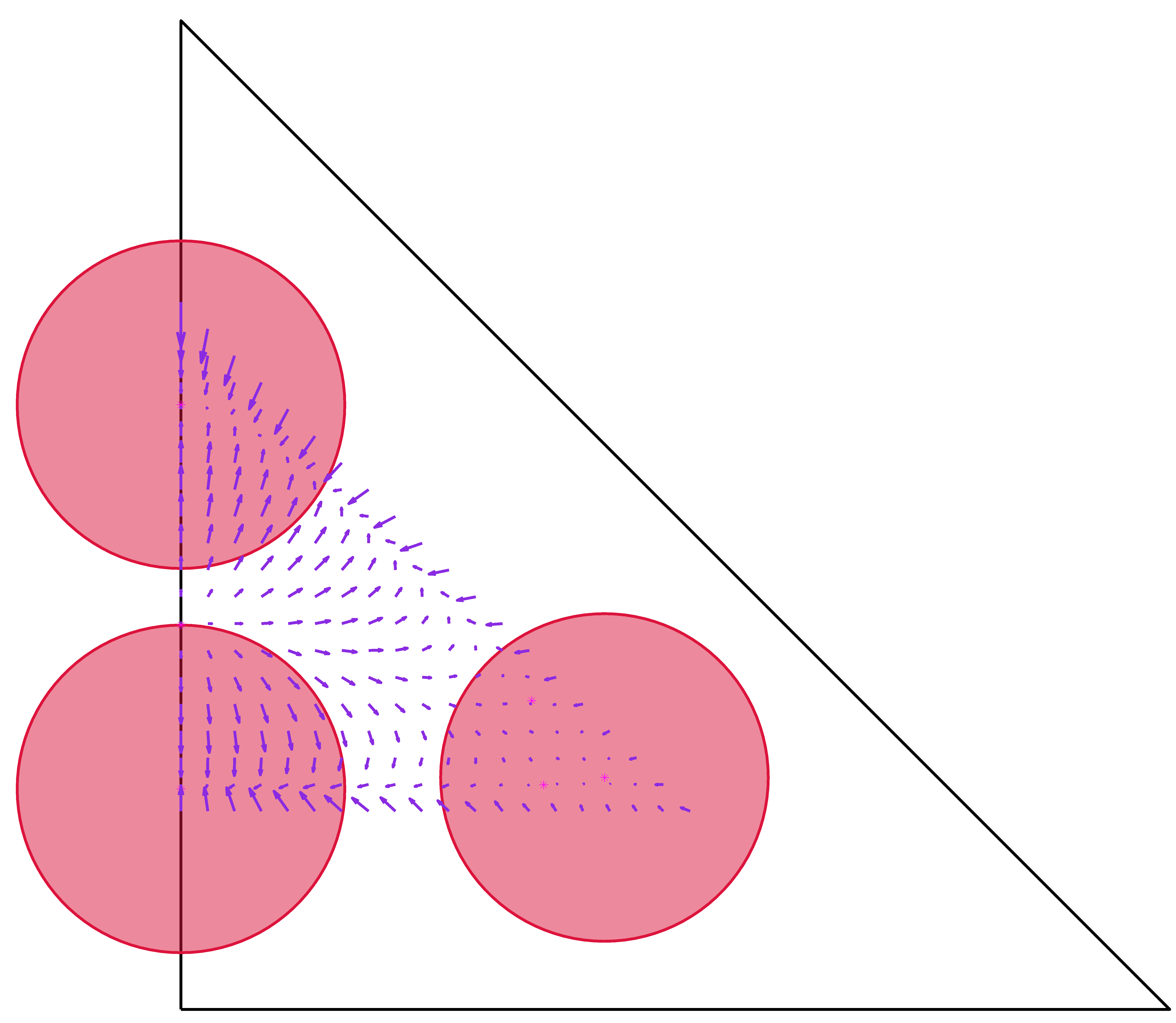}\label{fig:fig6_4}}
\subfloat[]{\includegraphics[width=.29\linewidth]{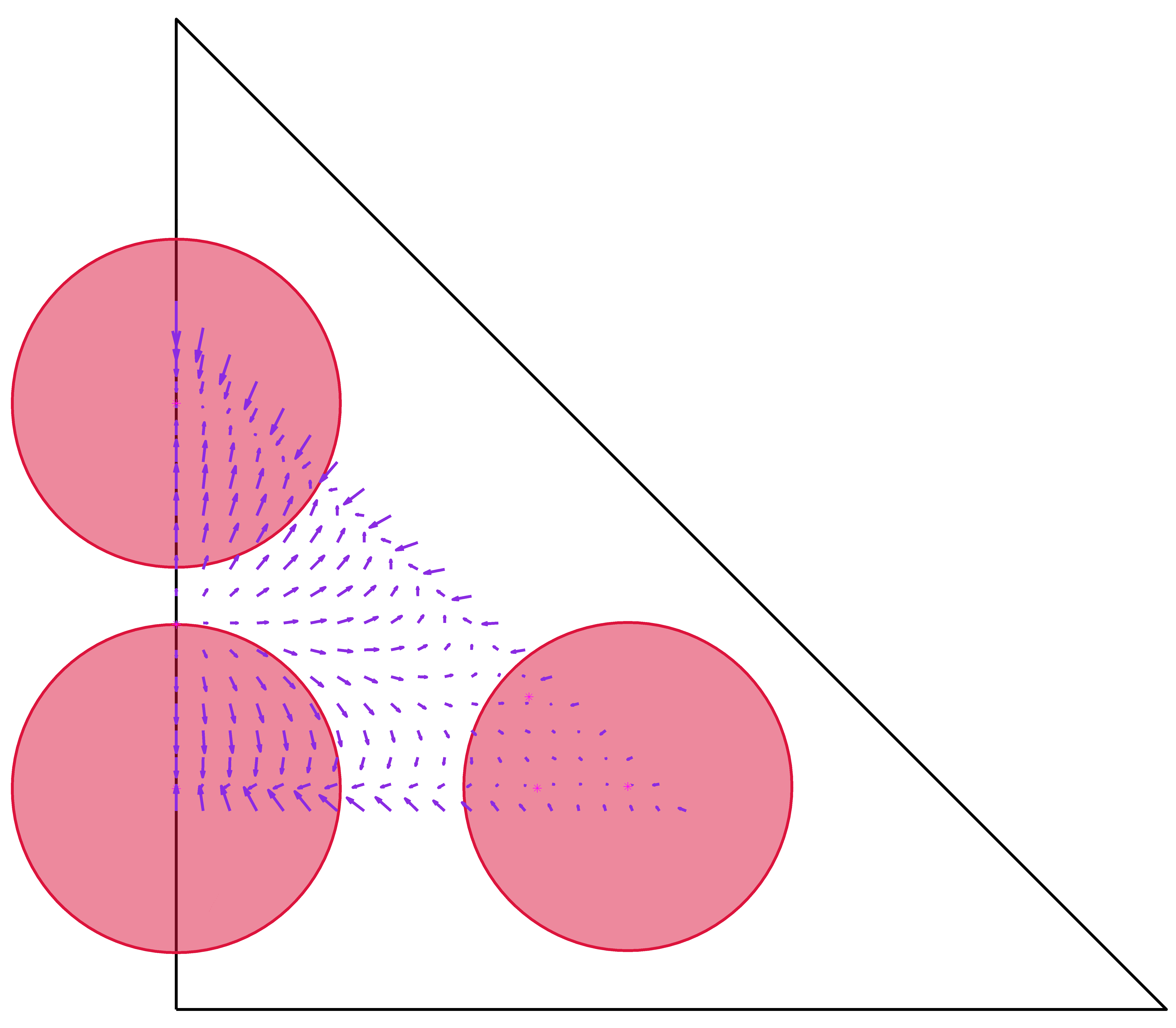}\label{fig:fig6_5}}
\subfloat[]{\includegraphics[width=.29\linewidth]{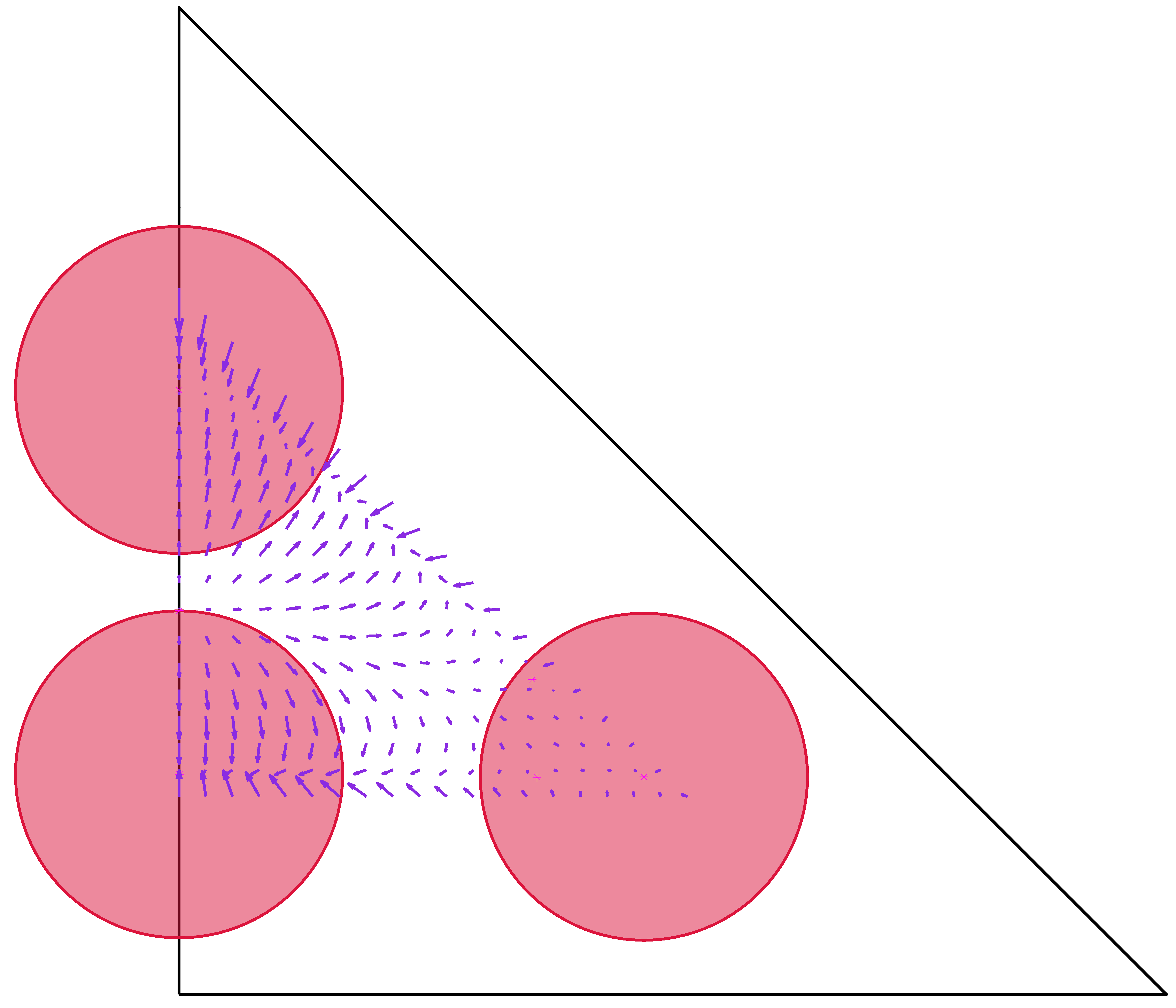}\label{fig:fig6_6}}
\hfill
\subfloat[]{\includegraphics[width=.3\linewidth]{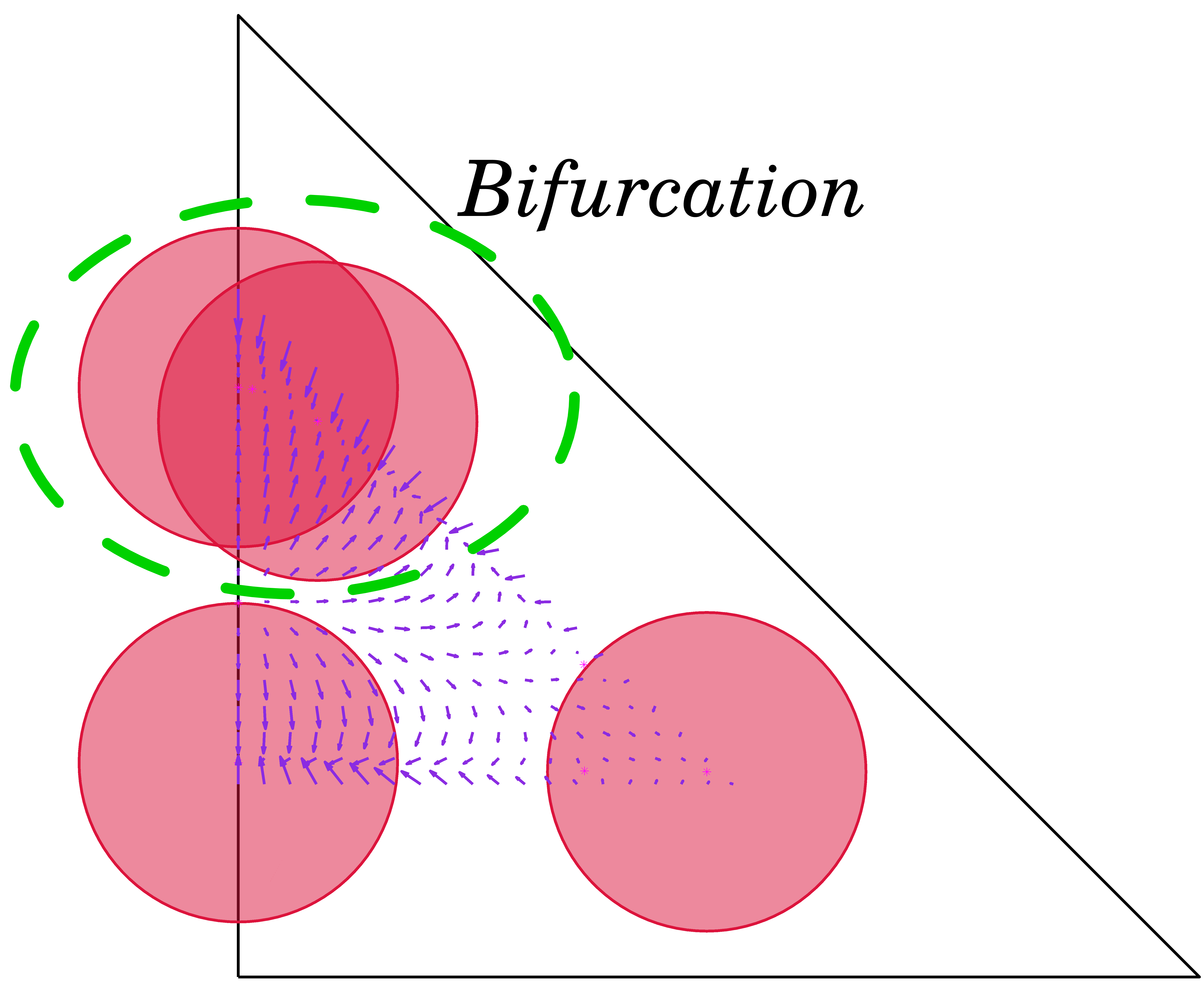}\label{fig:fig6_7}}
\subfloat[]{\includegraphics[width=.29\linewidth]{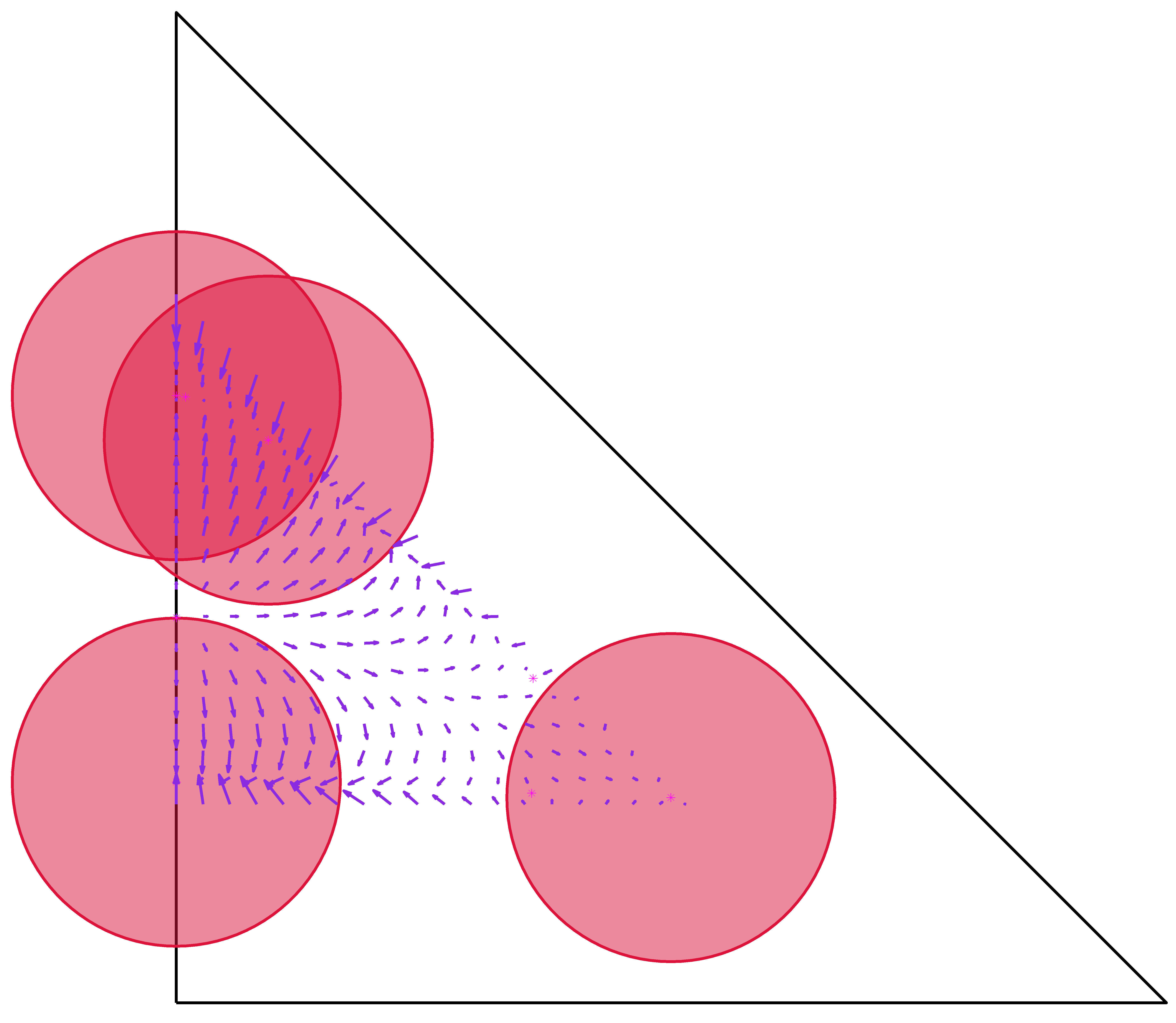}\label{fig:fig6_8}}
\subfloat[]{\includegraphics[width=.29\linewidth]{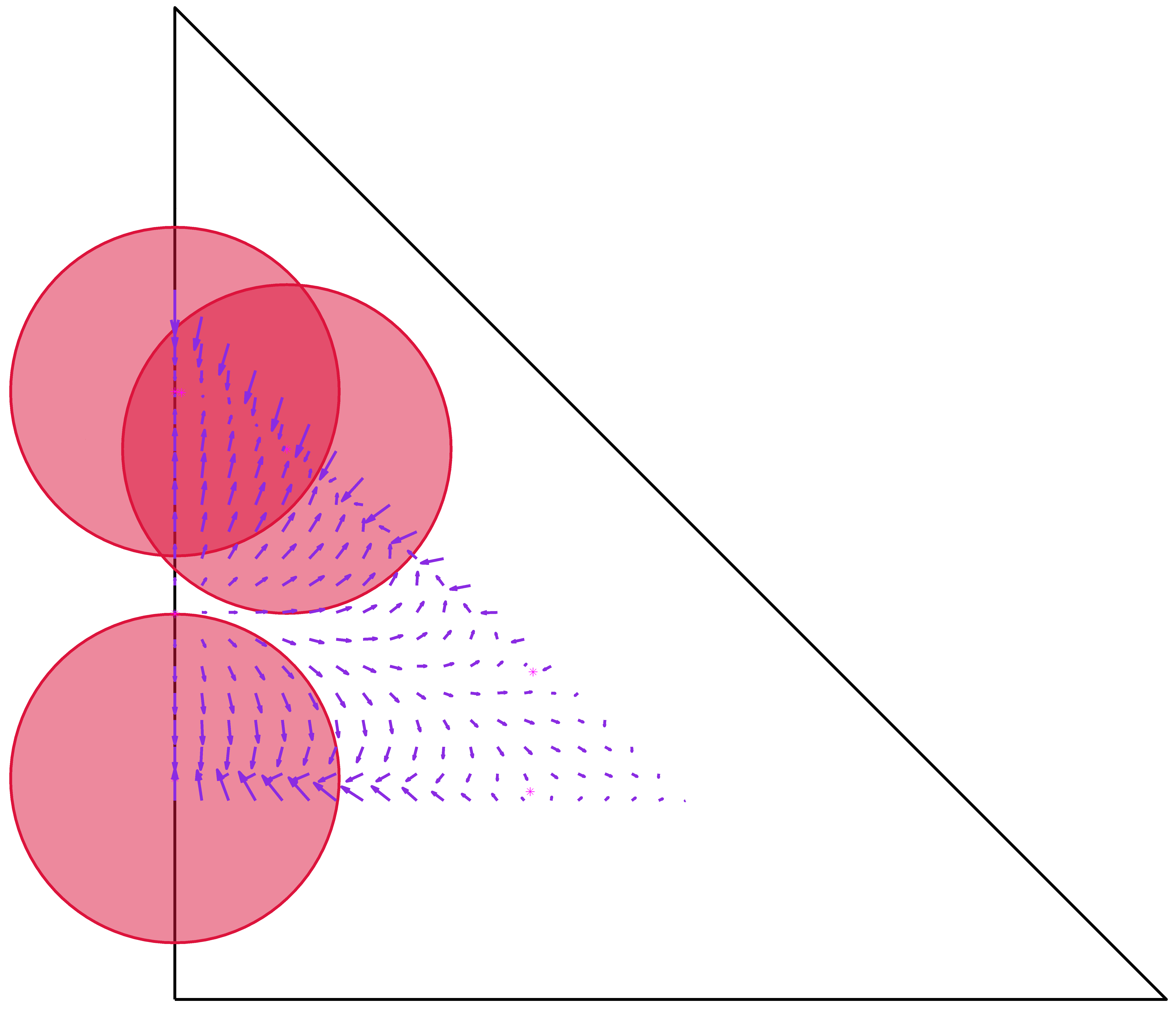}\label{fig:fig6_9}}
\hfill
\subfloat[]{\includegraphics[width=.29\linewidth]{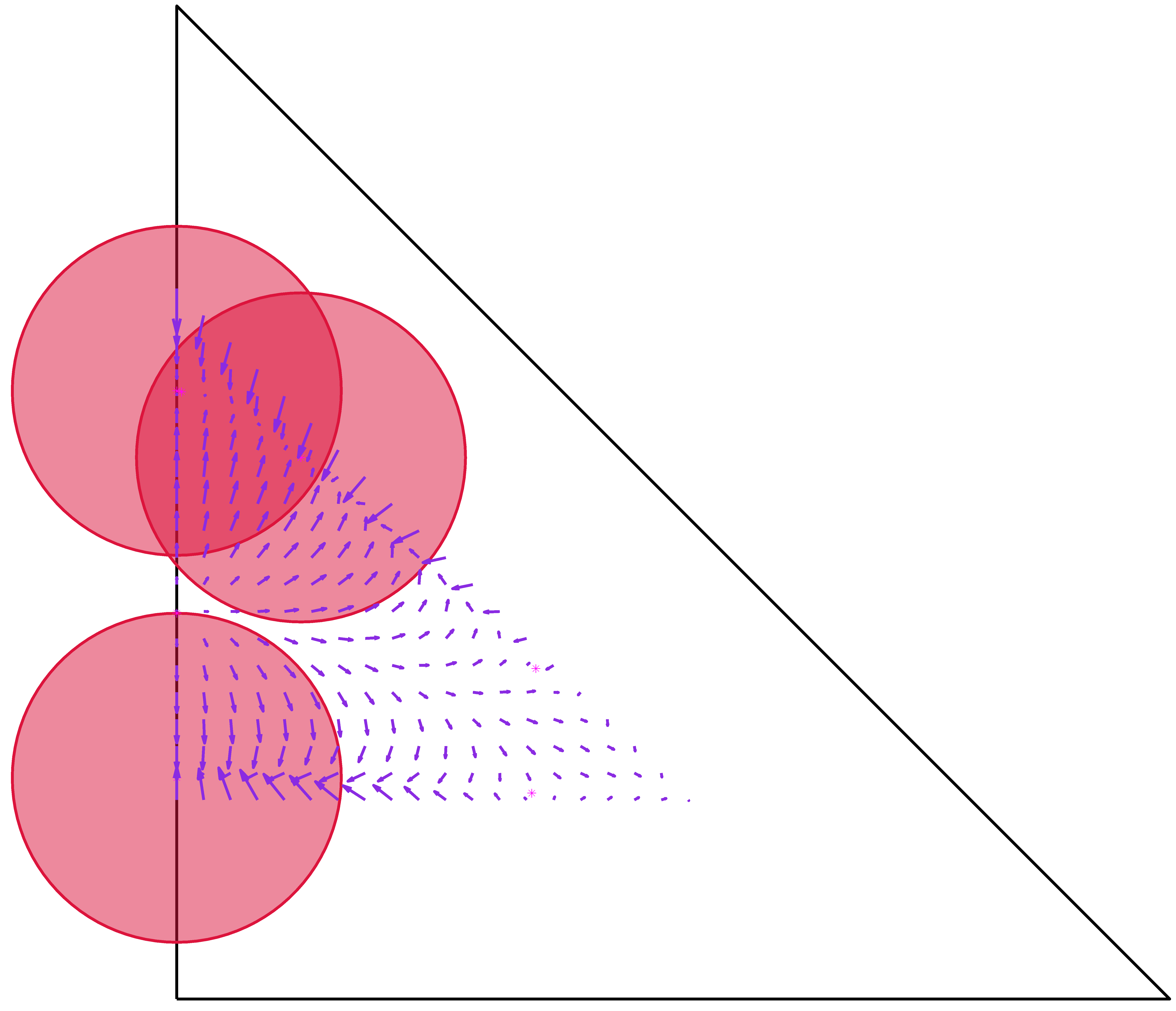}\label{fig:fig6_10}}
\subfloat[]{\includegraphics[width=.29\linewidth]{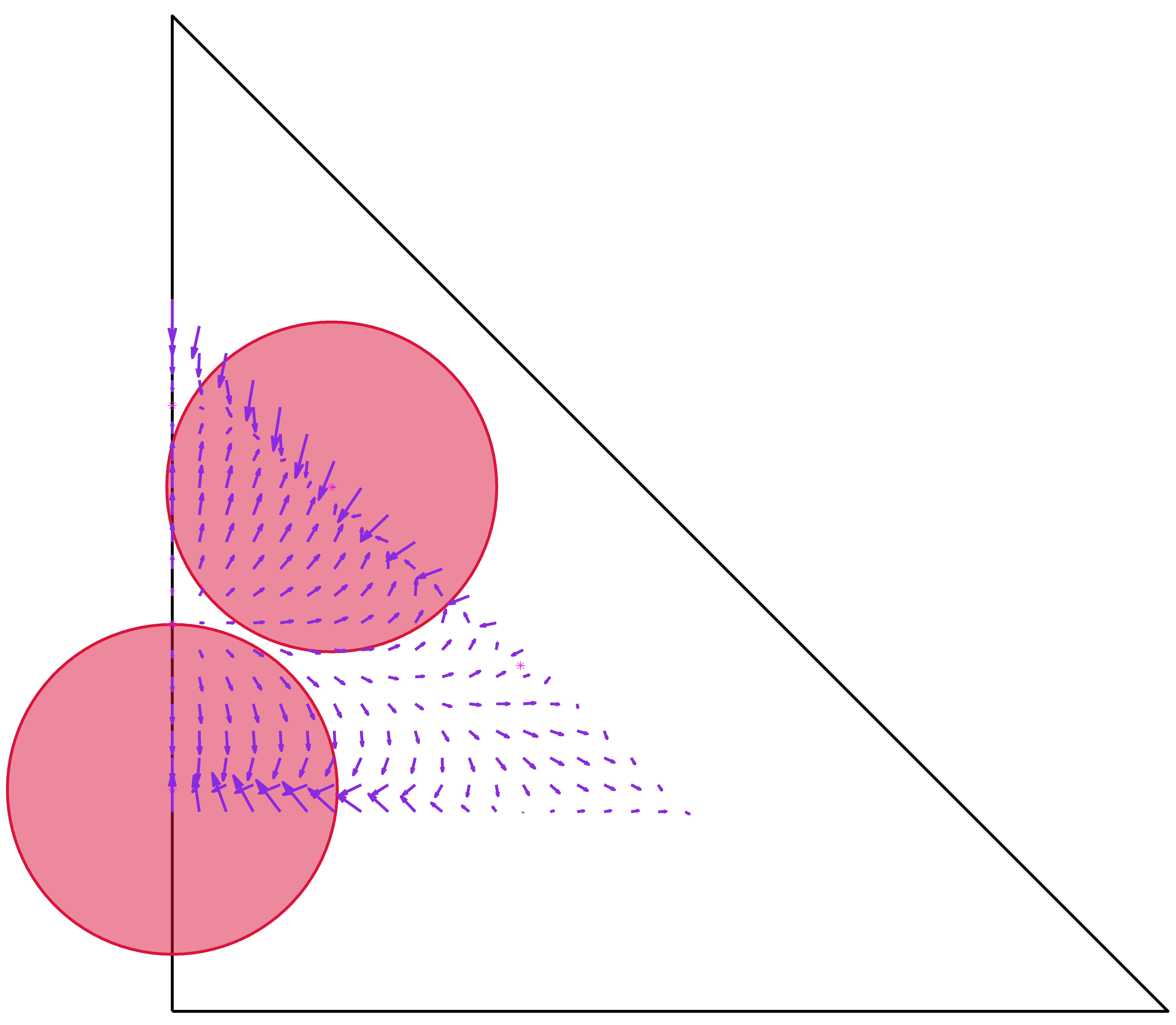}\label{fig:fig6_11}}
\subfloat[]{\includegraphics[width=.29\linewidth]{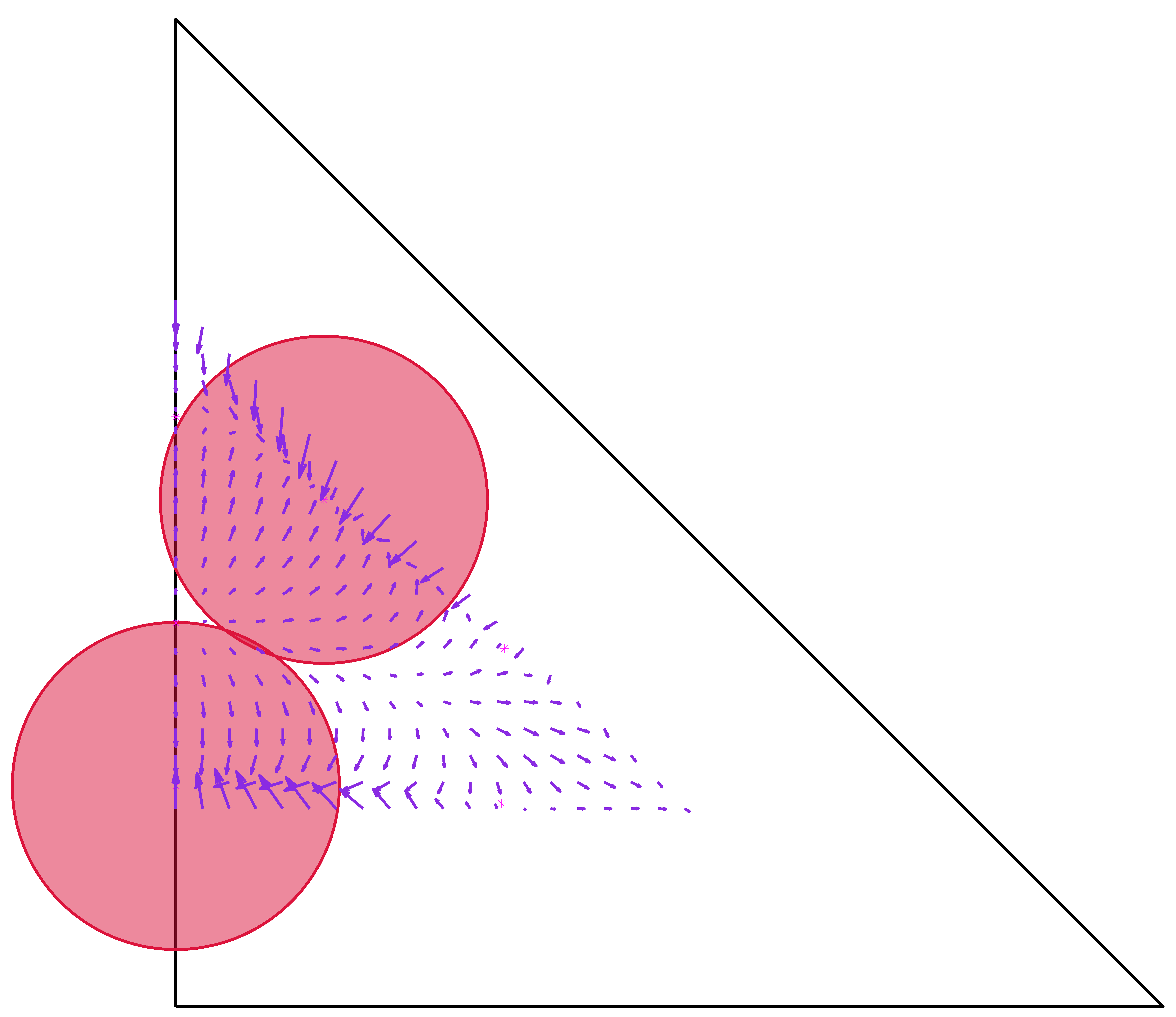}\label{fig:fig6_12}}
\caption{\textit{Focusing patterns}: for {\color{magenta} $\frac{a}{H}$}$ = 0.4$ in a $90^\circ$-channel at ${\color{magenta} Re} = $ \protect\subref{fig:fig6_1} 20 $\vert$ \protect\subref{fig:fig6_2} 60 $\vert$ \protect\subref{fig:fig6_3} 70 $\vert$ \protect\subref{fig:fig6_4} 80 $\vert$ \protect\subref{fig:fig6_5} 90 $\vert$ \protect\subref{fig:fig6_6} 100 $\vert$ \protect\subref{fig:fig6_7} 120 $\vert$  \protect\subref{fig:fig6_8} 130 $\vert$ \protect\subref{fig:fig6_9} 140 $\vert$ \protect\subref{fig:fig6_10} 150 $\vert$ \protect\subref{fig:fig6_11} 200 $\vert$ \protect\subref{fig:fig6_12} 250 (violet arrows represent force-maps, magenta-asterisks represent all equilibrium locations, and red-circles represent the particle (to scale))}
\label{fig:fig6}
\end{figure}

\begin{table}
\centering
\resizebox{\columnwidth}{!}{%
\begin{tabular}{||c|c|c|c|c|c|c|c||} 
 \hline
 {\color{magenta} $Re$} & \textbf{No.} & $\bm{\frac{y}{a}}$ & $\bm{\frac{z}{a}}$ & $\bm{\lambda_{1,norm}}$ & $\bm{\lambda_{2,norm}}$ & $\bm{\lambda_{3,norm}}$ & $\bm{\lambda_{4,norm}}$\\ 
 \hline
 \multirow{3}{*}{\textbf{20}} & 1 & 0.0000 & -0.3380 & -0.0324 & -0.9000 + $i$0.3058 & -0.9000 - $i$0.3058 & -1.7676\\ \cline{2-8}
 & 2 & 0.0000 & 0.8338 & -0.0679 & -0.3680 & -1.4320 & -1.7321\\ \cline{2-8}
 & 3 & 0.9244 & 0.0102 & -0.0321 & -0.1391 & -1.6609 & -1.7679\\ \cline{2-8}
 \hline
 \multirow{2}{*}{\textbf{60}} & 1 & 0.0000 & -0.3328 & -0.6330 & -1.1670 & -0.9000 + $i$1.5952 & -0.9000 - $i$1.5952\\ \cline{2-8}
 & 2 & 0.0000 & 0.8367 & -0.6544 & -1.1456 & -0.9000 + $i$2.4799 & -0.9000 - $i$2.4799\\ \cline{2-8}
 \hline
 \multirow{3}{*}{\textbf{70}} & 1 & 0.0000 & -0.3328 & -0.9000 + $i$0.4969 & -0.9000 - $i$0.4969 & -0.9000 + $i$2.9222 & -0.9000 - $i$2.9222\\ \cline{2-8}
 & 2 & 0.0000 & 0.8384 & -0.9000 + $i$0.3031 & -0.9000 - $i$0.3031 & -0.9000 + $i$1.8577 & -0.9000 - $i$1.8577\\ \cline{2-8}
 & 3 & 1.1749 & -0.2773 & -0.2764 + $i$0.2373 & -0.2764 - $i$0.2373 & -1.5236 + $i$0.2373 & -1.5236 - $i$0.2373\\ \cline{2-8}
 \hline
 \multirow{3}{*}{\textbf{80}} & 1 & 0.0000 & -0.3328 & -0.9000 + $i$0.7760 & -0.9000 - $i$0.7760 & -0.9000 + $i$3.3699 & -0.9000 - $i$3.3699\\ \cline{2-8}
 & 2 & 0.0000 & 0.8400 & -0.9000 + $i$0.4786 & -0.9000 - $i$0.4786 & -0.9000 + $i$2.0975 & -0.9000 - $i$2.0975\\ \cline{2-8}
 & 3 & 1.2925 & -0.298 & -0.0653 & -0.5510 & -1.2490 & -1.7347\\ \cline{2-8}
 \hline
 \multirow{3}{*}{\textbf{90}} & 1 & 0.0000 & -0.3329 & -0.9000 + $i$1.0009 & -0.9000 - $i$1.0009 & -0.9000 + $i$3.8176 & -0.9000 - $i$3.8176\\ \cline{2-8}
 & 2 & 0.0000 & 0.8415 & -0.9000 + $i$0.5661 & -0.9000 - $i$0.5661 & -0.9000 + $i$2.3215 & -0.9000 - $i$2.3215\\ \cline{2-8}
 & 3 & 1.3763 & -0.3267 & -0.9000 + $i$0.3501 & -0.9000 - $i$0.3501 & -0.9000 + $i$1.3437 & -0.9000 - $i$1.3437\\ \cline{2-8}
 \hline
 \multirow{3}{*}{\textbf{100}} & 1 & 0.0000 & -0.3329 & -0.9000 + $i$1.2022 & -0.9000 - $i$1.2022 & -0.9000 + $i$2.5333 & -0.9000 - $i$2.5333\\ \cline{2-8}
 & 2 & 0.0000 & 0.8427 & -0.9000 + $i$0.5978 & -0.9000 - $i$0.5978 & -0.9000 + $i$2.5333 & -0.9000 - $i$2.5333\\ \cline{2-8}
 & 3 & 1.4207 & -0.3402 & -0.9000 + $i$1.2489 & -0.9000 - $i$1.2489 & -0.9000 + $i$1.8060 &  -0.9000 - $i$1.8060\\ \cline{2-8}
 \hline
 \multirow{4}{*}{\textbf{120}} & 1 & 0.0000 & -0.3333 & -0.9000 + $i$1.5746 & -0.9000 - $i$1.5746 & -0.9000 + $i$2.9334 & -0.9000 - $i$2.9334\\ \cline{2-8}
 & 2 & 0.0000 & 0.8440 & -0.9000 + $i$0.4706 & -0.9000 - $i$0.4706 & -0.9000 + $i$2.9334 & -0.9000 - $i$2.9334\\ \cline{2-8}
 & 3 & 0.2494 & 0.7384 & -0.9000 + $i$0.5405 & -0.9000 - $i$0.5405 & -0.9000 + $i$3.5364 & -0.9000 - $i$3.5364\\ \cline{2-8}
 & 4 & 1.4707 & -0.3621 & -0.9000 + $i$1.4770 & -0.9000 - $i$1.4770 & -0.9000 + $i$2.1845 & -0.9000 - $i$2.1845\\ \cline{2-8}
 \hline
 \multirow{4}{*}{\textbf{130}} & 1 & 0.0000 & -0.3334 & -0.9000 + $i$1.7496 & -0.9000 -$i$1.7496 & -0.9000 + $i$5.6338 & -0.9000 - $i$5.6338\\ \cline{2-8}
 & 2 & 0.0000 & 0.8441 & -0.9000 + $i$0.1653 & -0.9000 - $i$0.1653 & -0.9000 + $i$3.1244 & -0.9000 - $i$3.1244\\ \cline{2-8}
 & 3 & 0.2803 & 0.7088 & -0.9000 + $i$0.7720 & -0.9000 - $i$0.7720 & -0.9000 + $i$3.7920 & -0.9000 - $i$3.7920\\ \cline{2-8}
 & 4 & 1.5075 & -0.3806 & -0.9000 + $i$1.0459 & -0.9000 - $i$1.0459 & -0.9000 + $i$2.4369i & -0.9000 - $i$2.4369\\ \cline{2-8}
 \hline
 \multirow{3}{*}{\textbf{140}} & 1 & 0.0000 & -0.3335 & -0.9000 + $i$1.9212 & -0.9000 - $i$1.9212 & -0.9000 + $i$6.1019 & -0.9000 - $i$6.1019\\ \cline{2-8}
 & 2 & 0.0000 & 0.8436 & -0.3822 & -1.4178 & -0.9000 + $i$3.3130 & -0.9000 - $i$3.3130\\ \cline{2-8}
 & 3 & 0.3407 & 0.6689 & -0.6411 & -1.1589 & -0.9000 + $i$5.2319 & -0.9000 - $i$5.2319\\ \cline{2-8}
 \hline
 \multirow{3}{*}{\textbf{150}} & 1 & 0.0000 & -0.3336 & -0.9000 + $i$2.0947 & -0.9000 - $i$2.0947 & -0.9000 + $i$6.5793 & -0.9000 - $i$6.5793\\ \cline{2-8}
 & 2 & 0.0000 & 0.8425 & -0.0775 & -1.7725 & -0.9000 + $i$3.4995 & -0.9000 - $i$3.4995\\ \cline{2-8}
 & 3 & 0.3775 & 0.6400 & -0.9000 + $i$0.8474 & -0.9000 - $i$0.8474 & -0.9000 + $i$5.0415 & -0.9000 - $i$5.0415\\ \cline{2-8}
 \hline
 \multirow{2}{*}{\textbf{200}} & 1 & 0.0000 & -0.3330 & -0.9000 + $i$2.9914 & -0.9000 - $i$2.9914 & -0.9000 + $i$9.1508 & -0.9000 - $i$9.1508\\ \cline{2-8}
 & 2 & 0.4831 & 0.5840 & -0.9000 + $i$1.7910 & -0.9000 - $i$1.7910 & -0.9000 + $i$6.4721 & -0.9000 + $i$6.4721\\ \cline{2-8}
 \hline
 \multirow{2}{*}{\textbf{250}} & 1 & 0.0000 & -0.3314 & -0.0900 + $i$0.4179i & -0.0900 - $i$0.4179i & -0.0900 + $i$1.2064i & -0.0900 - $i$1.2064i\\ \cline{2-8}
 & 2 & 0.4524 & 0.5430 & -0.0900 + $i$0.4126i & -0.0900 - $i$0.4126i & -0.0900 + $i$0.6936i & -0.0900 - $i$0.6936i\\ \cline{2-8}
 \hline
\end{tabular}
}
\caption{\textit{Normalized eigenvalues}: for stable points for isosceles right triangular-channel with {\color{magenta} $\frac{a}{H}$}$ = 0.4$ (origin taken at the centroid of the channel)}
\label{tab:tab3}
\end{table}
\justify
From FIG. \ref{fig:fig6}, it is seen that stable points are observed at the top-centre, bottom-centre, and right-corner with {\color{magenta} $Re$} $< 120$ (except 60) (FIG. \ref{fig:fig6_1}-\ref{fig:fig6_6}). Furthermore, the stable focusing location at the right-corner is newly revealed by the algorithm, which has not been observed in experiments. For {\color{magenta} $Re$} $= 60$, the focusing pattern comprises the top and bottom mid-plane focusing positions, which is an exact match with experiments. For {\color{magenta} $Re$} $\ge 120$, in addition to all of the above stable points, we see that there is an off-centre stable location at the top. Thus, {\color{magenta} $Re$} $= 120$ is inferred to be the bifurcation-{\color{magenta} $Re$}  for the present case. Additionally, the right-corner focusing location at lower {\color{magenta} $Re$} completely vanishes at higher flow-speeds (FIG. \ref{fig:fig6_9}-\ref{fig:fig6_10}), and the bifurcated focusing pattern agrees with experiments for ${\color{magenta} Re} \ge 150$ (FIG. \ref{fig:fig6_11}-\ref{fig:fig6_12}). Hence it is confirmed that the predictions follow the general trend of focusing patterns, i.e., 2-point centreline focusing (top, bottom) for lower {\color{magenta} $Re$}, while higher {\color{magenta} $Re$} exhibits 3-point focusing (top off-centre, and bottom-centre). TAB. \ref{tab:tab3} gives eigenvalues for the attractor points at various {\color{magenta} $Re$}'s. It is interesting to note the nature of these stable points, as is evident from variations with {\color{magenta} $Re$}. Locations with purely real eigenvalues would exhibit an exponential decay in perturbations whereas those with an imaginary component would possess oscillatory components in velocity and displacements. It is also seen that for region away from the bifurcation-{\color{magenta} $Re$}, the focusing locations tend to have purely real components whereas the oscillatory components have a strong presence in the near-vicinity of bifurcation. Additionally, different stable points in a focusing pattern may have different types of stable eigenvalues (purely real/complex). Physically, this implies that particles would have unique ways of focusing to these locations and different responses to impulsive perturbations in flow (eg. pressure-jumps, variable cross-section etc.).

We next address the following issues:

\begin{itemize}
    \item \textbf{A}: {\color{magenta} $Re$}-matching for onset of bifurcation with experiments (at ${\color{magenta} Re} = 80$) and, 
    \item \textbf{B}: accounting for the presence of miscellaneous stable points which are not observed in experiments (for eg. the right-corner positions)
\end{itemize}

\begin{figure}[!htpb]
\centering
\subfloat[]{\includegraphics[width=.26\linewidth]{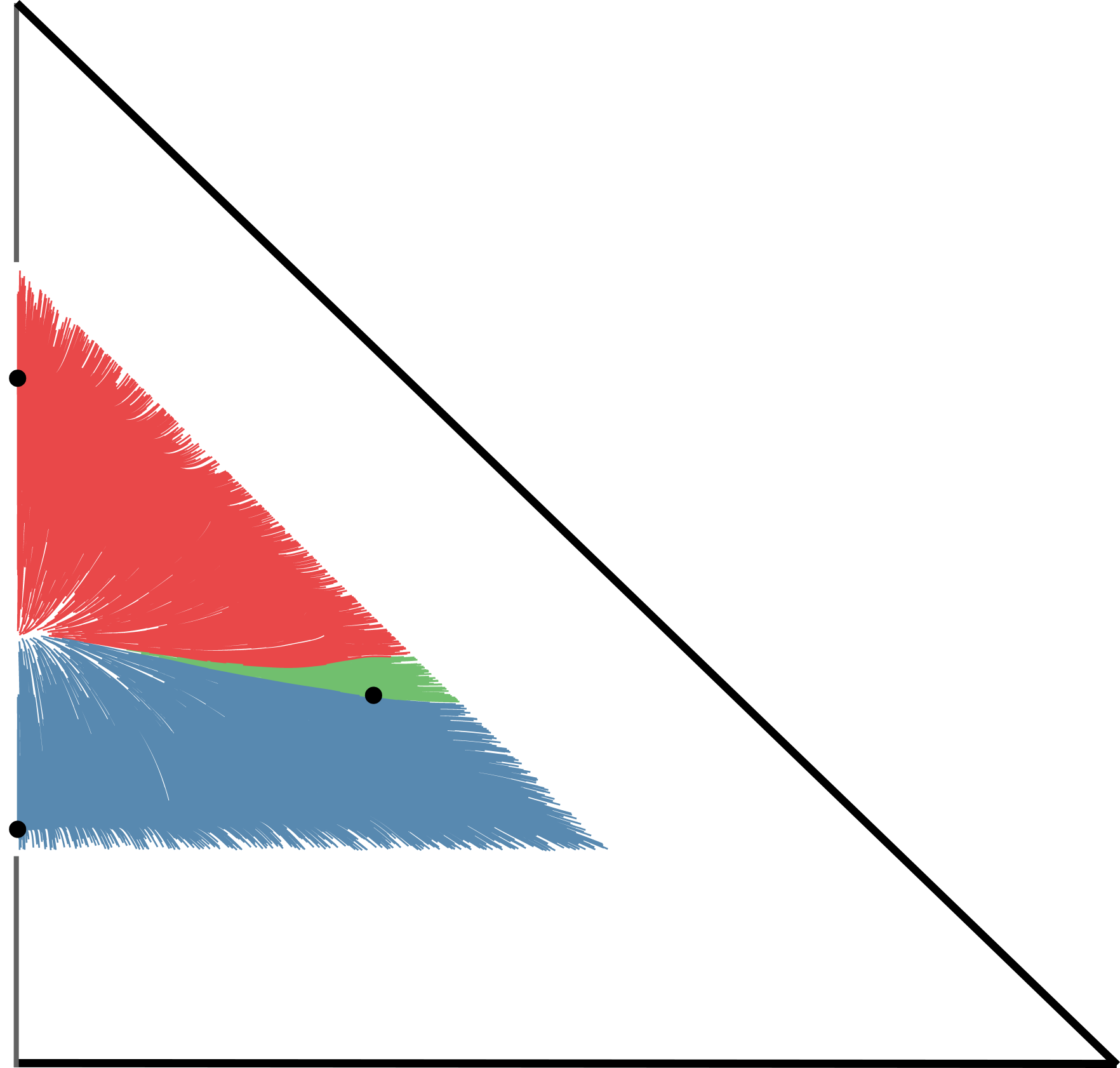}\label{fig:fig7_1}}
\subfloat[]{\includegraphics[width=.26\linewidth]{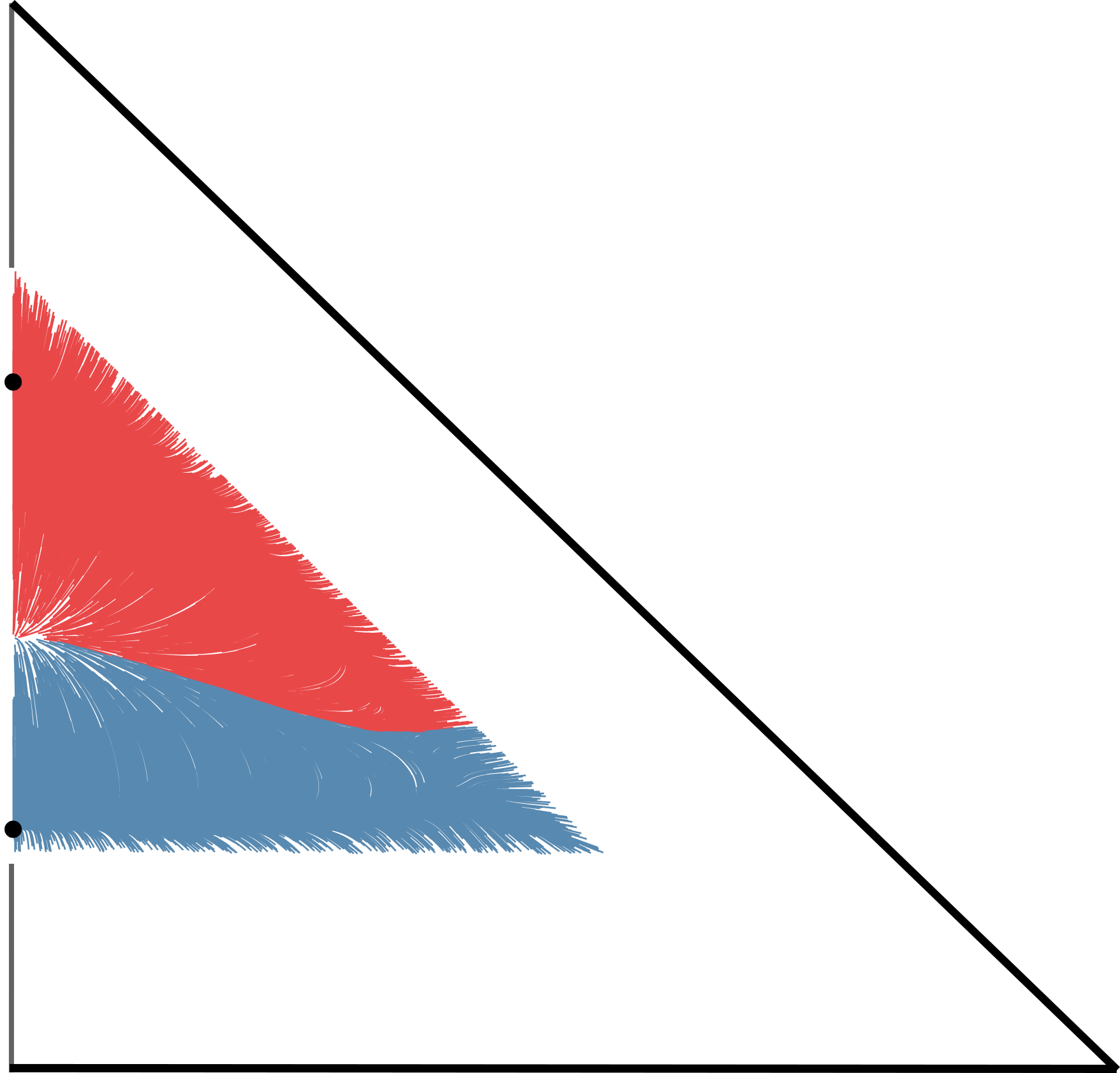}\label{fig:fig7_2}}
\subfloat[]{\includegraphics[width=.26\linewidth]{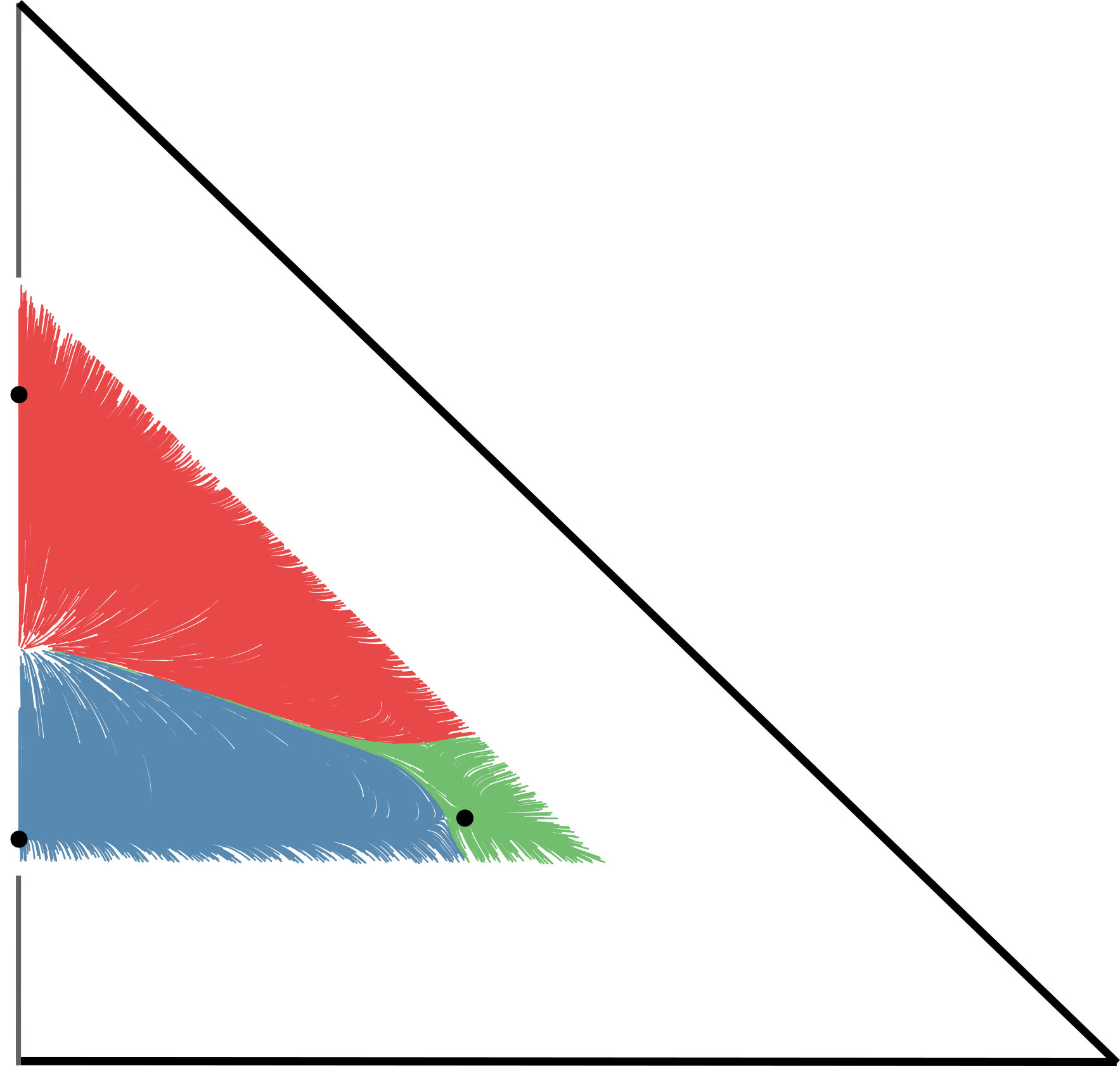}\label{fig:fig7_3}}
\hfill
\subfloat[]{\includegraphics[width=.26\linewidth]{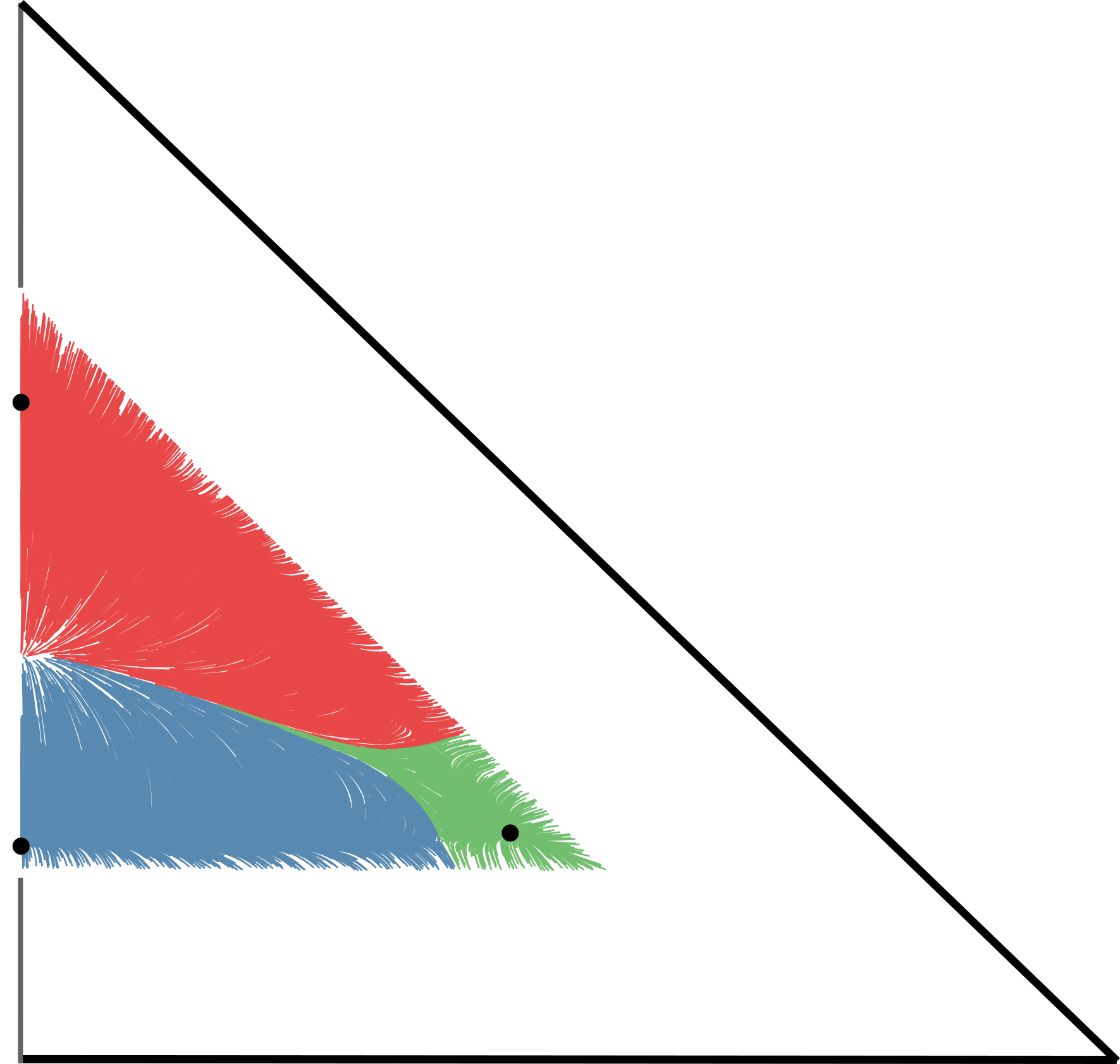}\label{fig:fig7_4}}
\subfloat[]{\includegraphics[width=.26\linewidth]{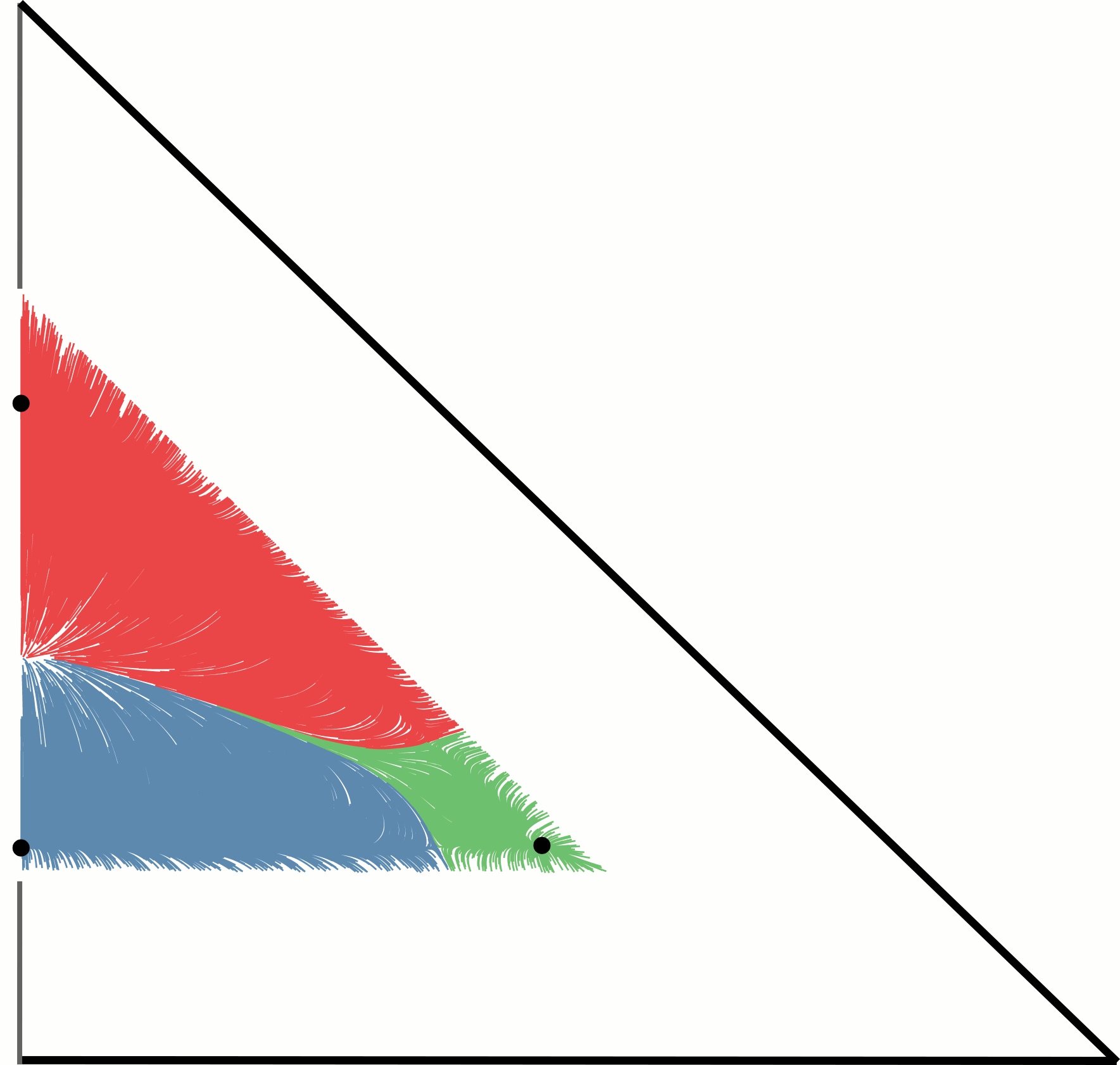}\label{fig:fig7_5}}
\subfloat[]{\includegraphics[width=.26\linewidth]{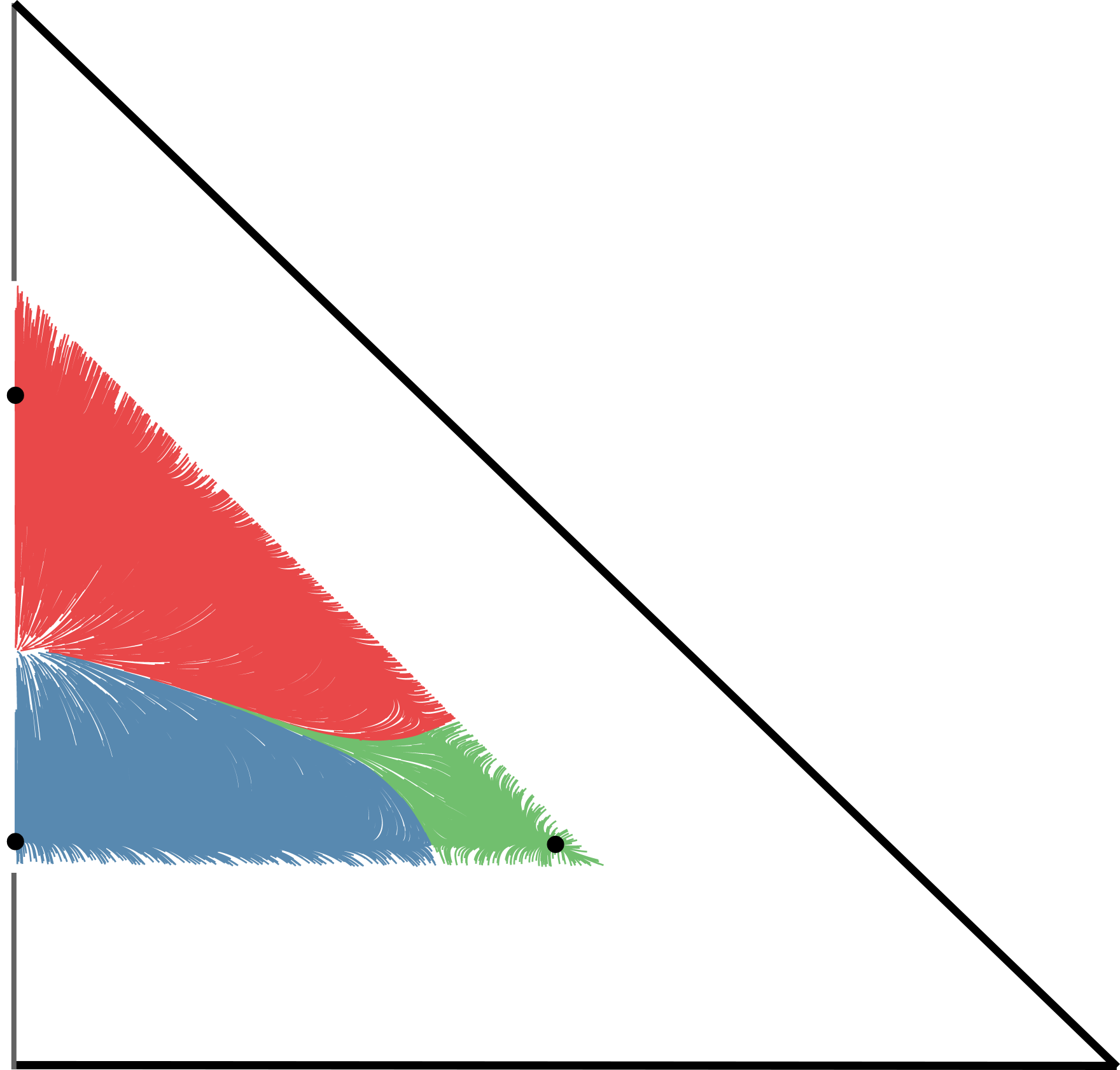}\label{fig:fig7_6}}
\hfill
\subfloat[]{\includegraphics[width=.26\linewidth]{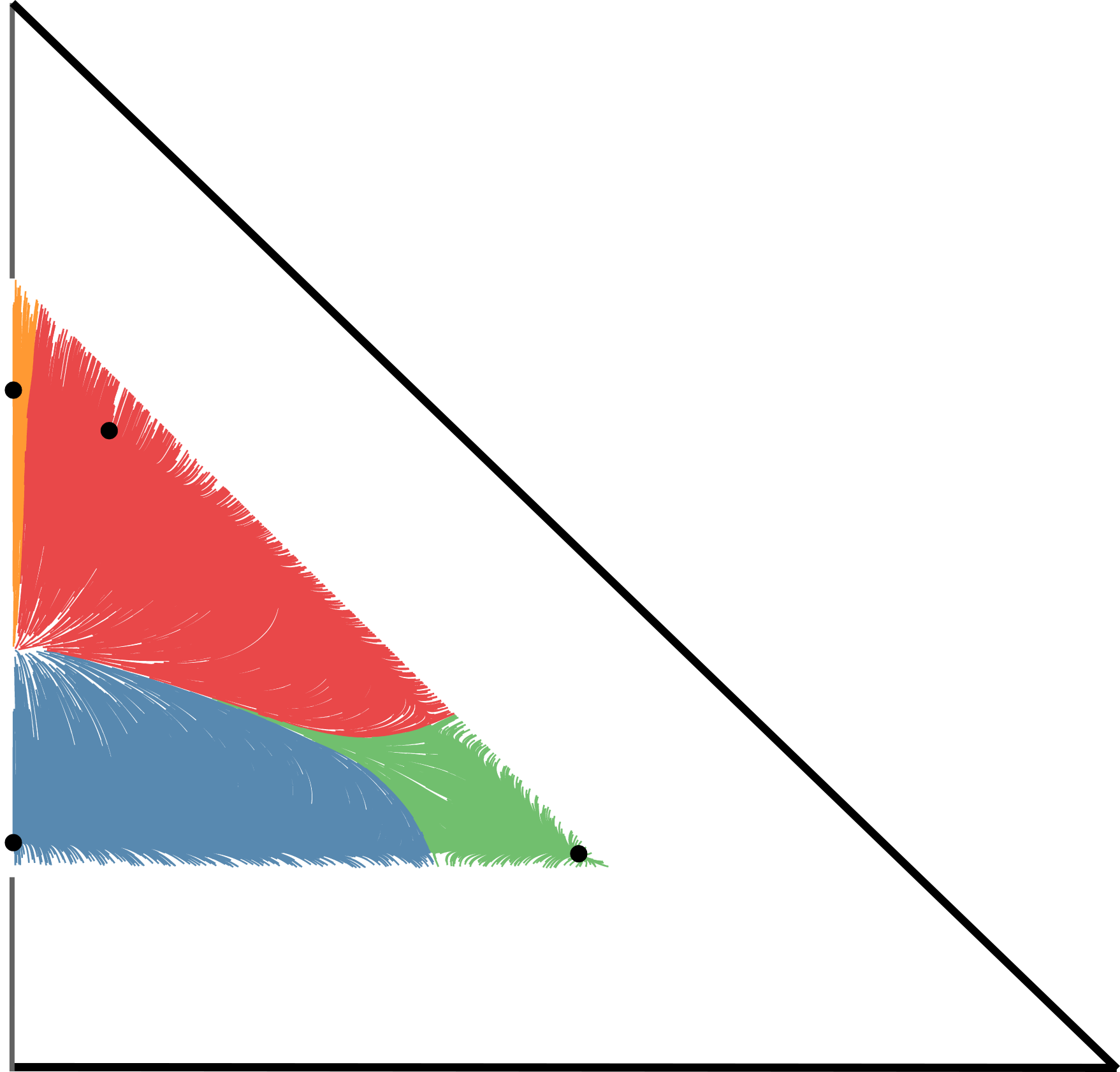}\label{fig:fig7_7}}
\subfloat[]{\includegraphics[width=.26\linewidth]{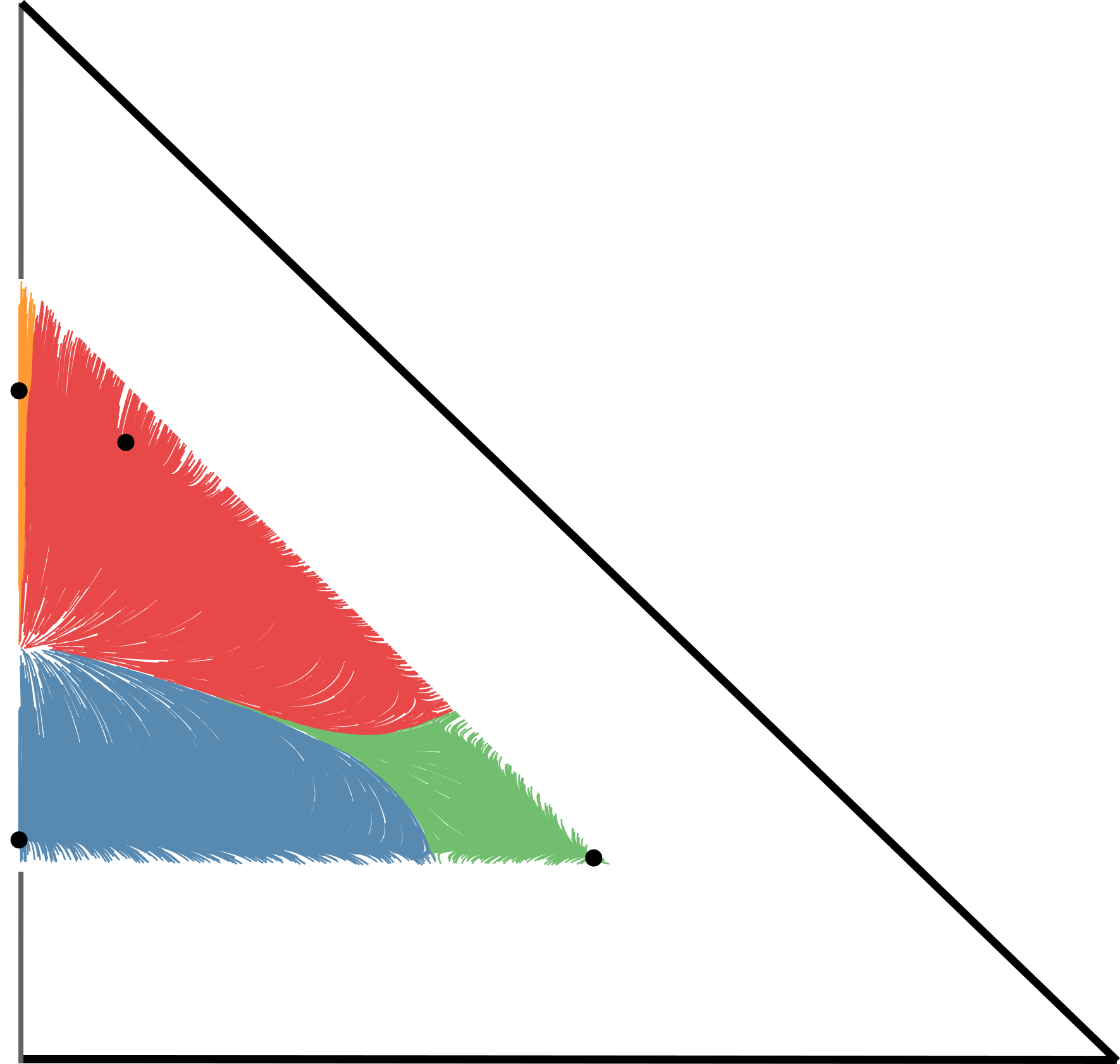}\label{fig:fig7_8}}
\subfloat[]{\includegraphics[width=.26\linewidth]{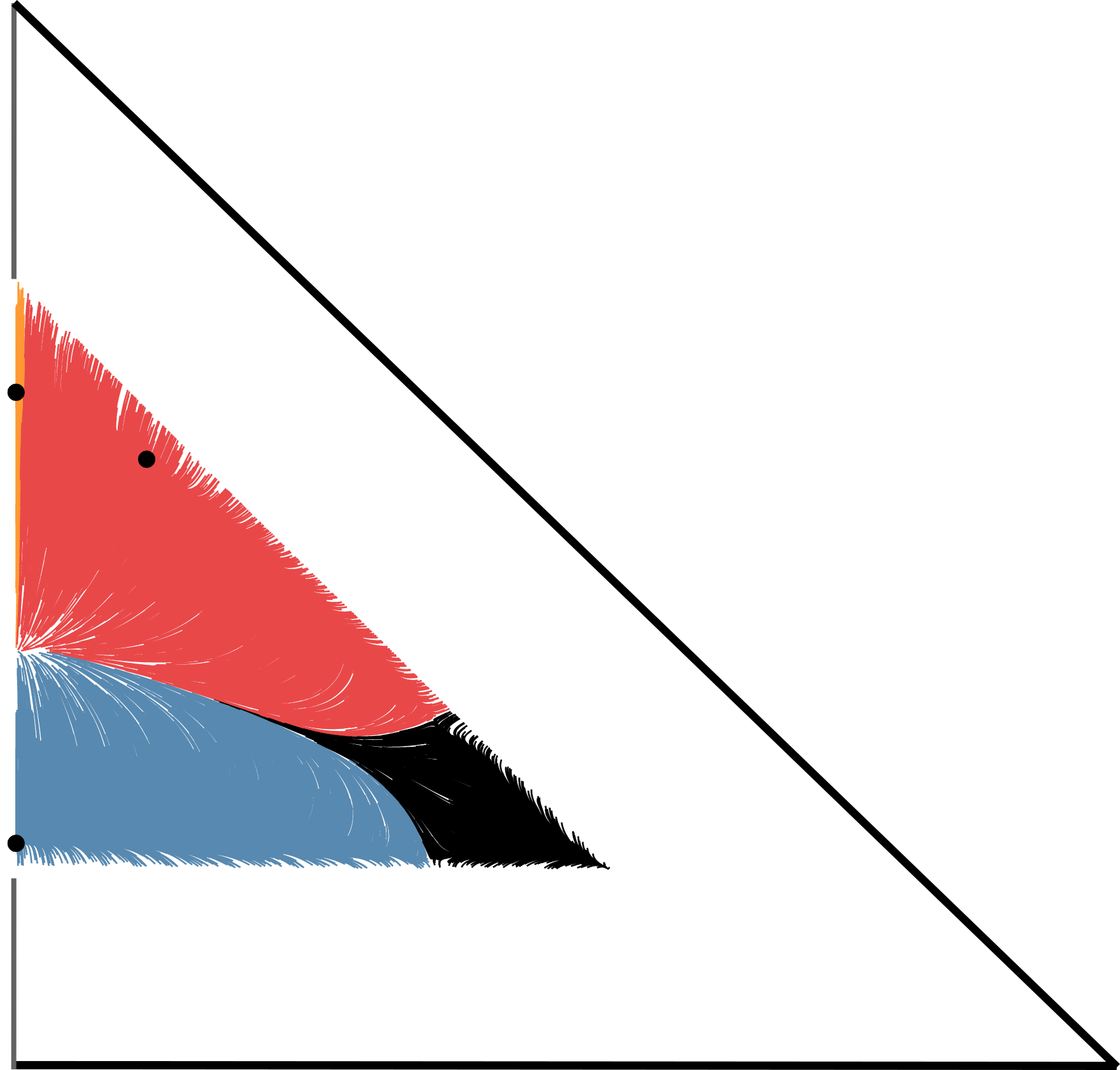}\label{fig:fig7_9}}
\hfill
\subfloat[]{\includegraphics[width=.26\linewidth]{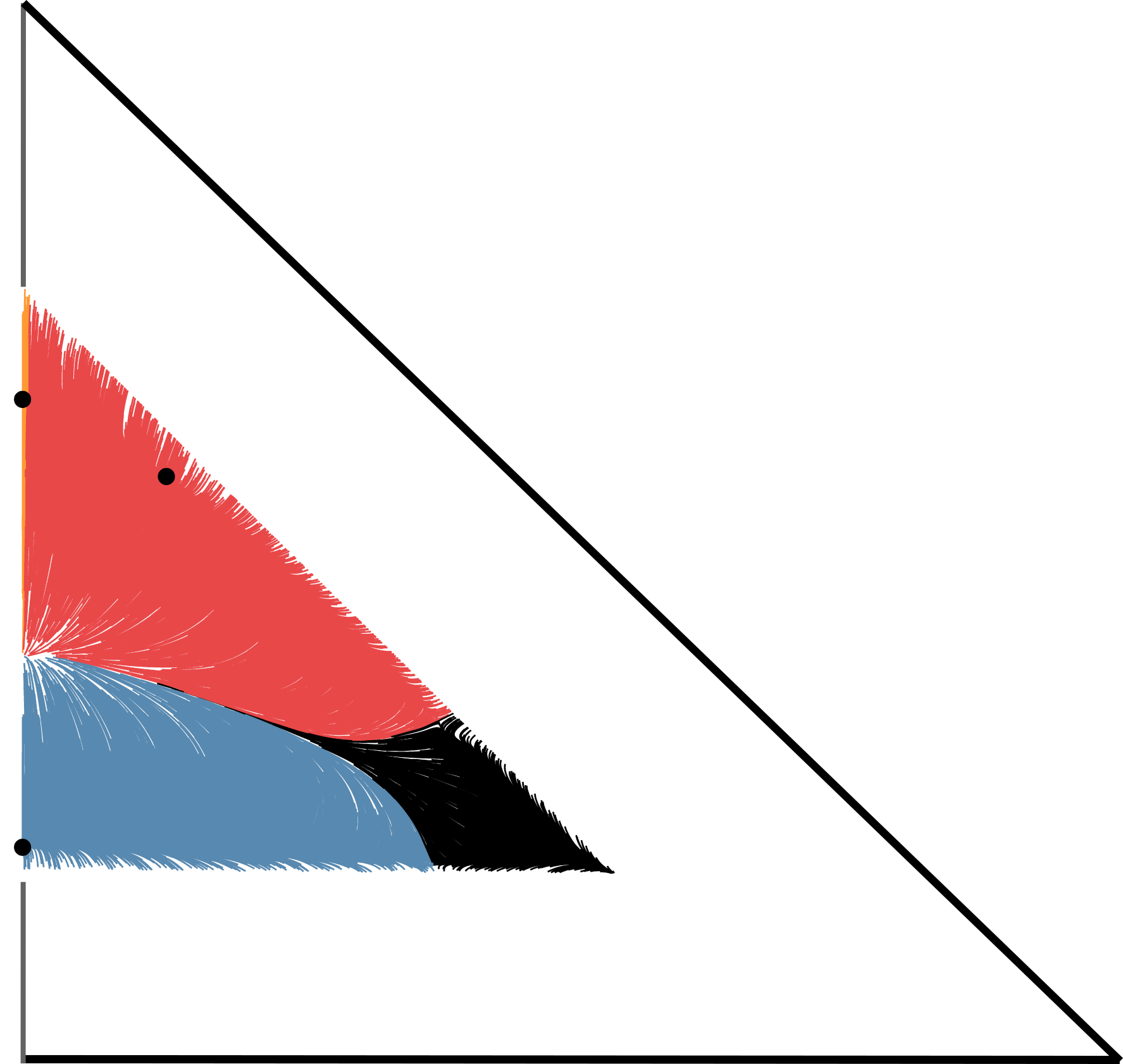}\label{fig:fig7_10}}
\subfloat[]{\includegraphics[width=.26\linewidth]{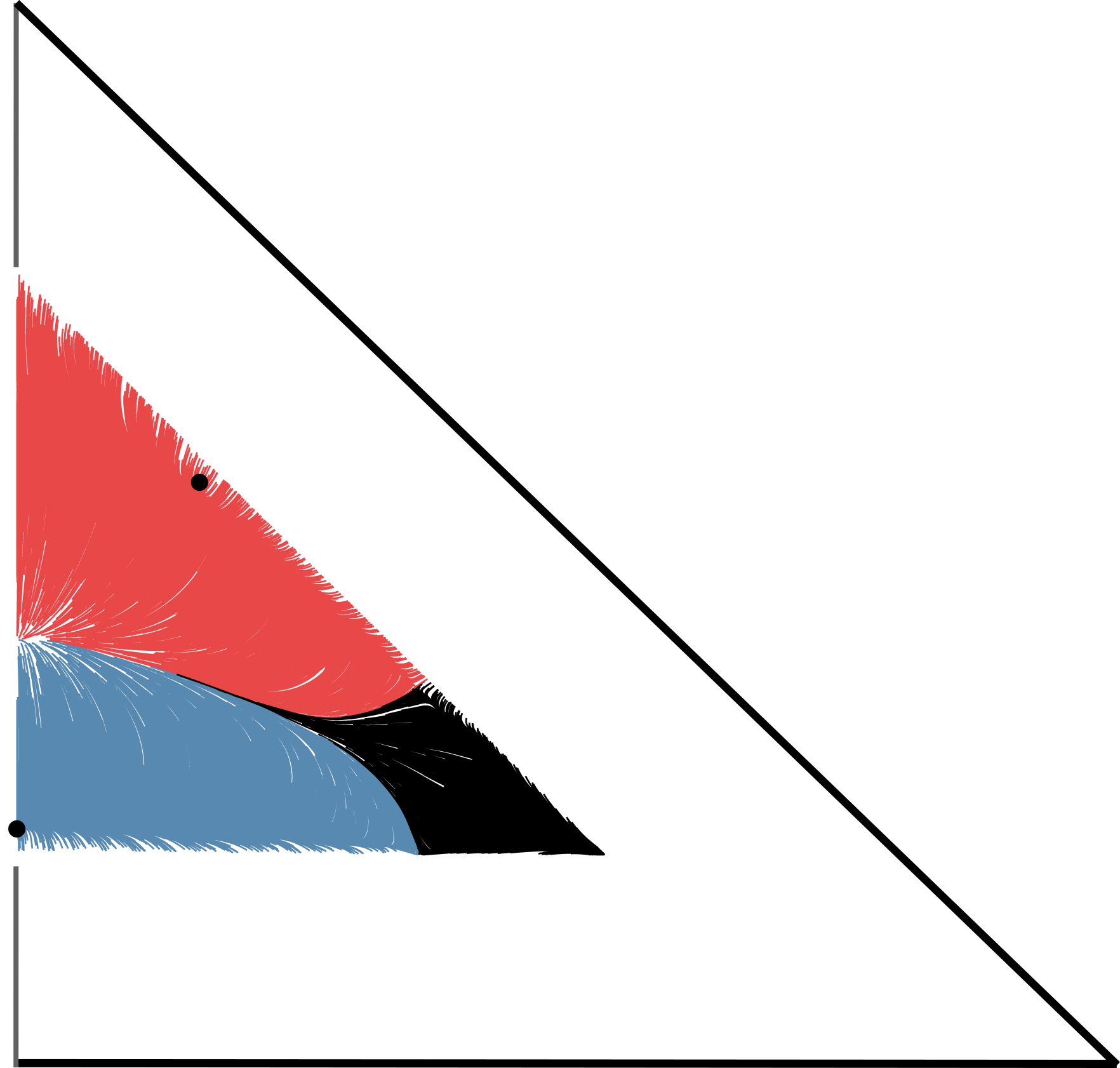}\label{fig:fig7_11}}
\subfloat[]{\includegraphics[width=.26\linewidth]{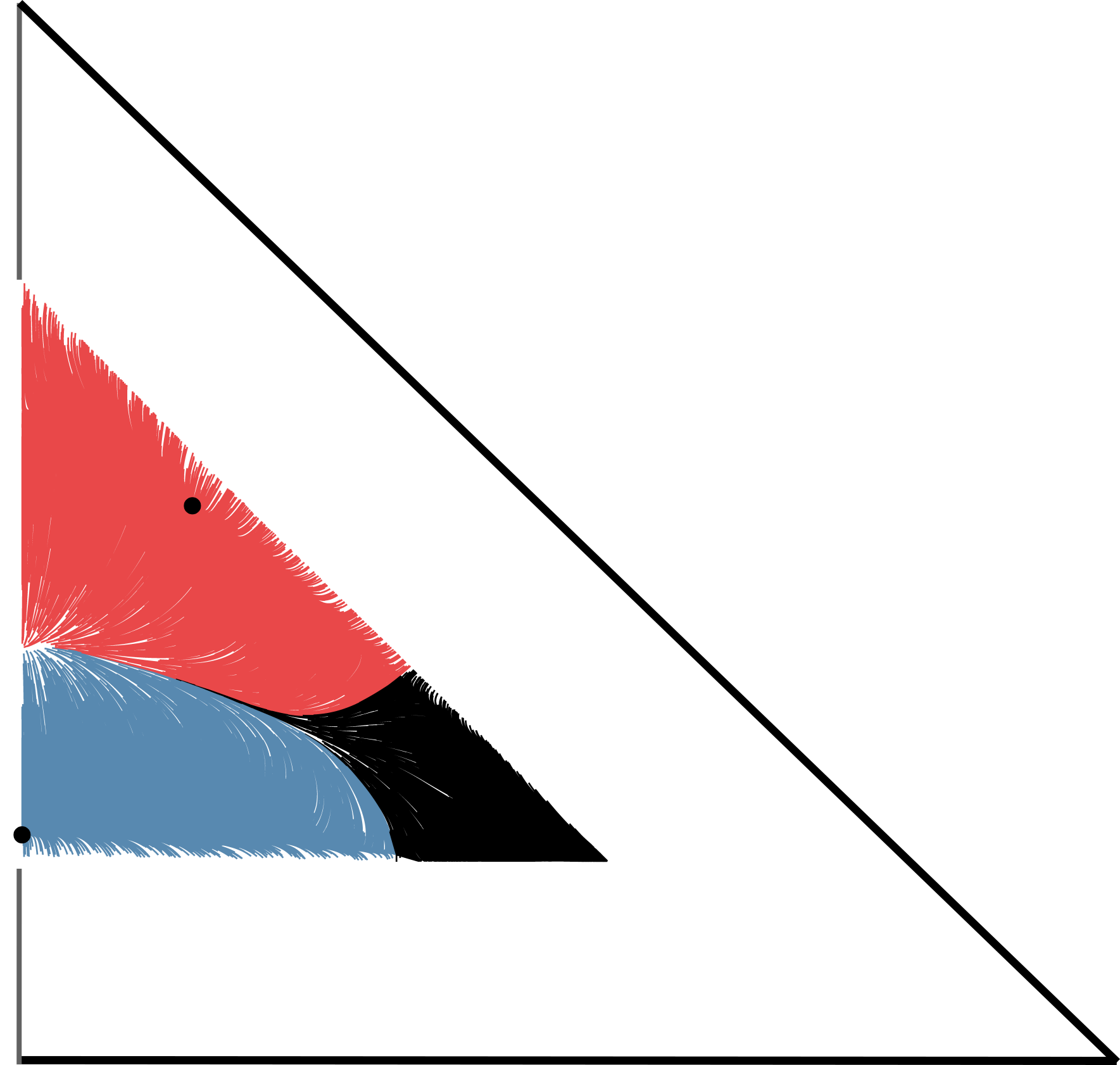}\label{fig:fig7_12}}
\caption{\textit{Basins of attraction}: for {\color{magenta} $\frac{a}{H}$}$ = 0.4$ in a $90^\circ$-channel at ${\color{magenta} Re} = $ \protect\subref{fig:fig7_1} 20 $\vert$ \protect\subref{fig:fig7_2} 60 $\vert$ \protect\subref{fig:fig7_3} 70 $\vert$ \protect\subref{fig:fig7_4} 80 $\vert$ \protect\subref{fig:fig7_5} 90 $\vert$ \protect\subref{fig:fig7_6} 100 $\vert$ \protect\subref{fig:fig7_7} 120 $\vert$  \protect\subref{fig:fig7_8} 130 $\vert$ \protect\subref{fig:fig7_9} 140 $\vert$ \protect\subref{fig:fig7_10} 150 $\vert$ \protect\subref{fig:fig7_11} 200 $\vert$ \protect\subref{fig:fig7_12} 250 (basins demarcated by color; the black dots represent stable locations; note that the basins shown here span ONLY the sampled region, which is smaller than the half-channel cross-sectional area, and black lines represent the half-channel boundary (sliced top-down, to scale))}
\label{fig:fig7}
\end{figure}

We first examine issue \textbf{B} by consulting the basins of attraction for the cases under consideration, as shown in FIG \ref{fig:fig7}. For all the cases (wherever applicable), it is seen that the basin of attraction for the right-corner point is significantly smaller compared to the top-centre, top off-centre, and bottom-centre focusing points. In addition, the basin of attraction for the corner focusing-location is confined to a narrow sliver in the bulk-region of the flow, and a wider zone towards the corner. Since the likelihood of releasing particles in either of these two regions for the corner-basin is low, we conclude that the corner focusing location is not realized in experiments. The black-regions for FIG. \ref{fig:fig7_9}, \ref{fig:fig7_10}, \ref{fig:fig7_11}, and \ref{fig:fig7_12} represent possible basins for a stable point which lies outside the sampled region. This can be seen from the fact that the corner stable point moves closer to the corner with increasing {\color{magenta} $Re$} (from FIG. \ref{fig:fig7_3}-\ref{fig:fig7_8}). For low- to moderate- {\color{magenta} $Re$} ($< 120$), the focusing pattern comprises mainly of the top-centre and bottom-centre positions. By applying a similar argument to the higher {\color{magenta} $Re$} ($\ge 120$), we eliminate the contribution of the corner location to the focusing patterns. Additionally, we are also able to eliminate the presence of the top-centre focusing location due to a negligible basin of attraction, and thus, the final focusing pattern is a 3-centred pattern with 1 location at the bottom-centre, and the other 2 being the off-centre top locations. Thus, the final focusing patterns are seen to be a 2-centred pattern for moderate {\color{magenta} $Re$} ($< 120$), or a 3-centred pattern for higher {\color{magenta} $Re$} ($\ge 120$). Even though one stable point might have a faster decay-rate and more stable than another location, the basins of attraction are the primary ``guiding'' factors in governing the focusing pattern \textit{before} focusing has been achieved as these are global measures of the force-field whereas the eigenvalues of each stable point are pertinent only up to a neighborhood around the particle, and are useful measures that govern stability to perturbation \textit{after} particles are focused. This transition agrees with experiments on a broad level, although the precise {\color{magenta} $Re$} for matching bifurcation in simulations (${\color{magenta} Re} = 120$) needs further investigation, which is deferred until  \cref{sssec:convergence}.

\centering
\paragraph*{Explaining the bifurcation}
\begin{figure}[!htpb]
\centering
\subfloat[]{\includegraphics[width=.7\linewidth]{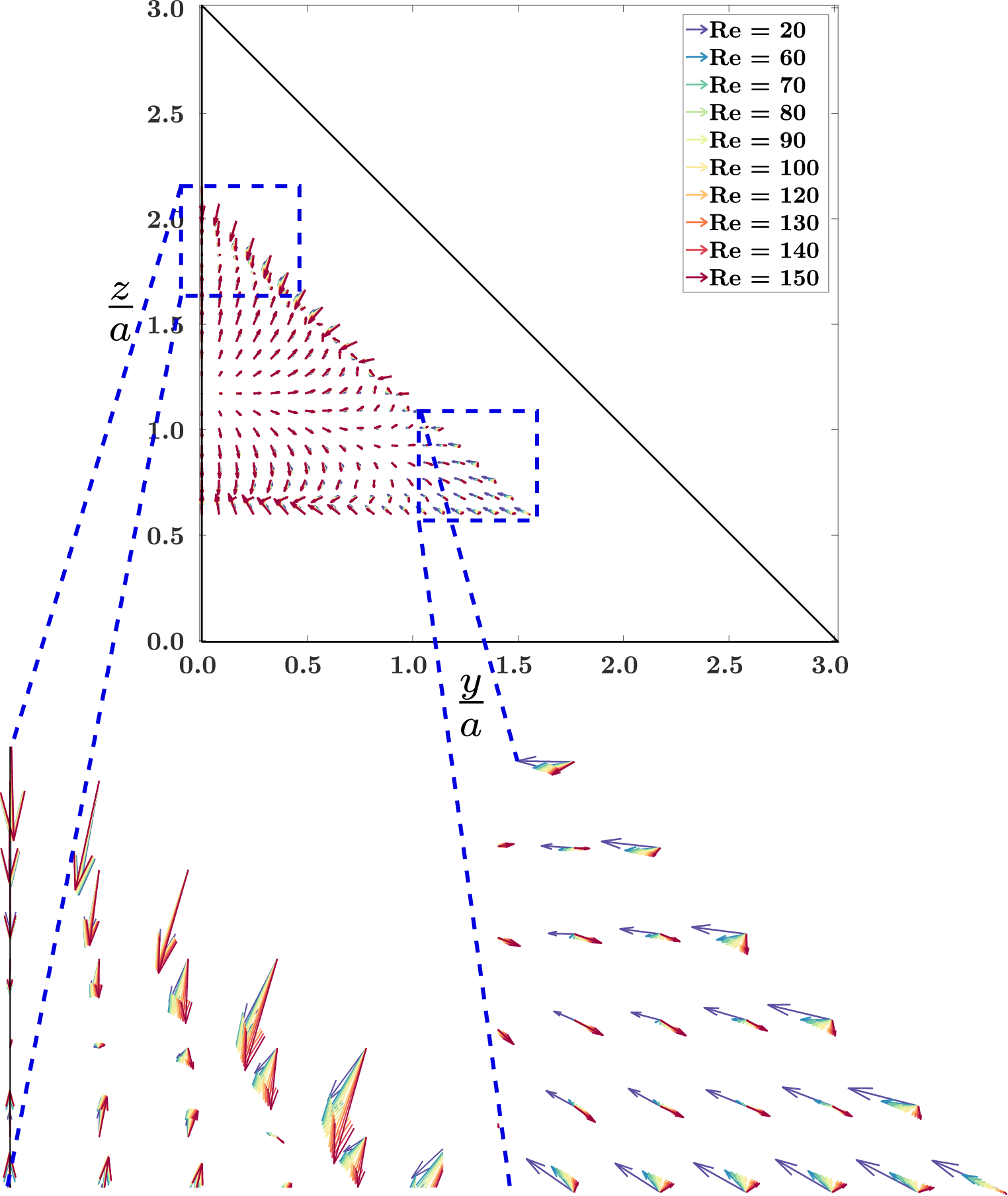}\label{fig:fig8_1}}
\newline
\subfloat[]{\includegraphics[width=.45\linewidth]{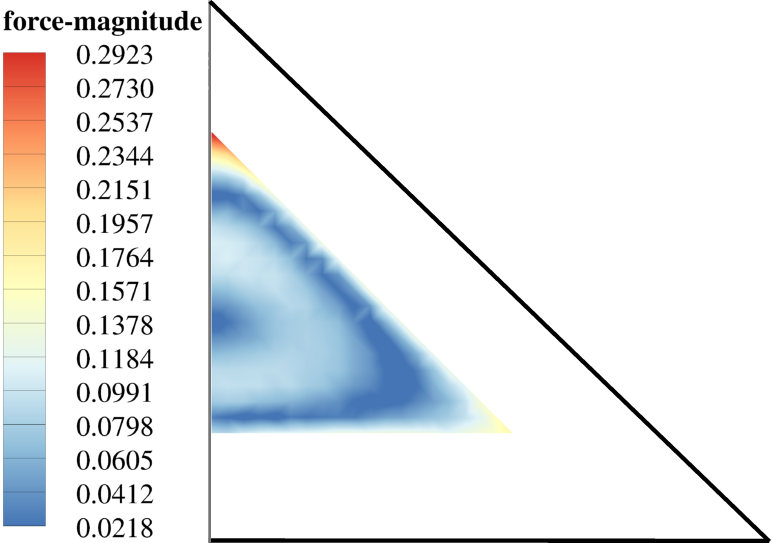}\label{fig:fig8_2}}
\subfloat[]{\includegraphics[width=.45\linewidth]{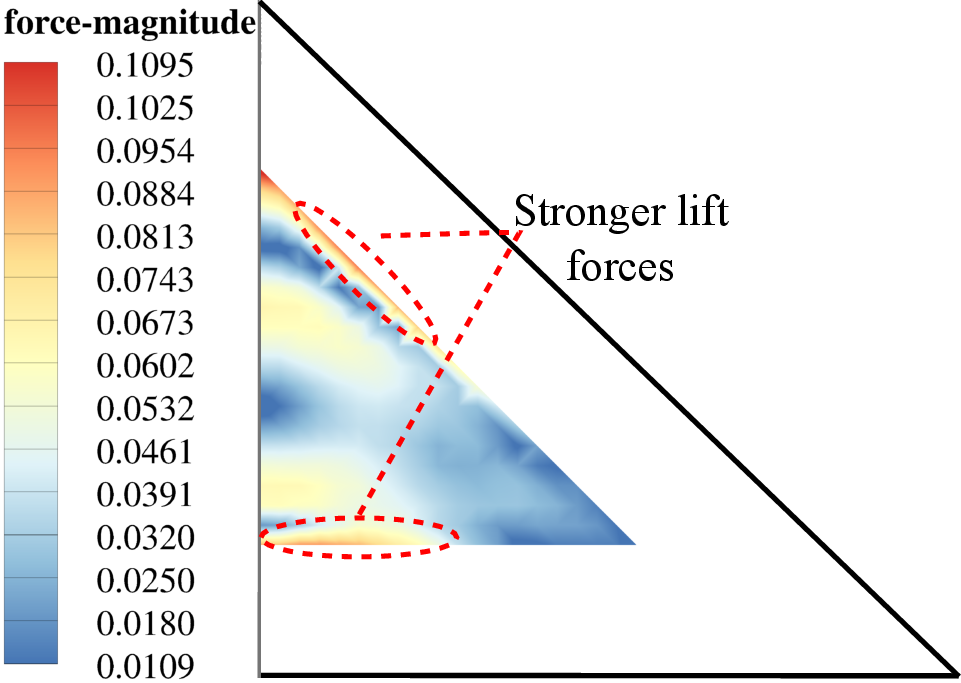}\label{fig:fig8_3}}
\caption{\textit{Force-maps}: for {\color{magenta} $\frac{a}{H}$}$ = 0.4$ in a $90^\circ$-channel \protect\subref{fig:fig8_1} overlayed for sampled-region (with insets for top and corner focusing locations, and black lines represent the half-channel boundary (sliced top-down, to scale)) $\vert$ \textit{Magnitude of lift-forces}: at ${\color{magenta} Re} = $ \protect\subref{fig:fig8_2} 20 $\vert$ \protect\subref{fig:fig8_3} 150}
\label{fig:fig8}
\end{figure}

\justify
Force-maps for the particle at various {\color{magenta} $Re$}'s are shown in FIG. \ref{fig:fig8_1}. While the trends remain similar in the bulk of the channel, it is interesting to note the change in forces around the top and side corners. Firstly, the bottom-centre focusing position appears to be unconditionally stable owing to the forces creating a ``sink" for solution trajectories at that location. Secondly, we see that at the top focusing position, at low-{\color{magenta} $Re$}, the forces similarly create a stable node. At high-{\color{magenta} $Re$}, however, the forces along the z-axis tend to introduce a saddle-point. This destabilizing-effect is thought to be the primary cause of the bifurcation, akin to the that seen in high aspect-ratio rectangular channels, where a decrease in {\color{magenta} $Re$} destabilizes the focusing locations along the short-faces. Lastly, the reversal of forces at the right-corner point can be seen at high-{\color{magenta} $Re$}, which is again testament to the increased inertial lifts. However, the destabilizing-forces at the top focusing location are counter-acted by a corresponding net-force along the side-walls at the off-centre locations, and this leads to stabilization of the off-centre locations (FIG. \ref{fig:fig8_3}), which is not seen for low-{\color{magenta} $Re$}.

\subsubsection{Shifting of focusing position after bifurcation}

\begin{figure}
\centering 
\subfloat[]{\includegraphics[width=.45\linewidth]{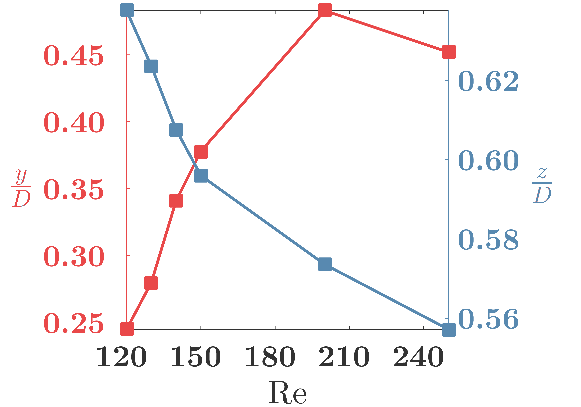}\label{fig:fig9_1}}
\hfill
\subfloat[]{\includegraphics[width=1\linewidth]{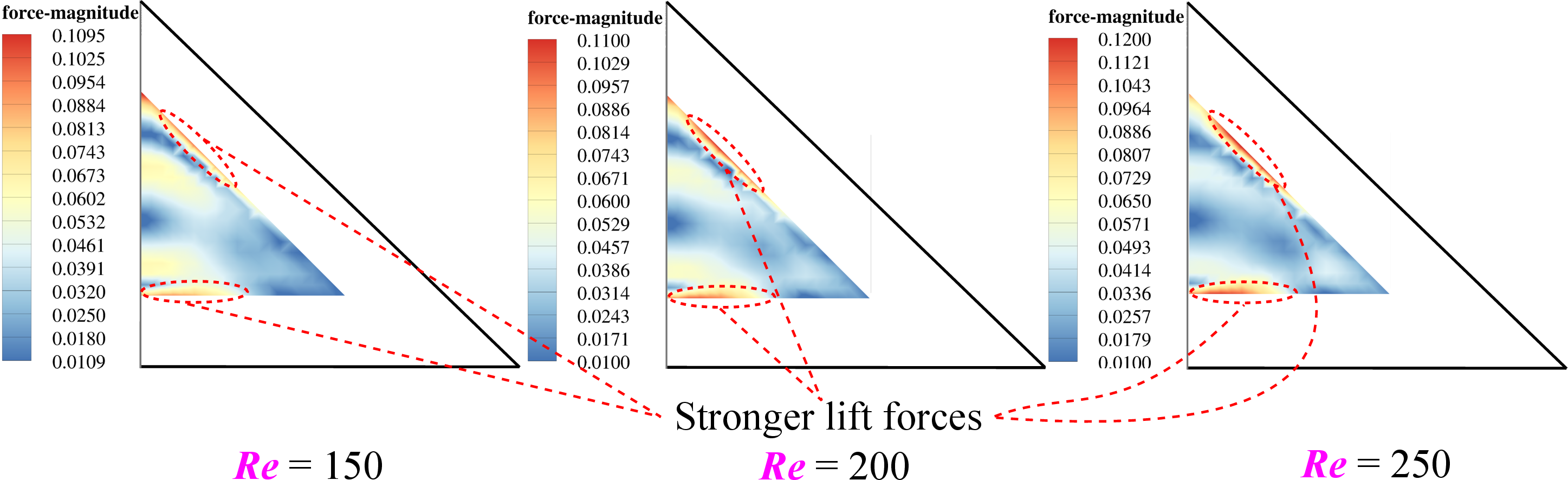}\label{fig:fig9_2}}
\caption{\textit{Off-centre shifting}: \protect\subref{fig:fig9_1} Y and Z coordinates of the bifurcated top off-centre location as function of {\color{magenta} $Re$}, after bifurcation (the red curve represents the z-coordinate, and the blue curve represents the y-coordinate) $\vert$ \protect\subref{fig:fig9_2} \textit{Map} of the magnitude of net lift-forces with varying ${\color{magenta} Re}$ (black lines represent the half-channel boundary (sliced top-down, to scale))}
\label{fig:fig9}
\end{figure}

Upon bifurcation into a 3-centred focusing pattern, particles have been experimentally observed to move downwards and away from the symmetry-plane with an increase in {\color{magenta} $Re$}, and was counter-intuitive to standard results from rectangular/circular channels   \cite{amini2014inertial}, where an increase in {\color{magenta} $Re$} shifted particles closer to top-walls. Firstly, we validate this observation (FIG. \ref{fig:fig9_1}) by noting that the stable off-centre location moves downward along the side-wall of the channel. As seen in the previous section, an increase in {\color{magenta} $Re$} destabilizes the top-centre focusing position due to the combined effect of shear-gradient and wall-lift forces. As seen in FIG. \ref{fig:fig8_3} and \ref{fig:fig9_2}, the lift-magnitude is prominent in regions where the particles actually focus, and this implies a balance between the competing forces. Shifting of particles towards top-walls was primarily due to increase in shear-gradient force being larger than wall-lift forces in rectangular channels. Applying the same argument, wall-lift force increases more than the shear-gradient lift as {\color{magenta} $Re$} increased in the current case. With increase in {\color{magenta} $Re$}, stronger wall-lifts are achieved further down the side-walls, while the shear-gradient lift does not change significantly along the side wall, which results in the downward shifting of the off-centre focusing location. However, thus far we have only qualitatively dealt with the bifurcation trends in the focusing pattern, and focusing shifts thereafter. We now discuss our approach to identify the exact bifurcation point in our simulations. 

\subsubsection{Convergence of focusing patterns: ${\color{magenta} \frac{a}{H}} = 0.4, 60^\circ$-channel, {\color{magenta} $Re$} $= 20$} \label{sssec:convergence}
\begin{figure}
\centering 
\subfloat[]{\includegraphics[width=.22\linewidth]{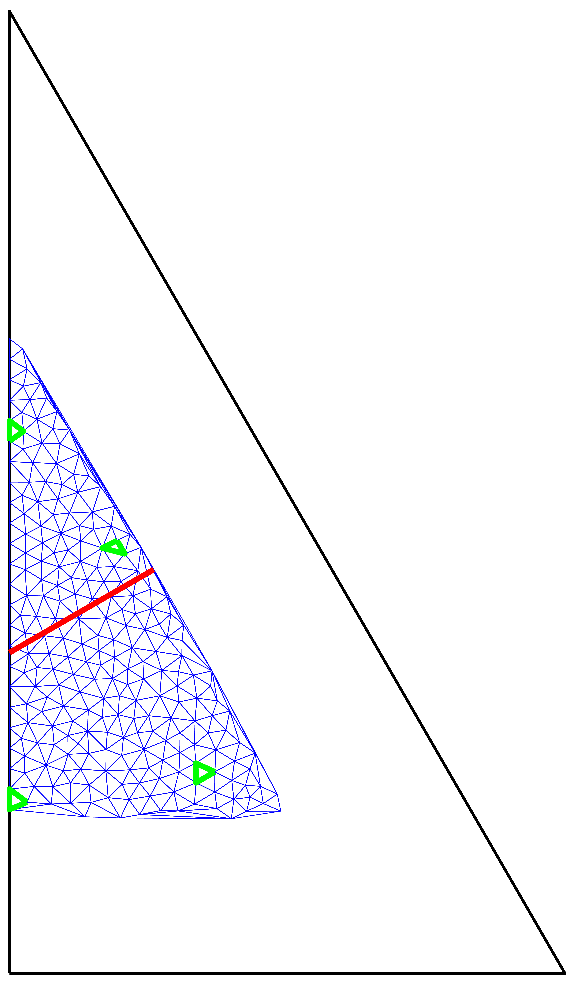}\label{fig:fig10_1}}
\subfloat[]{\includegraphics[width=.22\linewidth]{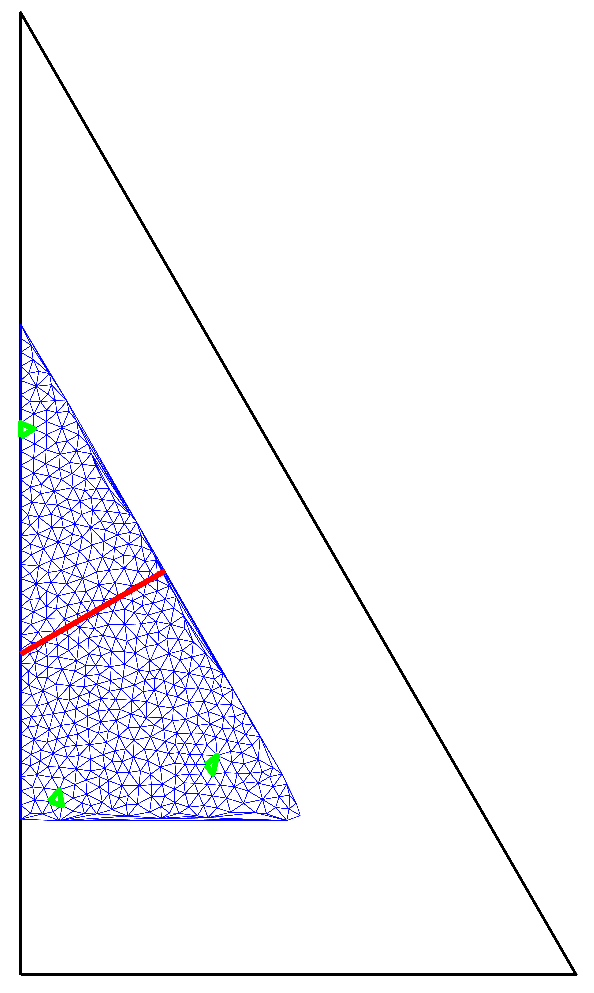}\label{fig:fig10_2}}
\subfloat[]{\includegraphics[width=.22\linewidth]{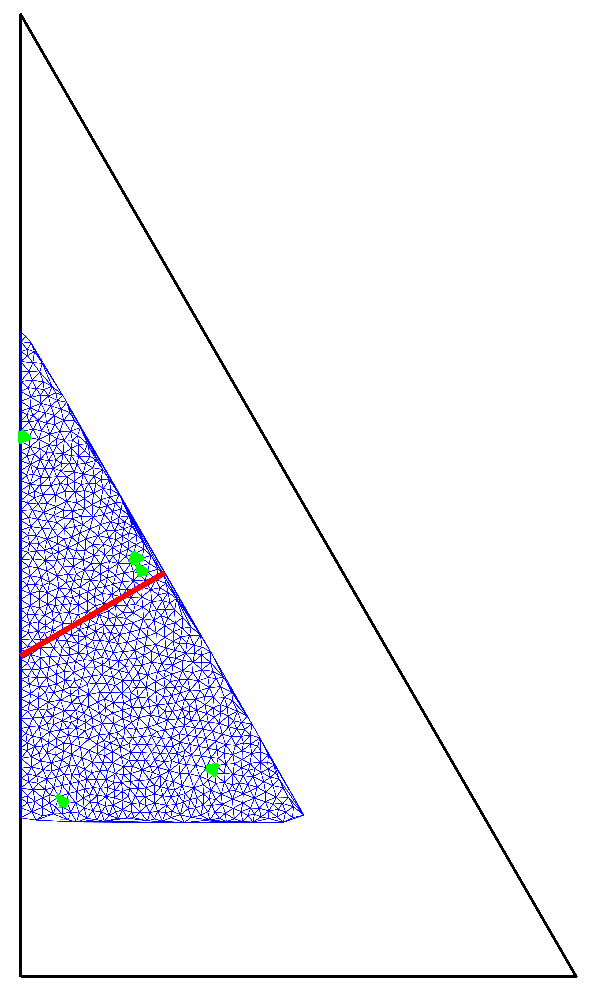}\label{fig:fig10_3}}
\subfloat[]{\includegraphics[width=.22\linewidth]{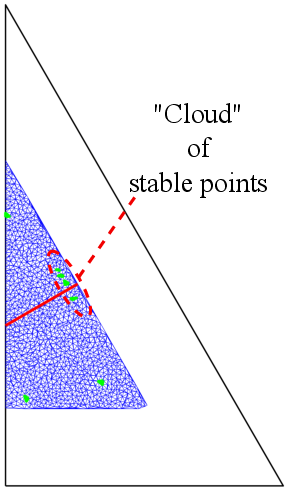}\label{fig:fig10_4}}
\caption{\textit{Force-map convergence}: Focusing patterns for a $60^\circ$-channel, ${\color{magenta} \frac{a}{H}} = 0.4$, {\color{magenta} $Re$} $= 20$ with \protect\subref{fig:fig10_1} 200 particle-locations (coarse) $\vert$ \protect\subref{fig:fig10_2} 500 particle-locations (medium) $\vert$ \protect\subref{fig:fig10_3} 1000 particle-locations (fine) $\vert$ \protect\subref{fig:fig10_4} 1700 particle-locations (finest) (the blue region represents the sampled half-channel particle-locations/triangulation, green-triangles represent elements containing stable points, the solid red line represents a mirror-plane, and black lines represent the half-channel boundary (sliced top-down, to scale))}
\label{fig:fig10}
\end{figure}

We revisit issue \textbf{A} from   \cref{sec:spb} to confirm the onset of bifurcation around ${\color{magenta} Re} = 80$. The prediction of the algorithm directly depends on the input force-maps, and for this reason, we explore convergence in the force-map refinement by monitoring the convergence in the focusing pattern predicted. To do so, we choose an over-sampled region for a $60^\circ$-triangular channel, for a particle-channel size-ratio of, ${\color{magenta} \frac{a}{H}} = 0.4$, and ${\color{magenta} Re} = 20$. We refer to the sampling as over-sampled because for an equilateral triangular-configuration, a unique sampling would entail only ${\frac{1}{3}}^{rd}$ (both mirror and rotational) of the domain whereas we choose a $\frac{1}{2}$-domain here (mirror only). The motivation to do so lies in the fact that the algorithm should ideally produce a focusing pattern which obeys the symmetry in the underlying geometry. Thus by oversampling, we are able to identify that refinement which gives a focusing pattern that satisfies the symmetry of the system. We pick force-map refinements of 200 (coarse), 500 (medium), 1000 (fine), and 1700 (finest) particle-locations, the corresponding focusing patterns for which are shown in FIG. \ref{fig:fig10}.

\justify
For the coarse-refinement (FIG. \ref{fig:fig10_1}), we find that the top and the right-corner position have symmetric consistency, whereas the bottom and the off-centre side focusing position do not respect symmetry with reference to each other since to respect symmetry, either the bottom focusing position would have to lie off-centre, or the off-centre side position should lie on the symmetry-plane (solid red line). For the medium-refinement (FIG. \ref{fig:fig10_2}), the top and right-corner positions satisfy symmetry-requirements, whereas the bottom focusing position has shifted off-centre. However, the off-centre side position previously present has completely vanished, and this instantly violates the underlying symmetry. This also suggests the need for additional refinement to arrive at a converged focusing pattern. With further refinement (FIG. \ref{fig:fig10_3}) the off-centre side focusing position now re-appears in the pattern along with the bottom off-centre position, the top-centre position, and the right-corner position, the latter two of which maintain symmetry. However, one peculiar feature with the fine-map is that the off-centre side focusing position has split into two closely-spaced stable points, instead of being present as just one point. While this point-pair is present on the same side of the symmetry-plane (marked in red) as the bottom off-centre point, there is rotational symmetry among these points. However, the presence of the side focusing locations on one side of the symmetry-plane requires that they also on the other side of the symmetry-plane (marked in red), and since that is not observed here, we discard this refinement as well. Finally, one more level of refinement (FIG. \ref{fig:fig10_4}) produces a focusing pattern which is consistent with itself symmetrically. The top-centre and the right-corner locations maintain their symmetry as with the other refinements. This time, the off-centre side focusing position has split into additional points, which lie on either side of the symmetry-plane (marked in red). This ``cloud of stable points'' now agrees with the bottom off-centre location in terms of rotational symmetry, and with itself in terms of mirror-symmetry. 
\par
\begin{figure}
\centering 
\includegraphics[width=1.\linewidth]{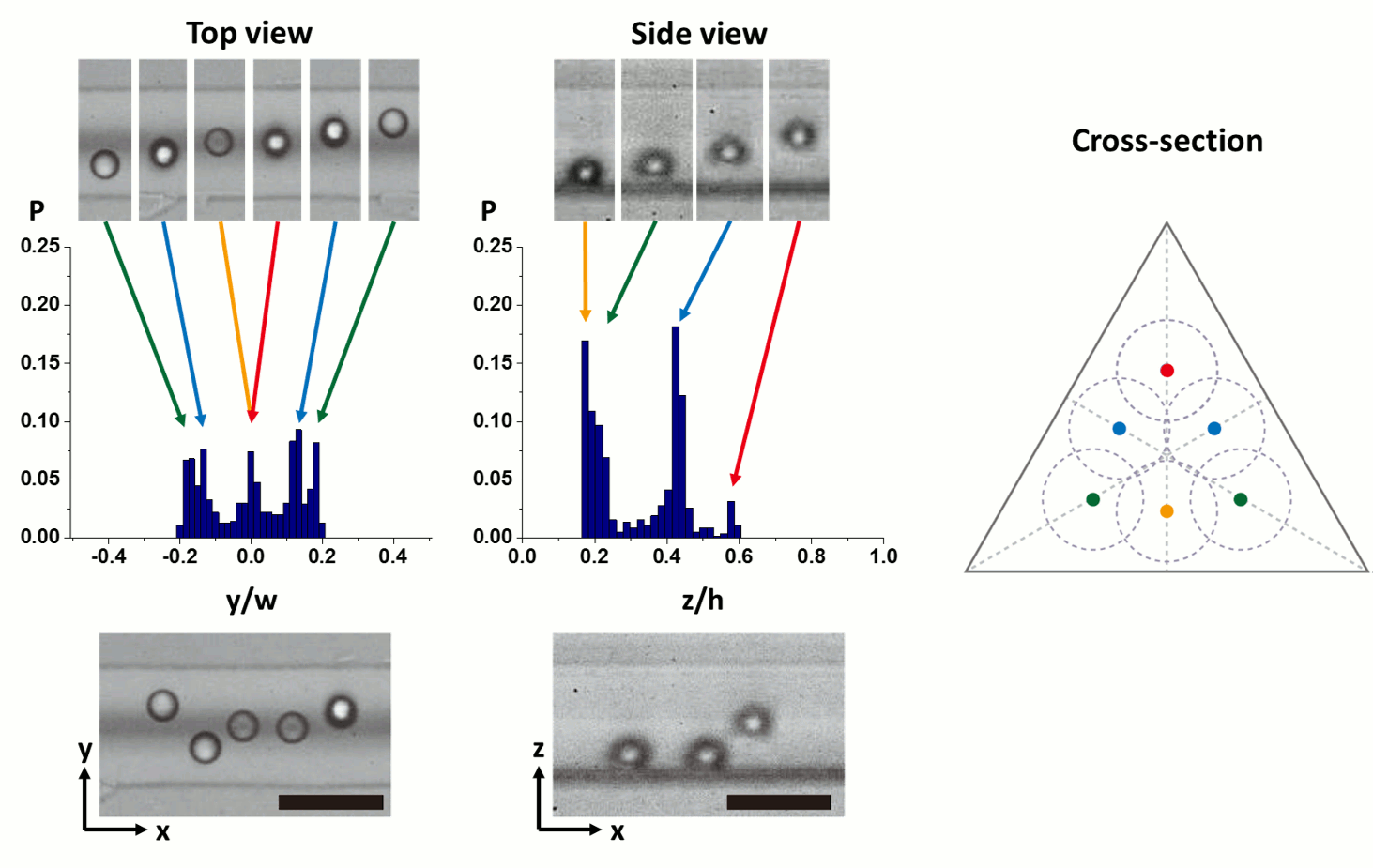}
\caption{\textit{Experimental focusing patterns}: Statistics of particle positions in the equilateral triangular channel (${\color{magenta}\frac{a}{H}}=0.43$) and high-speed capture images indicating each peak position at {${\color{magenta}Re}=20$} from the top view, and side view (scale bar = $50 \mu$m). The schematic represents reconstructed focusing positions in the cross-section. The representative images of the particles at different focusing positions are shown at the top of the statistics. The particles located at different focal depths can be distinguished from the images.
}
\label{fig:fig12}
\end{figure}

This example serves to demonstrate that although mirror and rotational-symmetry may not be present in every configuration, we still need to maintain a certain level of refinement in order to resolve for all stable-points/point-clouds. We obtain a reliable prediction for this example using 1700 particle-locations, which is easily 10-20 times that used in previous studies  \cite{DiCarlo2009},  \cite{liu2015inertial},  \cite{Kim2016},  \cite{amini2014inertial}. This implies the possibility that the force-maps used thus far in literature may not have been entirely accurate except for use in visual interpretation. It is also worth pointing out that particle-scatter plots from conventional confocal microscopy techniques \cite{Kim2016} for experiments usually produces clusters of particles. It is often thought that the broadening of the focusing position is due to the limitations by non-ideal experimental conditions including particle size-variation, defective channel or flow conditions, and limited channel length (for particles to reach equilibrium position). However, the finding of the ``cloud" of stable focusing positions suggests that there is an area (or a line) where the sum of lift forces is vanishingly small and multiple equilibrium positions exist within. Such an area would be observed as the broadening of focusing position in actual experiments. FIG. \ref{fig:fig12} shows experimental findings for this configuration, which support those observed numerically. Focusing positions appear at each corner and the centre of walls in the equilateral triangular channel because of $120^\circ$ rotational symmetry. We found five peaks from the top view and three peaks from the side view. Particle- images with various locations corresponding to each peak and focusing positions in different focal planes can be easily distinguished by the color of particles. Although the bottom corner and bottom face focusing positions are different in the y-axis, they appear as one broad peak in particle distribution from the side view due to close distance in the z-axis, and the ``clouds" observed numerically are evidenced by distributions about corresponding peaks for the side-face focusing positions (blue dots). In addition to refinement, we see that features such as ``clouds" of stable points could possibly explain the early bifurcation in experimental focusing patterns for the $90^\circ$-channel cases considered in   \cref{sec:0.4_90_20_250}. So we increase the refinement in our test-cases to check our hypothesis of the presence of a similar feature there, and this will conclude if the early bifurcation is actually indeed a ``cloud" of stable points around the top-centre position. 

\begin{figure}[!htpb]
\centering 
\subfloat[]{\includegraphics[width=.4\linewidth]{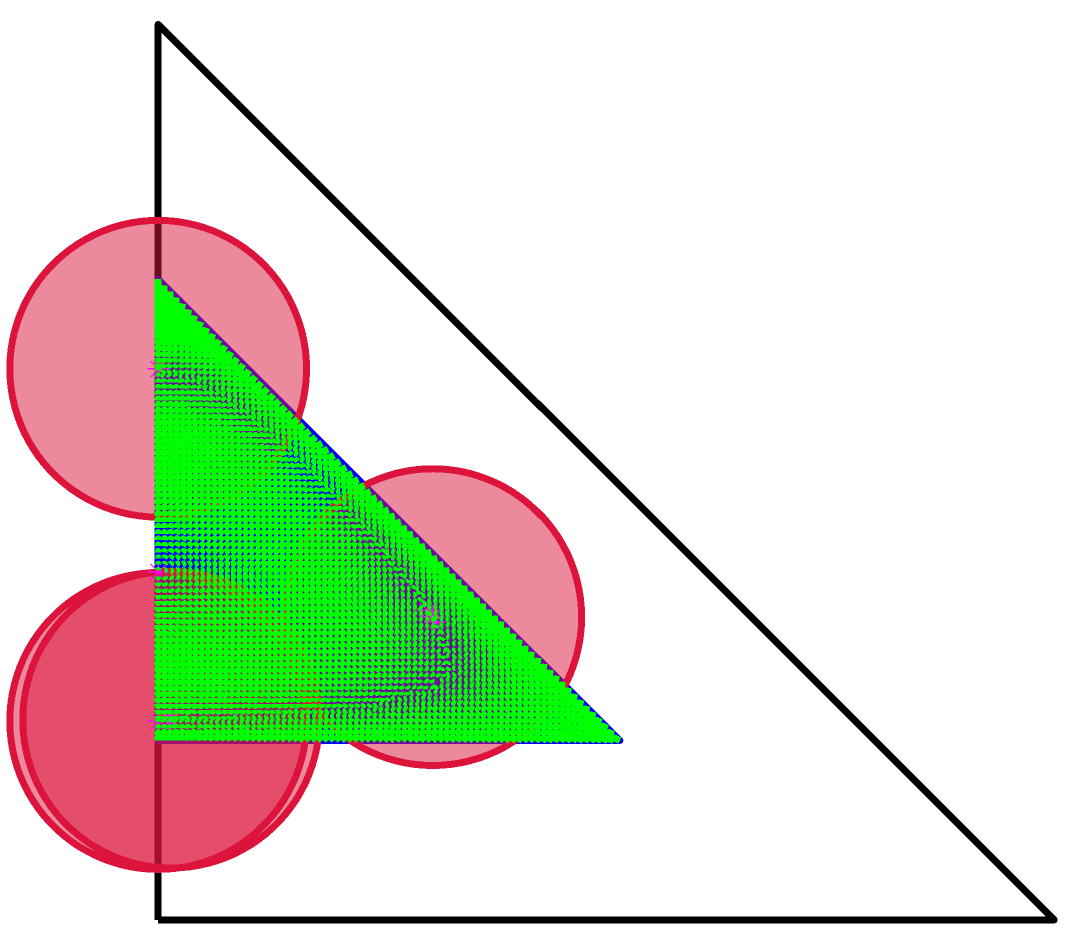}\label{fig:fig11_1}}
\hfill
\subfloat[]{\includegraphics[width=.58\linewidth]{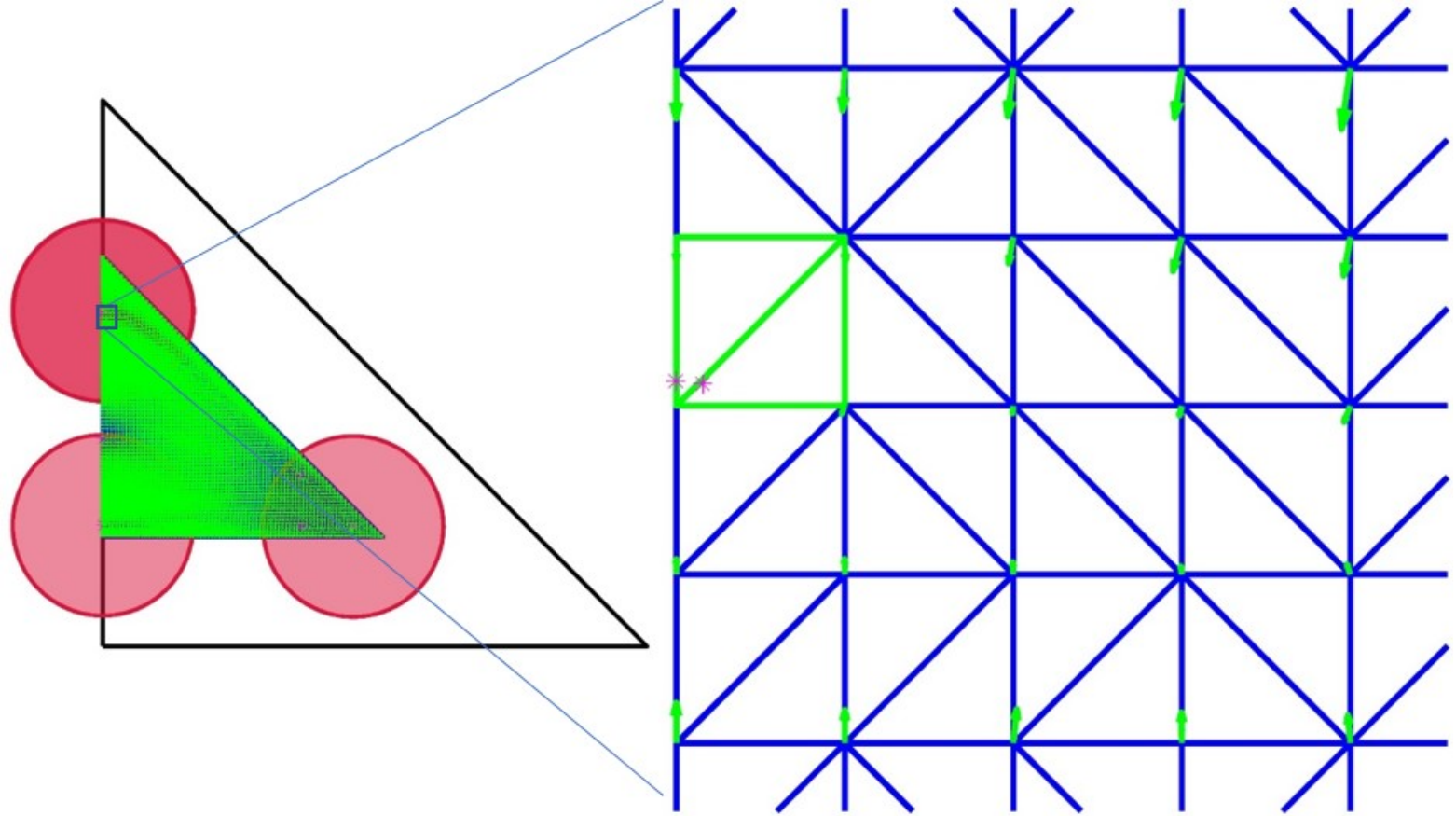}\label{fig:fig11_2}}
\subfloat[]{\includegraphics[width=.3\linewidth]{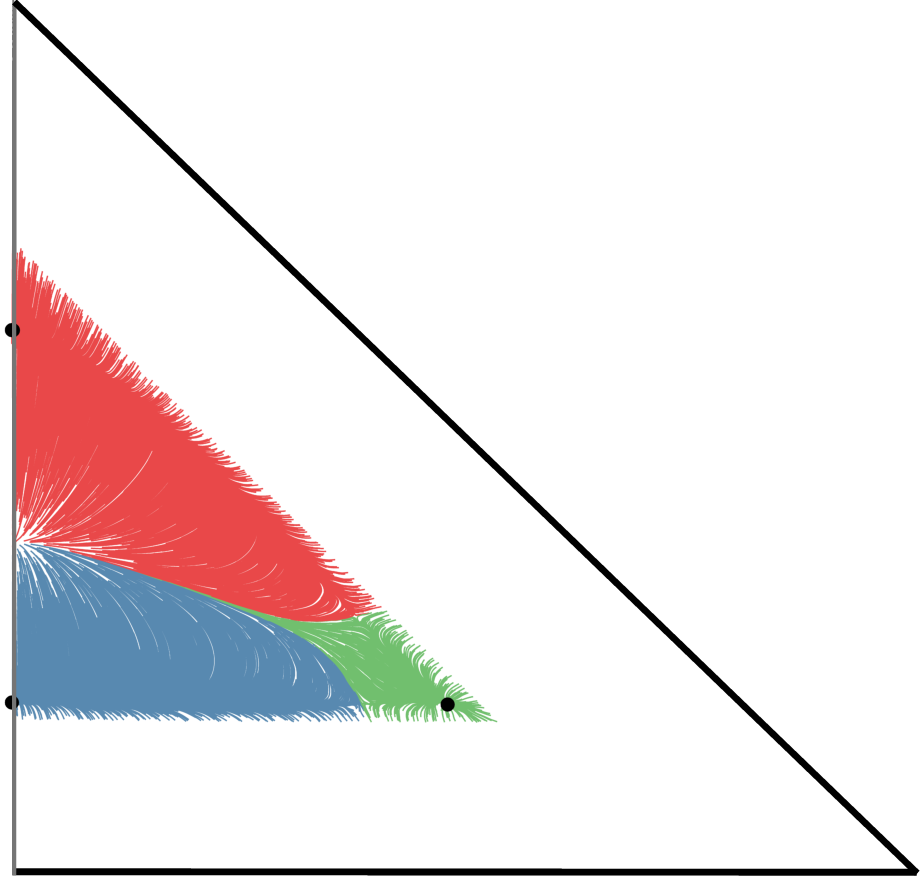}\label{fig:fig11_3}}
\hfill
\subfloat[]{\includegraphics[width=.58\linewidth]{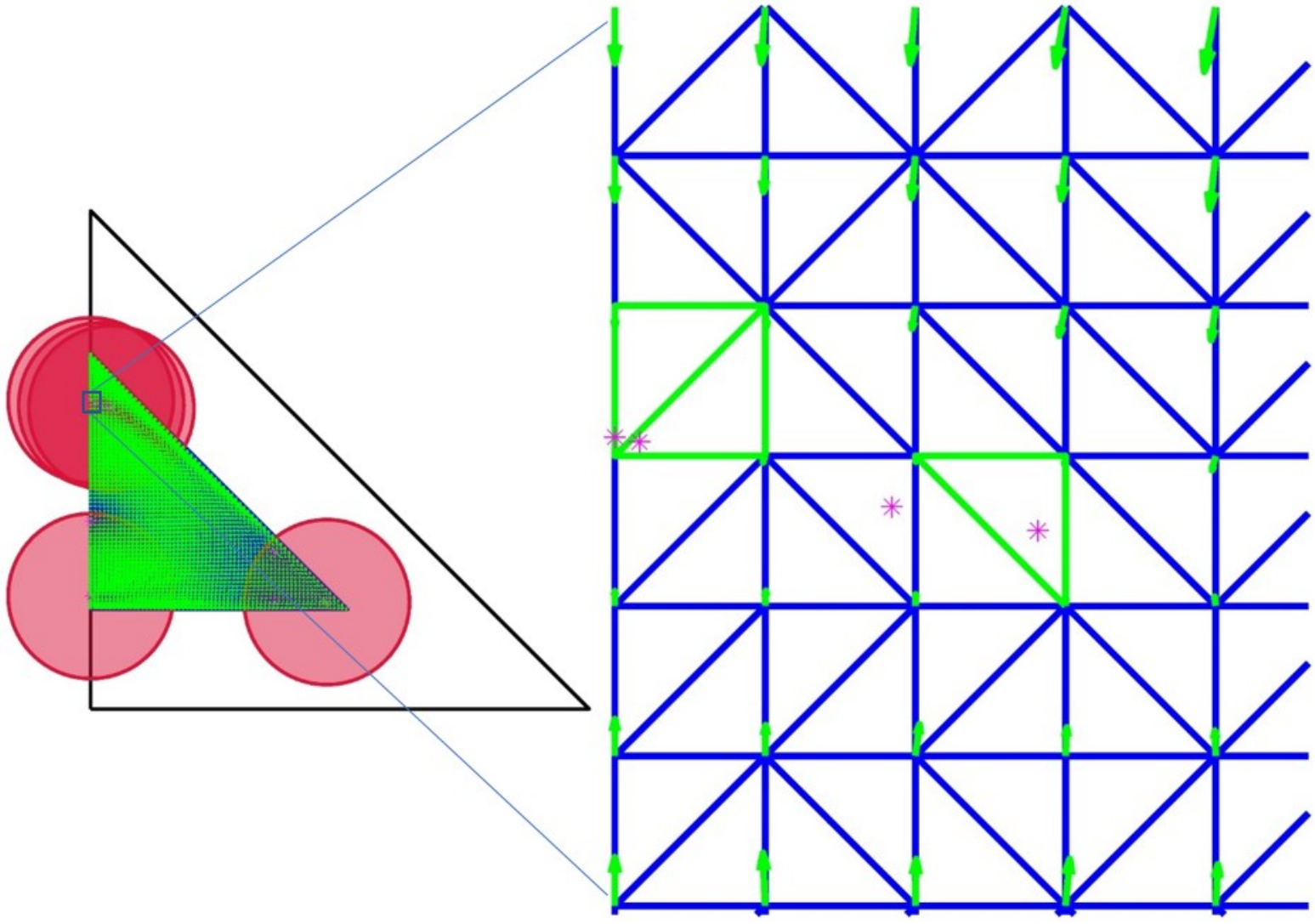}\label{fig:fig11_4}}
\subfloat[]{\includegraphics[width=.3\linewidth]{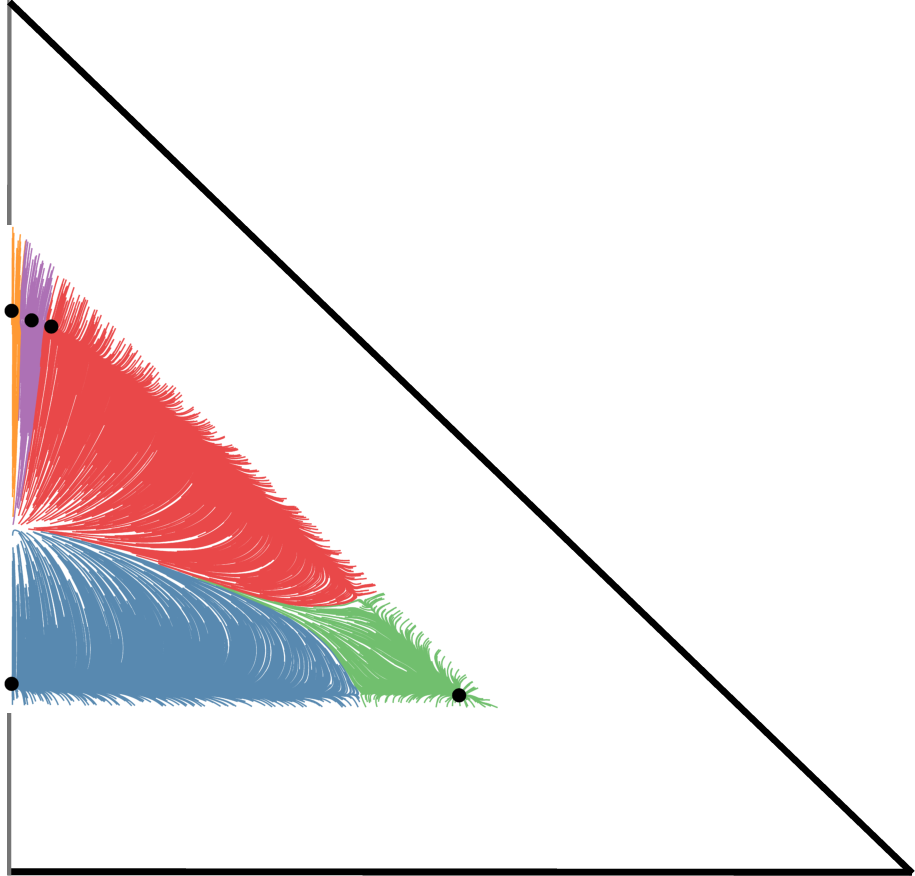}\label{fig:fig11_5}}
\caption{\textit{Revised bifurcation with refined force-maps}: for {\color{magenta} $\frac{a}{H}$}$ = 0.4$ in a $90^\circ$-channel ${\color{magenta} Re} =$ \protect\subref{fig:fig11_1} 20 (focusing pattern) $\vert$ \protect\subref{fig:fig11_2} 90 (focusing pattern) $\vert$ \protect\subref{fig:fig11_3} 90 (basins of attraction) \protect\subref{fig:fig11_4} 100 (focusing pattern) $\vert$ \protect\subref{fig:fig11_5} 100 (basins of attraction) (green arrows represent force-maps, the blue region represents the sampled half-channel particle-locations/triangulation, magenta-asterisks represent all equilibrium locations, green-triangles represent elements containing stable points, red circles represent the particle, and black lines represent the half-channel boundary (sliced top-down, to scale); basins demarcated by color - the black dots represent stable locations; note that the basins shown here span ONLY the sampled region, which is smaller than the half-channel cross-sectional area)}
\label{fig:fig11}
\end{figure}

\justify
FIG. \ref{fig:fig11} shows a second campaign of focusing pattern-predictions for the $90^\circ$-channel, for ${\color{magenta} \frac{a}{H}} = 0.4$, and ${\color{magenta} Re} = 20, 90,$ and $100$, with refinement commensurate to the $60^\circ$-channel-case employed for testing convergence, with ${\color{magenta} \frac{a}{H}} = 0.4$, and ${\color{magenta} Re} = 20$. For the refined force-maps, we see the onset of bifurcation at ${\color{magenta} Re} = 90$ (FIG. \ref{fig:fig11_2}-\ref{fig:fig11_3}), which is drastically different than the bifurcation-point of ${\color{magenta} Re} = 120$ from the coarse-maps. This observation is substantially closer to the bifurcation-point seen in experiments. Although it is not apparent from the basins of attraction for ${\color{magenta} Re} = 90$ (FIG. \ref{fig:fig11_3}), it can be seen for ${\color{magenta} Re} = 100$ that the off-centre top focusing location has the largest basin (FIG. \ref{fig:fig11_5}). Our hypothesis that the early bifurcation could be explained by a ``cloud'' of stable points thus stands confirmed, and further refinement might reveal bifurcation at a much lower ${\color{magenta} Re}$ and/or a wider spread within the ``cloud'', but the idea was to establish the need for a certain level of refinement to observe and explain crucial and intricate effects such as bifurcation with reasonable accuracy. This novel feature of ``clouds'' of stable points in conjunction with basins of attraction serves to tune numerical predictions with a higher degree of confidence for real-world experiments. For ${\color{magenta} Re} = 20$ (FIG. \ref{fig:fig11_1}), however, we see that the focusing pattern is similar to that obtained with a coarse-sampling (FIG. \ref{fig:fig6_1}). This also suggests that the off-centre focusing position so obtained is not a numerical artefact but a consequence of the underlying physics. Additionally, we observe that the focusing pattern-prediction seems to directly depend on the regime of focusing (near-bifurcation vs. far) - for cases like the $90^\circ$-channel with ${\color{magenta} \frac{a}{H}} = 0.4$, at ${\color{magenta} Re} \le 80$, and ${\color{magenta} Re} \ge 120$, the focusing pattern is well-converged using coarse refinements. But for {\color{magenta} $Re$} close to bifurcation ($= 90, 100$), higher refinements are necessary to capture the focusing trends well, so, for any general case, a convergence analysis is recommended. 

\justify
In this regard, the proposed stability algorithm for generating focusing patterns provides a valuable measure of convergence for the force-map sampling as there are no such metrics available till date. On another note, we hypothesize that bifurcation does not occur abruptly at a {\color{magenta} $Re$} but that it takes place gradually in two stages: the first stage ($90 \le {\color{magenta} Re} < 120$), where the top focusing-point splits up into a localized ``cloud'' of stable locations, and the second stage (${\color{magenta} Re} \ge 120$), where the off-centre focusing location has a well-bifurcated, distinct identity on the coarse-level (which may further have a ``stable'' cloud of its own). Finally, we note that the suggested algorithm is capable of comprehensively finding all possible equilibrium points which gives us possibility of manipulating flow parameters to stabilize unstable points as in a high-aspect rectangular channel, where increasing ${\color{magenta} Re}$ stabilizes the equilibrium points towards the short-faces. 

\section{Conclusions}\label{sec:concl}
We have formulated an automated, geometry-and-{\color{magenta} $Re$}-agnostic computational framework based on linear-stability analysis for predictions of hydrodynamic particle-focusing patterns. We validate our code against known experimental results of focusing in rectangular channels, and relatively recent findings on $90^\circ$-triangular channels, and later apply it to validate/explain focusing patterns observed in $90^\circ$-triangular channels for a range of {\color{magenta} $Re$}. The two main features of the focusing mainly dealt with in this regard are: the stable-point bifurcation for the top focusing position, as well as the downward shift of particles after bifurcation. From a numerical standpoint, we find that eigenvalues give us a local measure of the stability of a particular stable point, but the final pattern is governed by the basins of attraction, which is a global measure. The general trends of the bifurcation are well-matched with experiments upon including basins of attraction into our analysis, using standard force-map refinements. However, for a higher refinement, ``clouds" of stable points are detected, which could indicate a local spread in the focusing locations, and this argument figures well into explaining the bifurcation in addition to the underlying physics. The ``clouds" of stable points also serve to illustrate the idea that the bifurcation seen in the test-cases for the $90^\circ$-channel (particle-channel size-ratio of 0.4) happens gradually in two stages: for initial ${\color{magenta} Re}: 90-120$ the bifurcation appears as a ``cloud" of stable locations around the top-centre focusing location, whereas for ${\color{magenta} Re} \ge 120$, the bifurcated off-centre top position is distinct (which may have a local cloud of its own). We also think that a field-based approach for the lift-forces enables us to analyze the migration-effect better, since we can better quantify the directionality of forces, basins of attraction, which might be non-trivial for an intuitive analysis of non-rectangular geometries, where the channel walls are non-orthogonal. From a computational standpoint, the proposed algorithm utilizes fairly popular subroutines, and the work-flow discussed in this paper should be straightforward for implementation by the interested researcher for additional study. Additionally, we hope that our current attempt at a stand-alone tool is a first-step for calculation of focusing patterns in a large phase-space of cross-sectional geometries, particle-channel size-ratios, {\color{magenta} $Re$}'s, and so on, to create a library of focusing patterns. These should pave way for creating so-called transition-maps (governed by the basins of attraction for corresponding configurations), ultimately serving to design novel devices for generating tailored particle-streams. Lastly, we see the possibility of adopting global exploration-based metamodelling strategies, to reduce computational effort in case of high-refinements. 


\bibliographystyle{apsrev4-1}
%

\bibliography{main.bbl}

\pagebreak

\appendix
\renewcommand\thefigure{\thesection.\arabic{figure}}
\section{Triangulation and basis functions} \label{sec:app1}
\setcounter{figure}{0}

\begin{figure}
\centering
\includegraphics[width=1\linewidth]{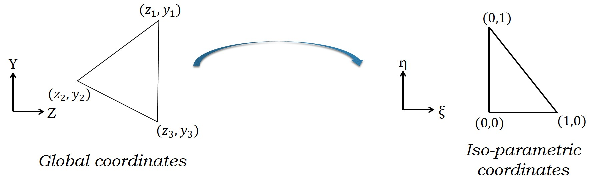}
\caption{\textit{Isoparametric representation:} conversion from global system to local system}
\label{fig:fig3}
\end{figure}

For each element in the force-map triangulation, linear basis functions, $N_i$, can be used to interpolate forces within the element (FIG. \ref{fig:fig3}) using pre-computed nodal forces, $\{(F_{z_1},F_{y_1}), (F_{z_2},F_{y_2}), (F_{z_3},F_{y_3})\}$, at the vertices, $\{(z_1,y_1), (z_2,y_2), (z_3,y_3)\}$, resp., as,

\begin{gather}
z(\xi,\eta) = {\sum_{i=1}^{3}z_iN_i(\xi,\eta)} = z_1\eta + z_2(1-\xi-\eta) + z_3\xi, \nonumber \\
y(\xi,\eta) = {\sum_{i=1}^{3}y_iN_i(\xi,\eta)} = y_1\eta + y_2(1-\xi-\eta) + y_3\xi, \nonumber \\
F_z(\xi,\eta) = {\sum_{i=1}^{3}F_{z_i}N_i(\xi,\eta)} = F_{z_1}\eta + F_{z_2}(1-\xi-\eta) + F_{z_3}\xi, \nonumber \\
F_y(\xi,\eta) = {\sum_{i=1}^{3}F_{y_i}N_i(\xi,\eta)} = F_{y_1}\eta + F_{y_2}(1-\xi-\eta) + F_{y_3}\xi \label{eq:eq11}
\end{gather}

\justify
where, the independent variables are represented in the isoparametric space rather than in the global space (FIG. \ref{fig:fig3}).

\section{Linearization around an equilibrium point} \label{sec:app2}

The linearization for perturbation, $\Delta\boldsymbol{X}$, about an equilibrium point, $\boldsymbol{X_0}$, can be arrived at by expanding Taylor series to first order as follows:

\begin{gather}
\boldsymbol{f(X_0}+\Delta\boldsymbol{X)} = \boldsymbol{f(X_0)} + \bigg(\frac{\mathrm{\partial}\boldsymbol{f}}{\mathrm{\partial}\boldsymbol{X}}\bigg)_{\boldsymbol{X_0}}\Delta\boldsymbol{X} + \underbrace{O(\Delta\boldsymbol{X}^2)}_{\text{higher-order terms}} \nonumber \\
\frac{\mathrm{d}\boldsymbol{X_0}}{\mathrm{d}t}+\frac{\mathrm{d(\Delta}\boldsymbol{X)}}{\mathrm{d}t} \approx \boldsymbol{f(X_0)} + \bigg(\frac{\mathrm{\partial}\boldsymbol{f}}{\mathrm{\partial}\boldsymbol{X}}\bigg)_{\boldsymbol{X_0}}\Delta\boldsymbol{X} \nonumber
\end{gather}

Thus,
\begin{gather}
\frac{\mathrm{d(\Delta}\boldsymbol{X)}}{\mathrm{d}t} \approx \bigg(\frac{\mathrm{\partial}\boldsymbol{f}}{\mathrm{\partial}\boldsymbol{X}}\bigg)_{\boldsymbol{X_0}}\Delta\boldsymbol{X} \nonumber
\end{gather}

\justify
We can always centre the origin at the equilibrium point without altering its stability, in which case the above system assumes the following form,

\begin{gather}
\Delta\boldsymbol{X} \equiv \boldsymbol{X} \nonumber \\
\frac{\mathrm{d}\boldsymbol{X}}{\mathrm{d}t} = \bigg(\frac{\mathrm{\partial}\boldsymbol{f}}{\mathrm{\partial}\boldsymbol{X}}\bigg)_{\boldsymbol{X_0}}\boldsymbol{X} \label{eq:eq4}
\end{gather}

\justify
Comparing equations \eqref{eq:eq3} and \eqref{eq:eq4}, we see that the Jacobian matrix, ${A}$, around any equilibrium point, $\boldsymbol{X_0}$, is given by:

\begin{gather}
{A} = \bigg(\frac{\mathrm{\partial}\boldsymbol{f}}{\mathrm{\partial}\boldsymbol{X}}\bigg)_{\boldsymbol{X_0}} \label{eq:eq5}
\end{gather}

\end{document}